\documentclass[smallcondensed]{svjour3}       

\usepackage{paralist}
\usepackage{booktabs}
\usepackage[linesnumbered,noend, noline]{styles/algorithm2e}
\usepackage{algpseudocode}
\usepackage{url}
\usepackage[gen]{eurosym}
\makeatletter
\g@addto@macro{\UrlBreaks}{\UrlOrds}
\makeatother
\usepackage{breakurl}
\newcommand{\urls}[1]{{\scriptsize\url{#1}}}

\def\it{\textit}
\def\tr{\textrm}
\newcommand{\ib}[1]{{\textbf {\textit { #1}}}}

\def\tt{\ft}
\usepackage{wrapfig}

\def\bf{\textbf}
\def\eq {Equation~}

\def\fig {Figure~}
\def\figs {Figures~}
\usepackage{subfig}

\usepackage{balance}  
\usepackage{graphics} 
\usepackage{booktabs}
\usepackage{ccicons}

\usepackage{url}
\usepackage{fancybox}
\usepackage{multirow}
\usepackage{flushend}
\usepackage{booktabs}
\usepackage{tabularx}
\usepackage{comment}
\usepackage{array}
\usepackage[flushleft]{threeparttable}
\usepackage{mdframed}
\graphicspath{{}{images/}{dia/}}
\DeclareGraphicsExtensions{.pdf,.png}

\usepackage{pgfplots}
\usepackage{pgfplotstable}

\usepgfplotslibrary{statistics}
\usepackage{pgfplots}
\usepackage{pgfplotstable}

\usepgfplotslibrary{statistics}

\usepackage{scalefnt}

\usepackage{tabularx}
\usepackage{lipsum}
\usepackage{array}

\usepackage{tcolorbox}
\usepackage{multirow}

\usetikzlibrary{shapes.gates.logic.US,trees,positioning,arrows,shadows}
\usetikzlibrary{shapes.multipart}
\usetikzlibrary{shapes,arrows}

\usepackage{}
\usepackage{styles/pgf-pie}

\newif\ifpienumberinlegend
\pgfkeys{/number in legend/.code=
    \expandafter\let\expandafter\ifpienumberinlegend
    \csname if#1\endcsname
    \ifpienumberinlegend

    \def\beforenumber##1\afternumber{}%
    \fi,
    /number in legend/.default=true
}

\makeatletter
\pgfplotsset{
    boxplot/hide outliers/.code={
        \def\pgfplotsplothandlerboxplot@outlier{}%
    }
}
\makeatother
\pgfplotsset{
    bar group size/.style 2 args={
        /pgf/bar shift={%
                -0.5*(#2*\pgfplotbarwidth + (#2-1)*\pgfkeysvalueof{/pgfplots/bar group skip})  + 
                (.5+#1)*\pgfplotbarwidth + #1*\pgfkeysvalueof{/pgfplots/bar group skip}},%
    },
    bar group skip/.initial=2pt,
    plot 0/.style={black,fill=black!10!white,mark=none},%
    plot 1/.style={black,fill=black!50!white,mark=none},%
    plot 2/.style={black,fill=black!80!white,mark=none},%
}

\usepackage{listings}
\usepackage{hyperref}

\newcounter{o}
\newcounter{d}
\newcounter{t}
\usepackage{tikz}
\usepackage{styles/pgf-pie}
\usetikzlibrary{positioning,shadows}
\newif\ifpienumberinlegend
\pgfkeys{/number in legend/.code=
    \expandafter\let\expandafter\ifpienumberinlegend
    \csname if#1\endcsname
    \ifpienumberinlegend

    \def\beforenumber##1\afternumber{}%
    \fi,
    /number in legend/.default=true
}
\definecolor{1c1}{RGB}{188,162,6}
\definecolor{1c2}{RGB}{137,129,80}
\definecolor{1c3}{RGB}{239,167,31}
\definecolor{1c4}{RGB}{88,194,241}
\definecolor{1c5}{RGB}{6,180,188}

\tikzset{mynode/.style={draw=white,solid,circle,fill=green,inner sep=1pt, thick,
text=black}}
\tikzset{arrow line/.style={dashed, line width= 2.5pt, color=#1}}

\usepackage[numbers]{natbib}

\def\bf{\textbf}
\def\eq {Equation~}

\def\fig {Figure~}
\def\figs {Figures~}
\def\tbl {Table~}

\def\sec {Section~}

\def\it{\textit}
\def\tr{\textrm}
\def\tt{\mct}

\newcommand{\qu}[1]{{\it{``#1''}}}

\usepackage{paralist}
\usepackage{hyperref}
\usepackage{soul}

\normalsize

\usepackage{enumitem}

\newcommand{\nd}{\vspace{1mm}\noindent}
\usepackage{caption}
\usepackage{tikz}

\usepackage[figuresright]{rotating}
\usepackage{pdflscape}
\lstdefinestyle{inlinecode}{basicstyle={\ttfamily\scriptsize\bfseries}}

\usepackage{tcolorbox}
\newcommand{\emt}[1]{\emph{``#1''}}
\usepackage{paralist}
\usepackage[outercaption]{sidecap}    
\usepackage{amssymb}
\usepackage{pifont}

\usepackage{enumitem}
\hypersetup{
    colorlinks=true,
    linkcolor=blue,
    filecolor=magenta,      
    urlcolor=cyan,
}
\newtcolorbox
{mybox}[2][]{colbacktitle=red!10!white,
colback=blue!10!white,coltitle=black!70!black,
title={#2},fonttitle=\bfseries,#1}

\usepackage{xcolor}
\definecolor{ao(english)}{rgb}{0.0, 0.5, 0.0}

\newcommand{\rev}[1]{#1}

\begin{document}
\title{An Empirical Study of Deep Learning Sentiment Detection Tools for Software Engineering in Cross-Platform Settings}
\titlerunning{Deep Learning Sentiment Detection Tools for SE in Cross-Platform Settings}
\author{Gias Uddin, Md Abdullah Al Alamin, and Ajoy Das}
\institute{Gias Uddin, Md Abdullah Al Alamin, and Ajoy Das \at
              DISA Lab, University of Calgary,
              \email{\{gias.uddin,mdabdullahal.alamin,ajoy.das\}@ucalgary.ca}}
\date{}
\maketitle

\sloppy

\begin{abstract}
Sentiment detection in software engineering (SE) has shown promise to support
diverse development activities. However, given the diversity of SE platforms
(e.g., code review, developer forum), SE-specific sentiment detection tools may
suffer in performance in cross-platform settings, i.e., trained in one dataset
created from a platform (e.g., code review) and then tested in dataset curated
from another platform (e.g., Jira issues). Recently deep learning (DL)-based
SE-specific sentiment detection tools are found to offer superior performance
than shallow machine learning (ML) based/rule-based tools. However, it is not
known how the DL tools perform in cross-platform settings.
In this paper, we study whether SE-specific DL sentiment detectors are more
effective than shallow ML-based/rule-based sentiment detectors in cross-platform
settings.
In three datasets, we study three DL tools (SEntiMoji, BERT4SEntiSE,
RNN4SentiSE) and compare those against three baselines: two shallow learning
tools (Senti4SD, SentiCR) and one rule-based tool (SentistrengthSE). The
datasets were previously used by Novielli et
al.\cite{Novielli-SEToolCrossPlatform-MSR2020} to study the cross-platform
performance of the baselines.
We find that \begin{inparaenum}[(1)]
    \item The deep learning SD tools for SE, BERT4SentiSE outperform other
    supervised tools in cross-platform settings in most cases, but then the tool
    is outperformed by the rule-based tool SentistrengthSE in most cases. 
    \item BERT4SentiSE outperforms SentistrengthSE by large margin in within
    platform settings across the three datasets and is only narrowly
    outperformed by SentiStrengthSE in four out of the six cross-platform
    settings. This finding offers hope for the feasibility to further improve a
    pre-trained transformer model like BERT4SentiSE in cross-platform settings.
    \item The two best-performing deep
    learning tools (BERT4SentiSE and SentiMoji) show varying level performance
    drop across the three datasets. We find that this inconsistency is mainly
    due to the ``subjectivity in annotation by the human coders''. \end{inparaenum} Further investigation shows that the six datasets are not very much different from 
    each other in terms of textual contents.
    Therefore,
    performance improvement for the studied supervised tools in
    cross-platform settings may require the fixing of the
    datasets (e.g., handling inconsistency in polarity labeling).
\end{abstract}

\section{Introduction}\label{sec:introduction}
Similar to Natural Language Processing (NLP), textual contents in software
engineering (SE) can broadly be divided into two types: facts and
opinions~\cite{liu-sentimentanalysis-handbookchapter-2010, Uddin-OpinionValue-TSE2019, Uddin-SurveyOpinion-TSE2019}. Facts are objective
expressions (e.g., ``I use this tool'') while opinions are subjective
expressions (e.g., ``I like this tool''). While opinions can have many forms,
the primary focus in SE research has been on sentiment detection, i.e.,
classifying a sentence/post to one of three polarity labels, positive, negative,
and neutral (i.e., absence of positive/negative polarity).

Research in SE finds that opinions are key determinants to many of the
activities in SE like analyzing developers burn
outs~\cite{Mika-MiningValenceBurnout-MSR2016}, assisting in developer API
selection~\cite{Uddin-OpinerEval-ASE2017,Uddin-OpinerReviewToolDemo-ASE2017},
determining the quality of shared knowledge in online
forums~\cite{Uddin-OpinerAPIUsageScenario-TOSEM2021,Uddin-UsageScenarioMining-IST2020,Uddin-MineAPIAspectReviews-PolyTech2017,Uddin-SurveyOpinion-TSE2019}.
As such, accurate detection of sentiment polarity in SE artifacts is essential,
so that the right information is collected to support analysis.

Sentiment detection for SE domain can be insufficiently accurate for not correctly understanding software domain specific jargons while using
off-the-shelf
tools~\cite{Jongeling-SentimentNegative-EMSE2017,Lin-SentimentDetectionNegativeResults-ICSE2018}.
As such, several SE-specific sentiment detection (denoted as `SD' for brevity
from hereon) tools are developed based on wide-ranging techniques like
\textit{rule-based} (e.g., SentistrengthSE~\cite{Islam-SentistrengthSE-MSR2017},
Opiner~\cite{Uddin-OpinerEval-ASE2017},
POME~\cite{Lin-PatternBasedOpinionMining-ICSE2019}), \textit{supervised shallow learning based}
(e.g., Senti4SD~\cite{Calefato-Senti4SD-EMSE2017},
SentiCR~\cite{Ahmed-SentiCRNIER-ASE2017}), and \textit{supervised deep learning based} (e.g.,
RNN4SentiSE~\cite{Biswas-SentiSEWordEmbedding-MSR2019},
BERT4SentiSE~\cite{Biswas-ReliableSentiSEBERT-ICSME2020},
SentiMoji~\cite{Chen-SentiEmoji-FSE2019}).

The SE-specific SD tools are found to outperform off-the-shelf SD tools in
several SE sentiment datasets~\cite{Chen-SentiEmoji-FSE2019,sentiCR_2017}. Each dataset is developed for a specific platform 
(e.g., Jira code reviews, Stack Overflow discussions, etc.).
The diversity of SE platforms can still make it challenging to apply a
SE-specific SD tool trained in one SE platform (e.g., Jira code review dataset) to apply to
another platform (e.g., Stack Overflow developer discussions). \rev{Recently, Novielli et
al.~\cite{Novielli-SEToolCrossPlatform-MSR2020,Novielli-SESentimentCP-EMSE2021} investigated several
supervised shallow learning SE-specific SD tools in cross-platform settings,
i.e., train in one SE sentiment dataset and then test on a different SE
sentiment dataset. They find that the supervised SD tools suffer a decrease in
performance in cross-platform settings. However, it is not known whether we
could observe similar results while using the SE-specific deep learning  SD tools
like BERT4SentiSE~\cite{Biswas-ReliableSentiSEBERT-ICSME2020}, SentiMoji~\cite{Chen-SentiEmoji-FSE2019}, and RNN4SentiSE~\cite{Biswas-SentiSEWordEmbedding-MSR2019}.}

\rev{An insight into the performance of SE-specific deep learning SD
tools in cross-platform settings and whether and how the deep learning tools outperform other tools is necessary for the following reasons: 
\begin{enumerate} 
\item \bf{Consistency in Superiority of Performance.} Novielli et al.~\cite{Novielli-SEToolCrossPlatform-MSR2020} reported that shallow machine learning SD tools for SE show a drop in 
their performance in cross-platform settings. 
  The deep learning SD tools for SE are found to be superior to the shallow learning SD tools in various studies for within-platform settings 
  (i.e., when trained and tested on an individual datasets)~\cite{Biswas-ReliableSentiSEBERT-ICSME2020,Chen-SentiEmoji-FSE2019}. 
  The training of a SD tool for various SE datasets can be difficult and sometimes impossible, when a labeled dataset for a new SE platform is not available. As such, it 
  is desirable to know whether the deep learning SD tools for SE can offer better performance than the shallow learning SD tools for SE in cross-platform settings, because then 
  the deep learning tools can still be preferable to the shallow learning SD tools in cross-platform settings, especially when the retraining of the tools in such cross-platform 
  settings may not be possible. As such,
a reasonable expectation is that the deep learning tools would also outperform the shallow learning SD tools in the cross-platform setting. 
Any deviation from this expectation may indicate issues with the underlying models or datasets like overfitting of the models for within-platform settings, which can
then contribute to their drop of performance for cross-platform settings.
  \item \bf{Pre-Training Setup of Models.}  By design, all the deep learning SD tools for SE that we studied are built based on philosophy of pre-training transformer models. This means that each of the tools 
  is already pre-trained on a large corpus of domain-independent textual datasets to learn the underlying general syntax and contexts of human conversations. 
  For example, BERT4SentiSE is built on top of the advanced
language-based pre-trained transformer model
BERT~\cite{Delvin-BERTArch-Arxiv2018}.
  The tools then are 
  trained based on SE specific datasets by adding extra layers on top of the pre-trained domain-independent layers. In contrast, the shallow 
  learning SD tools are not pre-trained, i.e., the tools learn about underlying contexts by simply training on the SE specific datasets. 
  This means that the deep learning SD tools are intuitively more knowledgeable than the shallow 
  learning SD tools, due to how they are pre-trained and trained. This dual setup of pre-training and training of 
  deep learning SD tools led us to assume that the deep learning SD tools 
  could offer better performance than the shallow learning SD tools in cross-platform settings.    
  This is because a benefit of such pre-trained models is
that they are already trained on a large amount of non-SE but cross-domain
datasets to learn language-specific embedding. Such insights can help the tools to offer
more precise classification in a relatively new domain with little training data. This can
also denote that such models may be more useful in cross-platform settings.
\end{enumerate}}

In this paper, we report the results of an empirical study that we conducted to
determine the effectiveness of three SE-specific Deep Learning (DL) SD tools
(BERT4SentiSE~\cite{Biswas-ReliableSentiSEBERT-ICSME2020},
SentiMoji~\cite{Chen-SentiEmoji-FSE2019},
RNN4SentiSE~\cite{Biswas-SentiSEWordEmbedding-MSR2019}) in three datasets:
GitHub~\cite{Novielli-SEToolCrossPlatform-MSR2020}, Jira~\cite{Ortu-AreBulliesMoreProductive-MSR205}, and Stack Overflow~\cite{Calefato-Senti4SD-EMSE2017}. 
These three datasets were also used by Novielli et al.~\cite{Novielli-SEToolCrossPlatform-MSR2020,Novielli-SESentimentCP-EMSE2021} 
to study the performance of SE shallow learning SD tools in cross-platform settings. 

For each of the DL-based SD tools, we checked six cross-platform settings, two for each of our three benchmark datasets 
(i.e., train on one of the datasets and test on the other two datasets). Our study offers several findings: \begin{inparaenum}[(1)]
\item BERT4SentiSE is the best performer in cross-platform settings, followed by SentiMoji. For each setting, we observed a drop in
performance in the tools in cross-platform settings compared to the within platform settings.
For example, BERT4SentiSE shows a Macro F1-score of 0.92 for GitHub dataset in within platform setting (i.e., trained on GitHub dataset too), but in cross-platform settings, its performance for GitHub dataset drops to 0.88 when the tool is trained on Stack Overflow (SO) dataset and to 0.65 when the tool is trained on the Jira dataset. The percentage differences in the drop in performance between within and cross-platform settings are the most for RNN4SentiSE (31\%  drop when trained in SO and then tested in Jira). The findings are consistent with Novielli et al.~\cite{Novielli-SEToolCrossPlatform-MSR2020}, who also observed a drop in performance in the shallow learning SD tools for SE in cross-platform settings. This observation motivated us to compare the performance of DL tools against non-DL tools based on the following research question.
\item We ran two best performing shallow learning SD tools from (Senti4SD and SentiCR) and rule-based
tools SentistrengthSE  Novielli et
al.~\cite{Novielli-SEToolCrossPlatform-MSR2020}. We find that among the shallow vs.
deep learning SD tools in cross-platform settings, BERT4SentiSE is the best
performer in five out of six settings. However, when we compare BERT4SentiSE
against rule-based SentistrengthSE in cross-platform settings, we find that
BERT4SentiSE outperforms SentistrengthSE in only two out of the six settings.
\item We picked 400
misclassified cases from our three datasets, where both BERT4SentiSE and
SentiMoji were wrong. We find a total of seven misclassification categories. The most
observed reason is `subjectivity in annotation', i.e., inconsistency in manual
labeling in the datasets. This means that there were inconsistencies between
human coders on the polarity label of two sentences that are similar (e.g.,
`Thank you' was labeled as `positive' by some coder and `neutral' by others in
multiple datasets). Such inconsistencies confuse the deep learning tools.
\item We find that the two
best-performing deep learning SD tools (BERT4SentiSE and SentiMoji) agree with
each other more than other tools' agreement among themselves. Indeed, when we
combine these two DL tools in a majority voting classifier (i.e., we label a
sentence by combining polarity labels from BERT4SentiSE and SentiMoji), in only
two out of six cases they outperform the standalone best performing tool in that
setting. Therefore, in cross-platform settings, it will require further
improvement of the tools other than simply taking the majority voting of the
tools, e.g., improving the datasets by fixing the inconsistency in data
labeling.
\end{inparaenum}

\nd\bf{Replication Package}: \url{https://tinyurl.com/yf8nwbv6}

\section{Studied Tools and Datasets}\label{sec:tools_datasets}
In this section, we describe the six SE-specific SD tools (\sec\ref{sec:studied-tools}) and the three SE sentiment datasets (\sec\ref{sec:studied-datasets}) that we studied in this paper.
\subsection{Studied Tools} \label{sec:studied-tools}
Our study focuses on three recently introduced SE-specific deep learning SD tools: \begin{inparaenum}
\item RNN4SentiSE~\cite{Biswas-SentiSEWordEmbedding-MSR2019}, 
\item SentiMoji~\cite{Chen-SentiEmoji-FSE2019}, and
\item BERT4SentiSE~\cite{Biswas-ReliableSentiSEBERT-ICSME2020}.
\end{inparaenum}
Recently, Zhang et al.~\cite{zhang_transformer_model_2020} showed that advanced pre-trained transformer models (PTM) like BERT can outperform shallow learning SE-specific SD tools. The tool BERT4SentiSE~\cite{Biswas-ReliableSentiSEBERT-ICSME2020} is introduced at the same time of Zhang et al.~\cite{zhang_transformer_model_2020}. We thus do not use any PTMs from Zhang et al.~\cite{zhang_transformer_model_2020}. Among the three tools, BERT4SentiSE~\cite{Biswas-ReliableSentiSEBERT-ICSME2020} is the latest SD tool in SE, and it is found to have outperformed all other SE-specific SD tools. 

We compare the performance of the three tools with two best performing shallow learning SE-specific SD tools in the cross-platform studies of Novielli et al.~\cite{Novielli-SEToolCrossPlatform-MSR2020}:  Senti4SD~\cite{Calefato-Senti4SD-EMSE2017} and SentiCR~\cite{Ahmed-SentiCRNIER-ASE2017}. In addition, given that the rule-based tool SentistrengthSE~\cite{Islam-SentistrengthSE-MSR2017} was found to be superior by Novielli et al.~\cite{Novielli-SEToolCrossPlatform-MSR2020}, we also compare its performance with the DL SD tools. So, our studied tools can be classified into three types
\begin{inparaenum}
\item \bf{Deep-learning based:} \textit{SentiMoji, BERT4SentiSE, RNN4SentiSE},
\item \bf{Shallow machine learning based:} \textit{Senti4SD, SentiCR},
\item \bf{Rule based:} \textit{SentiStrengthSE}.
\end{inparaenum} We now briefly describe each tool below.

\begin{inparaenum}

\item\bf{SentiMoji~\cite{Chen-SentiEmoji-FSE2019}}  is customized on DeepMoji~\cite{deepMoji_2017}, which uses word embedding learned from 56.5B tweets in two bi-LSTM layers and one attention layer~\cite{Chen-SentiEmoji-FSE2019}. SentiMoji learns vector representation of texts by leveraging how emojis are used and other texts in Twitter and GitHub. SentiMoji outperforms shallow and rule-based SD tools in the datasets of Lin et al.~\cite{Lin-SentimentDetectionNegativeResults-ICSE2018}. The hyper-parameters of SentiMoji are fine-tuned with one hundred million of GitHub-emoji~\cite{lu2018first} posts~\cite{Chen-SentiEmoji-FSE2019} using chain-thaw approach~\cite{deepMoji_2017}. At the final stage, the layers of DeepMoji are kept unchanged, and the final 64-dimensions of the softmax layer are updated with the training dataset. In this research, we use the best-performing architecture of SentiMoji to retrain on the datasets. The pre-trained version of SentiMoji uses SO posts as word embedding. It also utilizes tweets from Twitter to learn generalized and GitHub posts to learn SE-specific sentiment jargons.

\item\bf{BERT4SentiSE~\cite{Biswas-ReliableSentiSEBERT-ICSME2020}} is based on the BERT model that is developed by Devlin et al. at Google in 2018~\cite{Delvin-BERTArch-Arxiv2018}. It learns a contextual representation of words in a text by looking at the following and the preceding words. It is pre-trained on task-neutral languages. It uses transfer learning~\cite{ruder2019transfer} and attention-based architecture. Biswas et al.~\cite{Biswas-ReliableSentiSEBERT-ICSME2020} proposed a BERT-based pre-trained model, which so far achieved the best performance for SE datasets. In this study, we used the best model parameters reported by the original authors. We used the pre-trained BERT \footnote{\url{http://goo.gl/language/bert}} model consisting of 12 layers, encoder layer of 768, and it contained 110 Million parameters altogether.

\item\bf{RNN4SEntiSE~\cite{Biswas-SentiSEWordEmbedding-MSR2019}}  is an RNN-based model that uses generic word embedding from Google news data. The authors also reported that software domain-specific word embedding from stack overflow posts did not improve performance much. In this study, We use the Google news word embedding with the same LSTM based RNN configuration that the authors used in the original study. Similar to the original study, we used LSTM unit 30, drop out 0.2, batch size 30, and epochs 100 which provided the best results.

\item\bf{Senti4SD~\cite{Calefato-Senti4SD-EMSE2017}} is initially trained on 4K SO posts. It uses features as a bag of words (BoW), sentiment lexicons, and word embedding. The word embedding is created using Stack Overflow data dump to offer domain-specific information. For this study, we use the Python version~\cite{calefato2019emtk} of the tool. We also used the best hyperparameters that were used in the original study and provided the best results in our experiments. We used a full feature set like the original study, i.e., lexicon features, semantic features, and keyword features.

\item\bf{SentiCR~\cite{sentiCR_2017}} uses Gradient Boosting Tree (GBT) algorithm. It uses a bag of words as features. It leverages SMOTE~\cite{chawla2002smote} to handle a class imbalance by oversampling. Following Uddin et al.~\cite{Uddin-OpinionValue-TSE2019}, we train SentiCR to detect three polarity classes. In this study, we used the best configuration and hyper-parameters reported by the original authors. We used the GBT algorithm for the classification.

\item\bf{SentiStrengthSE~\cite{Islam-SentistrengthSE-MSR2017}} is developed on top of
SentiStrength~\cite{Thelwall-Sentistrength-ASICT2010} by introducing rules and
sentiment words specific to SE. Each negative word has a score
ranging from -2 to -5; a positive word has a score ranging from +2 to +5. The
polarity scores are \it{a priori}, i.e., they do not depend on the contextual
nature of the sentiment expressed in a unit.
It outputs both positive and negative
scores for an input text. The overall polarity score of
an input text is calculated using the algebraic sum of the positive and
negative scores. The text is labeled as `positive' if the sum of scores is
greater than 0, negative if the sum is less than 0, and neutral otherwise. The rules for this tool was developed and tested on the Jira dataset by Ortu et al.~\cite{Ortu-EmotionalSideJira-MSR2016}
\end{inparaenum}


\subsection{Studied Datasets}\label{sec:studied-datasets}
We study three datasets: Stack Overflow (SO)~\cite{Calefato-Senti4SD-EMSE2017}, Jira~\cite{Ortu-AreBulliesMoreProductive-MSR205}, and 
GitHub~\cite{Novielli-SEToolCrossPlatform-MSR2020}. These three datasets were also previously used by Novielli et al.~\cite{Novielli-SEToolCrossPlatform-MSR2020} in their cross-platform study. 
Table \ref{tab:benchmark_dataset} provides summary statistics of the datasets.
\begin{table}[h]
  \centering
   \caption{Descriptive statistics of the studied benchmark datasets}
    \begin{tabular}{lrrrr}
    \toprule{}
    \textbf{Dataset} & \textbf{\#Units} & \textbf{+VE} & \textbf{VN} & \textbf{-VE}\\
    \midrule
    GitHub~\cite{Novielli-SEToolCrossPlatform-MSR2020} & 7,122 & 2,013 (28\%) & 3,022 (43\%) & 2,087 (29\%) \\
    Jira~\cite{Ortu-AreBulliesMoreProductive-MSR205} & 5,869 & 1,128 (19\%) & 3,955 (67\%) & 786 (14\%) \\
    Stack Overflow~\cite{Calefato-Senti4SD-EMSE2017} & 4,423 & 1,527 (35\%) & 1,694 (38\%) & 1,202 (27\%)\\
    \bottomrule
    \end{tabular}%
  \label{tab:benchmark_dataset}%
\end{table}%

\begin{inparaenum}
\item\bf{Stack Overflow (SO) dataset~\cite{Calefato-Senti4SD-EMSE2017}} contains 4,423 SO posts, including questions, comments, and answers manually annotated by twelve trained coders. Each post was annotated by three raters and received the final polarity based on majority voting based on the guideline of Shaver framework~\cite{shaver1987emotion}. The dataset is quite well balanced 35\% positive emotions, 27\% negative emotions, 38\% neutral emotions, i.e., absence of emotions. The Cohen's~\cite{cohen1968weighted} Kappa, $k$, is 0.74 which signifies substantial inter-rater agreement~\cite{viera2005understanding}.

\item\bf{Jira dataset~\cite{Ortu-AreBulliesMoreProductive-MSR205}} contains 5,869 sentences and issue comments from different open-source software projects (e.g., Spring, Apache). Software developers originally annotated the dataset with six emotions (i.e., love, joy, anger, fear, sadness, surprise). To be consistent with other datasets, we followed similar studies by Novielli et al.~\cite{Novielli-SEToolCrossPlatform-MSR2020} and translate \qu{love} and \qu{joy} as a positive sentiment, \qu{anger} and \qu{sadness} as a negative sentiment. \qu{surprise} sentences are discarded because, based on the context, they can be either positive or negative. Finally, the absence of emotions is labeled as neutral. The dataset is not well balanced 19\% positive emotions, 14\% negative emotions, and 67\% neutral emotions. 

\item\bf{GitHub dataset~\cite{Novielli-SEToolCrossPlatform-MSR2020}} contains
7,122 pull requests and commit messages. Three authors annotated the dataset.
\rev{The raters were trained to label based on the emotion classification model proposed by Shaver et al.~\cite{Shaver-EmotionModel-SP1987}, which 
is tree-structured hierarchical classification of emotions. At top level, the hierarchy has six basic emotions: love, joy, surprise,
anger, sadness, and fear. The paper by Shaver et al.~\cite{Shaver-EmotionModel-SP1987} offers a detailed breakdown 
of keywords and emotion sub-types for each main level, which are easy to understand. 
In the GitHub dataset, positive polarity was labeled by coders when they detected the underlying emotion in a sentence/post as love or joy, while 
negative polarity was labeled when the underlying emotion was anger, sadness, or fear. The neutral label was provided in the absence of any detected emotions. 
For the sentences/posts containing the emotion ``surprise'', the underlying 
context was considered by the coders to label any of two sentiment polarity labels, i.e., positive, or negative.} It has around 28\% positive, 29\% negative, 43\%
neutral polarity labels.
\end{inparaenum}

\section{Empirical Study}\label{sec:result}
The major focus of our empirical study is to determine how well deep learning SE-specific SD tools perform in cross-platform settings compared to within platform settings. A within platform setting denotes the training and testing of an SD tool in the same dataset by splitting the dataset into training and testing sets. A cross-platform setting denotes the training of the tool in one dataset (e.g., GitHub) and the testing of it in another dataset (e.g., Jira/Stack Overflow). In particular, we answer five research questions (RQ):
\begin{enumerate}[label=\bf{RQ\arabic{*}.}, leftmargin=30pt]
  \item Can deep learning SD tools for SE perform equally well in both within and cross-platform settings? (\sec\ref{sec:rq0})
  \item Can the deep learning SD tools perform better than non-DL tools in cross-platform settings? (\sec\ref{sec:rq-cross-platform-performance})
  \item What are the error categories of the tools in the cross-platform settings? (\sec\ref{sec:rq-cross-platform-error})
  \item Can an ensemble of the deep learning and non-deep learning tools outperform individual tools in cross-platform settings? (\sec\ref{sec:cross-platform-agreement})
\end{enumerate}

\subsection{RQ$_1$ Can deep learning sentiment detection tools for software engineering perform equally well both within and in cross-platform settings?}\label{sec:rq0}

\subsubsection{Motivation} Research in SE finds that cross-domain SD tools can be insufficiently accurate for SE datasets. Such 
findings motivated the development of SE-specific SD tools.  
The recent advances in deep learning techniques influenced the development of several deep learning SD tools for SE. The most recent deep learning SD tools for SE
are BERT4SentiSE, SentiMoji, and RNN4SentSE. BERT4SentiSE and SentiMoji are found to have outperformed other SE-specific SD tools. However, 
the experimentation was conducted in within-platform settings. This means that the tools were trained and tested using the same dataset. 
 Recent research in SE also finds that the SE-specific shallow learning SD tools (Senti4SD and SentiCR) 
can suffer from a drop in performance in cross-platform settings, i.e., when trained in one SE dataset and tested on another SE dataset. 
It is thus necessary to know whether the tools show similar performance in cross-platform settings because that would 
indicate the broader applicability of the tools.
%

\subsubsection{Approach} \label{sec:rq0-approach}
We investigate the performance of three most recent deep learning SD tools for SE: BERT4SentiSE, RNN4SentiSE, and SentiMoji. 
We compare the performance of each tool between within and cross-platform settings. 
For each tool, we checked six cross-platform settings: two for each dataset
(i.e., train on a dataset and test on another dataset). We compute the performance of each tool in the settings as follows.
\begin{itemize}[leftmargin=10pt]

 \item \ib{Within platform.} We use k-fold cross validation~\cite{stone1974cross} to train and test a tool in a given dataset. 
In this study, we employ 10-fold cross-validation settings where the dataset is
 divided into ten folds, and each fold is used as a test set at some point. First,
 we divide each dataset into 10-folds using the Scikit
 StratifiedKFold \footnote{\url{https://scikit-learn.org/stable/modules/generated/sklearn.model_selection.StratifiedKFold.html}}
 technique, which ensures that each fold contains the same proportion of
 polarity classes as the original dataset.
 We train and test each supervised SD tool for each dataset in ten iterations.
 In the first iteration, we use fold 1 of the dataset as the test set and the remaining 9 folds as the training set. In the second iteration, we use fold 2 as the test set and the remaining 9 folds for training the supervised tool. This
 process continues ten times until all of the ten folds are used as a test set. Then we merge the predictions of the tool for each test fold and calculate the tool's overall performance on that dataset.

 \item \ib{Cross Platform.} We train each tool on one dataset (e.g., SO) and test it on the other two datasets (e.g., GitHub and Jira). 
 We report the performance of each tool per testing dataset. This means that for a given dataset under test, we check the correct and incorrect cases on all the 
 sentences/units of analysis in the dataset. Therefore, similar to the within-platform setting, we are checking an entire dataset for performance analysis in cross-platform 
 setting. In within-platform setting, we do ten iterations in a 10-fold cross-validation setup. For cross-platform 
 setting, we run the tool once in a train dataset (to create a trained model) and then run the trained model once on the other two test datasets.  
\end{itemize}

We report performance using three standard metrics of information
retrieval: precision ($P$), recall ($R$), and F1-score ($F1$)~\cite{sokolova2009systematic}. \textit{Precision} signifies the number of samples correctly predicted as positive out of total samples predicted as positive. \textit{Recall} signifies the number of samples correctly identified as positives out of total samples that are actually positive. \textit{$F1$ measure} is the harmonic average of precision ($P$) and recall ($R$). 
{
\begin{equation} \label{eq:metrics}
P  = \frac{\sum_{i=1}^{l} TP_i}{\sum_{i=1}^{l} TP_i+FP_i},~
R = \frac{\sum_{i=1}^{l} TP_i}{\sum_{i=1}^{l} TP_i+FN_i},~
F1 = 2*\frac{P*R}{P+R}
\end{equation}
} {$TP = $ \# of true positives, $FP =$ \# of false positives, $FN =$ \# of false negatives.}. 

In our multi-class sentiment problem, a $TP$, $FP$, and $FN$ are calculated for each class. For example, we have three possible sentiment classes (positive, neutral, and negative). 
For a particular class (e.g., neutral sentiment) $TP$ occurs when the tool predicts `neutral'. For the same class, $FN$ occurs when the tool predicts any other classes (e.g., positive or negative), and $TN$ occurs when the tool predicts `neutral' but the original label is not `neutral'.
We use the confusion matrix in \tbl\ref{tab:conmat-sentiment} and Equation~\cite{sokolova2009systematic} (\ref{eq:metrics}) following state of the art~\cite{Novielli-BenchmarkStudySentiSE-MSR2018,Calefato-Senti4SD-EMSE2017,Sebastiani-MachineLearningTextCategorization-ACMSurveys2002,Uddin-OpinionValue-TSE2019}. 
We report overall performance using micro and macro averages and by each
polarity class. The macro-average computes the metrics for each class and then
averages thus it treats each classes equally. On the other hand, the
micro-average aggregates the contribution of all classes. Micro-average is
influenced by the performance of the majority class~\cite{Sebastiani-MachineLearningTextCategorization-ACMSurveys2002}
because most number of examples belong to this class. We use Macro F1-score to
report the best tool, following standard
practices~\cite{Novielli-BenchmarkStudySentiSE-MSR2018,Manning-IRIntroBook-Cambridge2009}.
\begin{table}[t]
  \centering
  \caption{Confusion matrix for multi-class (three) sentiment detection}
    \begin{tabular}{crr|r|r}\toprule
          &       & \multicolumn{3}{c}{\textbf{Predicted}} \\
          \cmidrule{3-5}
          &       & \textbf{Positive (P)} & \textbf{Negative (N)} & \textbf{Neutral (O)} \\
          \cmidrule{2-5}
    \multirow{3}[0]{*}{\textbf{Actual}} & \textbf{Positive (P)} & TP$_P$ & FP$_N$, FN$_P$ & FP$_O$, FN$_P$ \\
          & \textbf{Negative (N)} & FP$_P$, FN$_N$ & TP$_N$ & FP$_O$, FN$_N$ \\
          & \textbf{Neutral (O)} & FP$_P$, FN$_O$ & FP$_N$, FN$_O$ & TP$_O$ \\
          \bottomrule
    \end{tabular}%
  \label{tab:conmat-sentiment}%
\end{table}%

\begin{table*}
\footnotesize
  \centering
  \caption{Performance of SE-specific sentiment analysis tools in cross-platform settings (SentiSE = SentistrengthSE).}
 \resizebox{\columnwidth}{!}{%
    \begin{tabular}{ll|rrr|rrr|rrr|rrr|rrr|rrr}
    \toprule{}

\multirow{3}{*}{\textbf{Setting}} &\multirow{3}{*}{\textbf{Polarity}} &\multicolumn{9}{c|}{\textbf{Deep-learning based SD Tools}} &\multicolumn{6}{c|}{\textbf{Shallow ML-based}} &\multicolumn{3}{c}{\textbf{Rule-based}} \\
\cmidrule{3-20}
& &\multicolumn{3}{c}{\textbf{SEntiMoji}} &\multicolumn{3}{c}{\textbf{BERT4SEntiSE}} &\multicolumn{3}{c}{\textbf{RNN4SentiSE}} &\multicolumn{3}{c}{\textbf{Senti4SD}} &\multicolumn{3}{c}{\textbf{SentiCR}} &\multicolumn{3}{c}{\textbf{SentiSE}} \\
\cmidrule{3-20}
& &\textbf{P} &\textbf{R} &\textbf{F1} &\textbf{P} &\textbf{R} &\textbf{F1} &\textbf{P} &\textbf{R} &\textbf{F1} &\textbf{P} &\textbf{R} &\textbf{F1} &\textbf{P} &\textbf{R} &\textbf{F1} &\textbf{P} &\textbf{R} &\textbf{F1} \\

    \midrule
\multicolumn{20}{c}{\bf{Test Dataset: GitHub}} \\
\cmidrule{2-20}    
 
     \multirow{5}{1.4cm}{Test:  GitHub Train: GitHub  (Within-platform)}
     & -NEG &  .89 & .85 & .87  & .89 & .91 & .90  & .68 & .69 & .68  & .90 & .89 & .90  & .87 & .69 & .77  & .81 & .71 & .76   \\ 
     & Neutral &  .87 & .92 & .89  & .92 & .90 & .91  & .73 & .66 & .69  & .89 & .92 & .91  & .77 & .91 & .84  & .74 & .86 & .79   \\ 
     & +POS &  .94 & .92 & .93  & .93 & .94 & .94  & .64 & .72 & .68  & .94 & .91 & .93  & .89 & .84 & .86  & .85 & .76 & .80   \\ 
     & Micro &  .90 & .90 & .90  & \textbf{.92} & \textbf{.92} & \textbf{.92}  & .68 & .68 & .68  & .91 & .91 & .91  & .83 & .83 & .83  & .79 & .79 & .79   \\ 
 & Macro &  .90 & .89 & .90  & \textbf{.91} & \textbf{.92} & \textbf{.92}  & .68 & .69 & .68  & \textbf{.91} & .91 & .91  & .84 & .81 & .82  & .80 & .78 & .78   \\ 

    \cmidrule{2-20}     

\multirow{5}{1.4cm}{Test:  GitHub Train: SO  (Cross-platform)}
 & -NEG &  .85 & .62 & .72  & .83 & .73 & .78  & .69 & .49 & .58  & .79 & .56 & .66  & .83 & .37 & .51  & .81 & .71 & .76   \\ 
 & Neutral &  .67 & .91 & .77  & .71 & .91 & .80  & .61 & .84 & .71  & .71 & .82 & .76  & .60 & .94 & .73  & .74 & .86 & .79   \\ 
 & +POS &  .88 & .65 & .75  & .88 & .63 & .73  & .78 & .57 & .66  & .76 & .81 & .78  & .86 & .62 & .72  & .85 & .76 & .80   \\ 
 & Micro &  .75 & .75 & .75  & .78 & .78 & .78  & .66 & .66 & .66  & .74 & .74 & .74  & .68 & .68 & .68  & \textbf{.79} & \textbf{.79} & \textbf{.79}   \\ 
 & Macro &  .80 & .73 & .75  & \textbf{.81} & .76 & .77  & .69 & .63 & .65  & .75 & .73 & .73  & .76 & .64 & .66  & .80 & \textbf{.78} & \textbf{.78}   \\ 

\cmidrule{2-20}

    \multirow{5}{1.4cm}{Test: GitHub Train: Jira  (Cross-platform)} 
     & -NEG &  .91 & .39 & .55  & .90 & .43 & .58  & .75 & .39 & .52  & .85 & .52 & .64  & .88 & .27 & .41  & .81 & .71 & .76   \\ 
 & Neutral &  .56 & .98 & .72  & .59 & .97 & .73  & .54 & .95 & .69  & .60 & .96 & .74  & .52 & .98 & .68  & .74 & .86 & .79   \\ 
 & +POS &  .94 & .45 & .60  & .91 & .51 & .65  & .80 & .30 & .44  & .90 & .46 & .61  & .89 & .38 & .53  & .85 & .76 & .80   \\ 
 & Micro &  .66 & .66 & .66  & .68 & .68 & .68  & .60 & .60 & .60  & .69 & .69 & .69  & .60 & .60 & .60  & \textbf{.79} & \textbf{.79} & \textbf{.79}   \\ 
 & Macro &  \textbf{.80} & .61 & .62  & \textbf{.80} & .64 & .65  & .70 & .55 & .55  & .78 & .65 & .67  & .76 & .54 & .54  & \textbf{.80} & \textbf{.78} & \textbf{.78}   \\ 


\midrule
\multicolumn{20}{c}{\bf{Test Dataset: Stack Overflow (SO)}} \\
\cmidrule{2-20}

    \multirow{5}{1.4cm}{Test: SO Train: SO  (Within-platform)} 
 & -NEG &  .84 & .83 & .83  & .86 & .85 & .85  & .74 & .73 & .74  & .80 & .83 & .82  & .80 & .72 & .76  & .76 & .76 & .76   \\ 
 & Neutral &  .83 & .84 & .84  & .86 & .85 & .85  & .75 & .77 & .76  & .84 & .79 & .81  & .79 & .82 & .81  & .73 & .79 & .75   \\ 
 & +POS &  .93 & .93 & .93  & .93 & .94 & .93  & .85 & .85 & .85  & .90 & .93 & .91  & .88 & .91 & .89  & .91 & .82 & .86   \\ 
 & Micro &  .87 & .87 & .87  & \textbf{.88} & \textbf{.88} & \textbf{.88}  & .78 & .78 & .78  & .85 & .85 & .85  & .82 & .82 & .82  & .79 & .79 & .79   \\ 
 & Macro &  .87 & .87 & .87  & \textbf{.88} & \textbf{.88} & \textbf{.88}  & .78 & .78 & .78  & .85 & .85 & .85  & .82 & .82 & .82  & .80 & .79 & .79   \\

    \cmidrule{2-20}
    
    \multirow{5}{1.4cm}{Test: SO Train: GitHub  (Cross-platform)} 
 & -NEG &  .82 & .76 & .79  & .80 & .86 & .83  & .66 & .60 & .63  & .69 & .76 & .72  & .64 & .49 & .55  & .76 & .76 & .76   \\ 
 & Neutral &  .75 & .85 & .80  & .84 & .80 & .82  & .61 & .64 & .63  & .73 & .76 & .75  & .65 & .85 & .74  & .73 & .79 & .75   \\ 
 & +POS &  .93 & .85 & .89  & .93 & .91 & .92  & .73 & .74 & .74  & .92 & .82 & .87  & .88 & .74 & .80  & .91 & .82 & .86   \\ 
 & Micro &  .83 & .83 & .83  & \textbf{.85} & \textbf{.85} & \textbf{.85}  & .67 & .67 & .67  & .78 & .78 & .78  & .71 & .71 & .71  & .79 & .79 & .79   \\ 
 & Macro &  .83 & .82 & .82  & \textbf{.85} & \textbf{.86} & \textbf{.85}  & .67 & .66 & .66  & .78 & .78 & .78  & .72 & .69 & .70  & .80 & .79 & .79   \\ 


    \cmidrule{2-20}

    \multirow{5}{1.4cm}{Test: SO Train: Jira  (Cross-platform)}
 & -NEG &  .93 & .31 & .47  & .88 & .55 & .68  & .66 & .31 & .42  & .75 & .30 & .43  & .67 & .22 & .33  & .76 & .76 & .76   \\ 
 & Neutral &  .54 & .98 & .70  & .63 & .94 & .76  & .50 & .91 & .65  & .54 & .93 & .69  & .50 & .95 & .66  & .73 & .79 & .75   \\ 
 & +POS &  .98 & .62 & .76  & .97 & .74 & .84  & .83 & .42 & .56  & .96 & .64 & .76  & .96 & .52 & .67  & .91 & .82 & .86   \\ 
 & Micro &  .67 & .67 & .67  & .76 & .76 & .76  & .58 & .58 & .58  & .66 & .66 & .66  & .60 & .60 & .60  & \textbf{.79} & \textbf{.79} & \textbf{.79}   \\ 
 & Macro &  .82 & .64 & .64  & \textbf{.83} & .74 & .76  & .66 & .55 & .54  & .75 & .62 & .63  & .71 & .56 & .55  & .80 & \textbf{.79} & \textbf{.79}   \\ 


    \midrule
\multicolumn{20}{c}{\bf{Test Dataset: Jira}} \\
\cmidrule{2-20}

    \multirow{5}{1.4cm}{Test: Jira Train: Jira  (Within-platform)} 
 & -NEG &  .84 & .68 & .75  & .82 & .76 & .79  & .69 & .46 & .55  & .71 & .55 & .62  & .79 & .66 & .72  & .68 & .73 & .71   \\ 
 & Neutral &  .89 & .91 & .90  & .90 & .90 & .90  & .83 & .89 & .86  & .84 & .88 & .86  & .89 & .89 & .89  & .92 & .82 & .87   \\ 
 & +POS &  .79 & .80 & .79  & .78 & .80 & .79  & .73 & .72 & .72  & .73 & .71 & .72  & .75 & .85 & .80  & .68 & .92 & .78   \\ 
 & Micro &  \textbf{.86} & \textbf{.86} & \textbf{.86}  & \textbf{.86} & \textbf{.86} & \textbf{.86}  & .80 & .80 & .80  & .81 & .81 & .81  & .85 & .85 & .85  & .83 & .83 & .83   \\ 
 & Macro &  \textbf{.84} & .80 & .81  & .83 & \textbf{.82} & \textbf{.83}  & .75 & .69 & .71  & .76 & .72 & .73  & .81 & .80 & .80  & .76 & \textbf{.82} & .79   \\ 

    \cmidrule{2-20}

    \multirow{5}{1.4cm}{Test: Jira Train: GitHub (Cross-platform)} 
 & -NEG &  .63 & .75 & .69  & .66 & .74 & .70  & .51 & .71 & .60  & .61 & .64 & .62  & .69 & .60 & .64  & .68 & .73 & .71   \\ 
 & Neutral &  .93 & .79 & .85  & .93 & .80 & .86  & .92 & .58 & .71  & .89 & .78 & .83  & .90 & .85 & .87  & .92 & .82 & .87   \\ 
 & +POS &  .66 & .91 & .77  & .65 & .91 & .76  & .45 & .92 & .61  & .62 & .86 & .72  & .69 & .89 & .78  & .68 & .92 & .78   \\ 
 & Micro &  .81 & .81 & .81  & .81 & .81 & .81  & .66 & .66 & .66  & .78 & .78 & .78  & .82 & .82 & .82  & \textbf{.83} & \textbf{.83} & \textbf{.83}   \\ 
 & Macro &  .74 & \textbf{.82} & .77  & .75 & \textbf{.82} & .77  & .63 & .74 & .64  & .71 & .76 & .73  & \textbf{.76} & .78 & .76  & \textbf{.76} & \textbf{.82} & \textbf{.79}   \\ 


    \cmidrule{2-20}

    \multirow{5}{1.4cm}{Test: Jira Train: SO (Cross-platform)}
 & -NEG &  .67 & .68 & .67  & .72 & .68 & .70  & .58 & .61 & .59  & .50 & .29 & .37  & .36 & .06 & .10  & .68 & .73 & .71   \\ 
 & Neutral &  .92 & .81 & .86  & .92 & .82 & .87  & .88 & .77 & .82  & .82 & .81 & .81  & .75 & .94 & .83  & .92 & .82 & .87   \\ 
 & +POS &  .67 & .93 & .78  & .65 & .93 & .76  & .62 & .89 & .73  & .59 & .80 & .68  & .71 & .49 & .58  & .68 & .92 & .78   \\ 
 & Micro &  .82 & .82 & .82  & .82 & .82 & .82  & .77 & .77 & .77  & .74 & .74 & .74  & .73 & .73 & .73  & \textbf{.83} & \textbf{.83} & \textbf{.83}   \\ 
 & Macro &  .75 & .81 & .77  & \textbf{.76} & .81 & .78  & .69 & .75 & .71  & .64 & .63 & .62  & .61 & .49 & .50  & \textbf{.76} & \textbf{.82} & \textbf{.79}   \\ 


    \bottomrule
    \end{tabular}%
}
  \label{tab:cross_platform_performance}%
\end{table*}%
\begin{table*}
  \centering
  \caption{Performance drop in SE-specific supervised sentiment analysis tools in cross-platform settings (SO = Stack Overflow)}
    \resizebox{\columnwidth}{!}{%
    \begin{tabular}{cl|rrr|rrr|rrr|rrr|rrr}
    \toprule{}

    
\multirow{3}{*}{\textbf{Train}} &\multirow{3}{*}{\textbf{Metric}} &\multicolumn{9}{c}{\textbf{Deep-learning based}} &\multicolumn{6}{c}{\textbf{Shallow ML-based}} \\
\cmidrule{3-17}
 & &\multicolumn{3}{c}{\textbf{SEntiMoji}} &\multicolumn{3}{c}{\textbf{BERT4SEntiSE}} &\multicolumn{3}{c}{\textbf{RNN4SentiSE}} & \multicolumn{3}{c}{\textbf{Senti4SD}} &\multicolumn{3}{c}{\textbf{SentiCR}} \\
\cmidrule{3-17}
\centering \textbf{Dataset} & &\textbf{P} &\textbf{R} &\textbf{F1} &\textbf{P} &\textbf{R} &\textbf{F1} &\textbf{P} &\textbf{R} &\textbf{F1} &\textbf{P} &\textbf{R} &\textbf{F1} &\textbf{P} &\textbf{R} &\textbf{F1} \\

    \midrule
    
    \multicolumn{17}{c}{\textbf{Test dataset: GitHub}} \\
    \midrule
 
    \multirow{2}{*}{\textbf{SO}} & $\triangledown$ Micro &  -16\% & -16\% & -16\% & -15\% & -15\% & -15\% & -3\% & -3\% & -3\% & -18\% & -18\% & -18\% & -17\% & -17\% & -17\%    \\ 
    &  $\triangledown$ Macro &  -11\% & -19\% & -17\% & -12\% & -18\% & -16\% & 2\% & -8\% & -5\% & -17\% & -20\% & -19\% & -10\% & -21\% & -20\%  \\ 
    \cmidrule{2-17}
    
    \multirow{2}{*}{\textbf{Jira}} &  $\triangledown$ Micro &  -27\% & -27\% & -27\% & -26\% & -26\% & -26\% & -12\% & -12\% & -12\% & -24\% & -24\% & -24\% & -27\% & -27\% & -27\%   \\ 
    &  $\triangledown$ Macro &  -11\% & -32\% & -31\% & -13\% & -31\% & -29\% & 2\% & -21\% & -20\% & -14\% & -29\% & -27\% & -9\% & -34\% & -34\%  \\ 
    \midrule

    \multicolumn{17}{c}{\textbf{Test dataset: Stack Overflow}} \\
    \midrule
    \multirow{2}{*}{\textbf{GitHub}} & $\triangledown$  Micro &  -5\% & -5\% & -5\% & -3\% & -3\% & -3\% & -15\% & -15\% & -15\% & -8\% & -8\% & -8\% & -13\% & -13\% & -13\%   \\ 
    & $\triangledown$  Macro &  -4\% & -5\% & -5\% & -3\% & -3\% & -3\% & -15\% & -15\% & -15\% & -7\% & -8\% & -8\% & -12\% & -15\% & -15\%  \\ 
    \cmidrule{2-17}
    \multirow{2}{*}{\textbf{Jira}} &  $\triangledown$ Micro &  -23\% & -23\% & -23\% & -13\% & -13\% & -13\% & -26\% & -26\% & -26\% & -22\% & -22\% & -22\% & -27\% & -27\% & -27\%   \\ 
    &  $\triangledown$ Macro &  -6\% & -27\% & -26\% & -6\% & -16\% & -14\% & -15\% & -30\% & -31\% & -11\% & -27\% & -26\% & -14\% & -31\% & -33\%  \\ 
    \midrule
    
    \multicolumn{17}{c}{\textbf{Test dataset: Jira}} \\
    \midrule
    \multirow{2}{*}{\textbf{GitHub}} &  $\triangledown$ Micro &  -6\% & -6\% & -6\% & -6\% & -6\% & -6\% & -17\% & -17\% & -17\% & -4\% & -4\% & -4\% & -3\% & -3\% & -3\%    \\ 
    &  $\triangledown$ Macro &  -11\% & 3\% & -5\% & -10\% & 0\% & -7\% & -16\% & 7\% & -10\% & -7\% & 7\% & -1\% & -6\% & -2\% & -5\%  \\ 
    \cmidrule{2-17}
    \multirow{2}{*}{\textbf{SO}} & $\triangledown$  Micro &  -5\% & -5\% & -5\% & -5\% & -5\% & -5\% & -4\% & -4\% & -4\% & -9\% & -9\% & -9\% & -14\% & -14\% & -14\%   \\ 
    &  $\triangledown$ Macro &  -10\% & 1\% & -5\% & -8\% & -1\% & -6\% & -7\% & 9\% & 1\% & -16\% & -12\% & -15\% & -25\% & -38\% & -37\%  \\ 

    \bottomrule
    \end{tabular}%
}
  \label{tab:cross_platform_performance_drop}%
\end{table*}%

\begin{table}[t]
  \centering
  \caption{Percentage of correctly classified sentences in within platform settings that were misclassified in cross-platform settings (each value is rounded to the nearest integer)}
  \resizebox{\columnwidth}{!}{%
    \begin{tabular}{ll|cc|cc|cc}\toprule
    {\textbf{SD}} & \textbf{\% Correct} & \multicolumn{2}{c}{\textbf{Test in GitHub}} & \multicolumn{2}{c}{\textbf{Test in SO}} & \multicolumn{2}{c}{\textbf{Test in Jira}} \\
    \cmidrule{3-8}
    {\textbf{Tool}} & \textbf{Misclassified} & {\textbf{Train SO}} & {\textbf{Train Jira}} & {\textbf{Train GitHub}} & {\textbf{Train Jira}} & {\textbf{Train GitHub}} & {\textbf{Train SO}} \\
    \midrule
    \multirow{3}[0]{*}{\textbf{BERT4SentiSE}} & \textbf{Overall} & 19    & 30    & 8     & 18    & 11    & 10 \\
          & \textbf{Severe} & 3     & 1     & 0     & 0     & 1     & 1 \\
          & \textbf{Mild} & 16    & 28    & 7     & 18    & 11    & 10 \\
          \midrule
    \multirow{3}[0]{*}{\textbf{SentiMoji}} & \textbf{Overall} & 19    & 30    & 10    & 29    & 11    & 10 \\
          & \textbf{Severe} & 1     & 1     & 1     & 0     & 0     & 0 \\
          & \textbf{Mild} & 18    & 29    & 9     & 29    & 11    & 10 \\
          \bottomrule
    \end{tabular}%
    }
  \label{tab:rq-error-analysis}%
\end{table}%

\subsubsection{Results} In Table \ref{tab:cross_platform_performance}, the third to fifth columns (SentiMoji, BERT4SentiSE, RNN4SentiSE) 
show the performance of the three deep learning SD tools for SE. The table is divided into three sections, each separated from the other section via a full-length horizontal line. For example, under the title `Test Dataset: GitHub', we show how each tool performs while using GitHub as the test dataset in within and cross-platform 
settings. This means that \begin{inparaenum}[(1)]
\item first we show how a tool performs within the GitHub dataset (i.e., using train-test splitting technique) and
\item second we show how each tool performed on GitHub dataset when they have been trained on the other two datasets, i.e., once trained on Stack Overflow (SO) and another trained on Jira dataset.
\end{inparaenum} Therefore, the first setup is within-platform, and the last two setups are cross-platform settings. 
For a given setting (i.e., within or cross-platform), we report our three performance metrics, Precision (P), Recall (R), and F1-measure (F1) for each of the three polarity classes (i.e., positive, 
negative, neutral). For each setting, we then report the overall performance of each tool using two metrics: Micro-avg and Macro-avg. We discuss significant observations below.   

\bf{Each deep learning SD tool for SE suffers a drop in performance in cross-platform settings compared to its within-platform performance.} There is a notable drop in performance of each of the deep learning SD tools from within to cross-platform settings. The drop, however, varies among the six cross-platform settings. The drop 
also varies among the three DL tools. In \tbl\ref{tab:cross_platform_performance_drop}, we show the percentage drop for each tool from within to cross-platform settings based on the two overall metrics: Macro and Micro avg. In terms of Macro-F1 score, RNN4SentiSE shows the largest drop of 31\% when trained in Jira and then tested in the SO dataset. That means that RNN4SentSE shows a Macro-F1 score of 0.78 tested on SO in within platform setting, but it shows a Macro F1-score of 0.54 for SO when trained in Jira. For the tool SentiMoji, the largest drop happens when the tool is trained in Jira and tested in GitHub (31\%). SentiMoji shows a Macro F1-score of 0.90 when trained and tested in GitHub (Within-platform), but it shows a Macro-F1 score of 0.62 when trained in Jira and then tested in GitHub. For the tool BERT4SentiSE, the largest drop in performance happens when the tool is trained in Jira and then tested in GitHub (29\%). BERT4SentiSE shows a Macro-F1 score of 0.92 when trained and tested in GitHub (Within-platform), but it shows a Macro F1-score of 0.65 when trained in Jira and tested in GitHub. Therefore, we find that three DL tools suffer the largest drop in performance when they are trained in the Jira dataset and then tested in the other two datasets.

\bf{BERT4SentiSE is the best performer among the three deep learning SD tools for SE both in within and cross-platform settings.} BERT4SentiSE shows Macro F1-scores of 0.92, 0.88, and 0.83 for within-platform settings in the three datasets: GitHub, SO, and Jira, respectively. The second best deep learning SD performer for within-platform settings is SentiMoji, which shows Macro F1-scores of 0.90, 0.87, and 0.81 in the three datasets: GitHub, SO, and Jira, respectively. RNN4SentiSE shows Macro F1-scores of 0.68, 0.78, 0.71 for the three datasets respectively. Similar to the close gaps in performance we observed between BERT4SentiSE and SentiMoji for within platform settings, the gaps in performance between the two tools remain close across the six cross-platform settings. For example, when trained in SO and tested in GitHub, BERT4SentiSE shows a Macro F1-score of 0.77, which SentiMoji shows a Macro F1-score of 0.75. In contrast, RNN4SentiSE shows a Macro F1-score of 0.65 for the same setup. Both BERT4SentiSE and SentiMoji are developed based on the principles of pre-trained models, i.e., the two tools were trained on a large corpus of non-SE datasets (e.g., Twitter, Wikipedia) to learn about general word embedding and representations. The architecture of RNN4SentiSE does not allow such pre-trained embedding. Pre-trained transformer models like BERT show superior performance in other natural language processing tasks~\cite{sun2019fine}. Therefore, it is intuitive that BERT4SentiSE and SentiMoji show less drop in performance than the other deep learning SD tool RNN4SentiSE.


\bf{The deep learning SD Tools suffer a drop in performance in cross-platform settings due mainly to their incorrect labeling of neutral sentences to non-neutral or vice-versa.} In \tbl\ref{tab:rq-error-analysis}, we show how the two best performing deep learning SD tools (BERT4SentiSE and SentiMoji) misclassify sentences in cross-platform settings that they correctly classified in within platform settings. 
For example, out of the sentences that are correctly classified by BERT4SentiSE in the GitHub dataset in within platform setting, 30\% of those were misclassified when trained in Jira and then tested in the same GitHub dataset. Only 1\% of those were severe misclassification, i.e., incorrectly predicted as positive/negative when the correct label was negative/positive. The rest (i.e., 29\%) of the misclassifications are mild, i.e., incorrectly predicted as neutral/non-neutral when the correct label was non-neutral/neutral. 
For example, BERT4SentiSE correctly predicts within platform settings ``that's because it's old API. :( (GitHub, oracle: negative)'' but mispredicts (neutral) in cross-platform settings when trained on Stack Overflow.
Similarly, SentiMoji correctly predicts within platform settings ``sorry for the delay (Jira, oracle: negative)'' but mispredicts (neutral) in cross-platform settings when trained on Stack Overflow.
The above observations denote that the deep learning SD tools suffer in cross-platform settings mainly due to mild errors and not due to severe errors in classifications. 



\begin{tcolorbox}[flushleft upper,boxrule=1pt,arc=0pt,left=0pt,right=0pt,top=0pt,bottom=0pt,colback=white,after=\ignorespacesafterend\par\noindent]
 \bf{RQ$_1$ Can deep learning sentiment detection tools for software engineering perform equally well for within and in cross-platform settings?} Each of three deep learning SD tools for SE suffers a drop in performance across our three studied datasets, with the most drop occurred when the tools were trained in the Jira dataset and then tested in the other two datasets. BERT4SentiSE is the best performing tool out of the three SD tools, followed closely by SentiMoji. The misclassifications in the cross-platform setting compared the within platform settings happened mainly due to mild errors, i.e., the tools misclassified neutral to non-neutral and vice-versa.
\end{tcolorbox}
\subsection{RQ$_2$ Do the deep learning SD tools perform better than non-deep learning SD tools in cross-platform settings?}\label{sec:rq-cross-platform-performance}
\subsubsection{Motivation}\label{sec:rq-cross-platform-performance-motivation}
Our findings from RQ1 show that all the three deep learning SD tools for SE in our study (BERT4SentiSE, SentiMoji, RNN4SentiSE) 
suffer from a drop in performance in cross-platform settings. This finding is consistent with those of Novielli et al.~\cite{Novielli-SEToolCrossPlatform-MSR2020} who 
observed a drop in performance in the shallow machine learning based SD tools (Senti4SD, SentiCR) in cross-platform settings. 
Given that the deep learning SD tools were previously found to have outperformed non-deep learning SD tools in SE in within-platform settings, 
it is thus necessary to know whether the tools, despite their drop in performance in cross-platform settings, 
can still outperform non-deep learning tools in cross-platform settings.
 

\subsubsection{Approach}\label{sec:rq-cross-platform-performance-approach} 
To compare the performance of the three deep learning SD tools against non-deep learning SD tools, we 
pick the best performing three non-deep learning SD tools from Novielli et al.~\cite{Novielli-SEToolCrossPlatform-MSR2020} in cross-platform 
settings: Senti4SD, SentiCR, and SentiStrengthSE. Out of the non-deep learning tools, Senti4SD and SentiCR are shallow machine learning-based, and SentiStrengthSE is rule-based. We compute the performance of the two shallow learning 
SD tools in our datasets following the same process that applied for the three deep learning tools in RQ$_1$. The approach 
is described in \sec\ref{sec:rq0-approach}, which is divided into within-platform and cross-platform settings. SentiStrengthSE is rule-based, and we do not need to train it on any dataset. Therefore, we simply run it once on each dataset and compute the performance metrics. 
We use the same performance metrics introduced in \sec\ref{sec:rq0} to report the performance.

 
\begin{table}[]
  \footnotesize
  \centering
  \caption{Agreement metrics between a tool and manual labeling.}
    \begin{tabular}{clrrrr}
    \toprule{}
    \multirow{2}{*}{\bf{Train}} & \multirow{2}{*}{\bf{Classifier }} & \multirow{2}{*}{\bf{K}} & \multirow{2}{*}{\textbf{Perfect}} & \multicolumn{2}{c}{Disagreement}\\
    \cmidrule{5-6}
    {Dataset} &  & {} & {\textbf{Agreement}} & \textbf{Severe} & \textbf{Mild} \\
    \midrule
\multicolumn{6}{c}{\textbf{Test dataset: GitHub}} \\
\midrule

\multirow{6}{*}{\parbox{2.5cm} {\centering GitHub \\ (Within-platform)}} 
 & SEntiMoji & 0.84 & 90\% & 1\% & 9\%  \\
 & BERT4SentiSE & 0.87 & 92\% & 1\% & 7\%  \\
 & RNN4SentiSE & 0.52 & 68\% & 7\% & 25\%  \\
 & Senti4SD & 0.86 & 91\% & 1\% & 8\%  \\
 & SentiCR & 0.73 & 83\% & 2\% & 15\%  \\
 & SentiStrengthSE & 0.67 & 79\% & 2\% & 19\%  \\
\cmidrule{2-6}

\multirow{6}{*}{\parbox{2.5cm} {\centering Stack Overflow \\ (Cross-platform)}}  
 & SEntiMoji & 0.61 & 75\% & 2\% & 23\%  \\
 & BERT4SentiSE & 0.65 & 78\% & 3\% & 19\%  \\
 & RNN4SentiSE & 0.47 & 66\% & 4\% & 30\%  \\
 & Senti4SD & 0.60 & 74\% & 4\% & 22\%  \\
 & SentiCR & 0.49 & 68\% & 2\% & 29\%  \\
 & SentiStrengthSE & 0.67 & 79\% & 2\% & 19\%  \\
\cmidrule{2-6}

\multirow{6}{*}{\parbox{2.5cm} {\centering Jira \\ (Cross-platform)}} 
 & SEntiMoji & 0.44 & 66\% & 1\% & 33\%  \\
 & BERT4SentiSE & 0.48 & 68\% & 2\% & 30\%  \\
 & RNN4SentiSE & 0.35 & 60\% & 4\% & 36\%  \\
 & Senti4SD & 0.50 & 69\% & 2\% & 28\%  \\
 & SentiCR & 0.34 & 60\% & 1\% & 39\%  \\
 & SentiStrengthSE & 0.67 & 79\% & 2\% & 19\%  \\
\midrule

\multicolumn{6}{c}{\textbf{Test dataset: Stack Overflow}} \\
\midrule

\multirow{6}{*}{\parbox{2.5cm} {\centering Stack Overflow (Within-platform)}}
 & SEntiMoji & 0.80 & 87\% & 1\% & 12\%  \\
 & BERT4SentiSE& 0.82 & 88\% & 1\% & 11\%  \\
 & RNN4SentiSE & 0.67 & 78\% & 3\% & 19\%  \\
 & Senti4SD & 0.77 & 85\% & 1\% & 14\%  \\
 & SentiCR & 0.73 & 82\% & 3\% & 15\%  \\
 & SentiStrengthSE & 0.68 & 79\% & 1\% & 20\%  \\

\cmidrule{2-6}

\multirow{6}{*}{\parbox{2.5cm} {\centering GitHub \\ (Cross-platform)}} 
 & SEntiMoji & 0.73 & 83\% & 1\% & 16\%  \\
 & BERT4SentiSE & 0.78 & 85\% & 1\% & 14\%  \\
 & RNN4SentiSE & 0.49 & 67\% & 4\% & 29\%  \\
 & Senti4SD & 0.67 & 78\% & 2\% & 20\%  \\
 & SentiCR & 0.56 & 71\% & 5\% & 23\%  \\
 & SentiStrengthSE & 0.68 & 79\% & 1\% & 20\%  \\
\cmidrule{2-6}

\multirow{6}{*}{\parbox{2.5cm} {\centering Jira \\ (Cross-platform)}} 
 & SEntiMoji & 0.48 & 67\% & 0\% & 33\%  \\
 & BERT4SentiSE & 0.63 & 76\% & 1\% & 23\%  \\
 & RNN4SentiSE & 0.34 & 58\% & 4\% & 38\%  \\
 & Senti4SD & 0.47 & 66\% & 1\% & 33\%  \\
 & SentiCR & 0.37 & 60\% & 2\% & 38\%  \\
 & SentiStrengthSE & 0.68 & 79\% & 1\% & 20\%  \\
\midrule

\multicolumn{6}{c}{\textbf{Test dataset: Jira}} \\
\midrule
\multirow{6}{*}{\parbox{2.5cm} {\centering Jira \\ (Within-platform)}}
 & SEntiMoji & 0.71 & 86\% & 0\% & 14\%  \\
 & BERT4SentiSE& 0.72 & 86\% & 0\% & 13\%  \\
 & RNN4SentiSE & 0.57 & 80\% & 1\% & 20\%  \\
 & Senti4SD & 0.59 & 81\% & 0\% & 19\%  \\
 & SentiCR & 0.69 & 85\% & 0\% & 15\%  \\
 & SentiStrengthSE & 0.67 & 83\% & 0\% & 17\%  \\
\cmidrule{2-6}

\multirow{6}{*}{\parbox{2.5cm} {\centering GitHub \\ (Cross-platform)}} 
 & SEntiMoji & 0.65 & 81\% & 1\% & 18\%  \\
 & BERT4SentiSE & 0.65 & 81\% & 1\% & 18\%  \\
 & RNN4SentiSE & 0.45 & 66\% & 2\% & 32\%  \\
 & Senti4SD & 0.58 & 78\% & 1\% & 21\%  \\
 & SentiCR & 0.65 & 82\% & 1\% & 17\%  \\
 & SentiStrengthSE & 0.67 & 83\% & 0\% & 17\%  \\
\cmidrule{2-6}

\multirow{6}{*}{\parbox{2.5cm} {\centering Stack Overflow \\ (Cross-platform)}} 
 & SEntiMoji & 0.65 & 82\% & 1\% & 18\%  \\
 & BERT4SentiSE & 0.66 & 82\% & 1\% & 17\%  \\
 & RNN4SentiSE & 0.57 & 77\% & 1\% & 22\%  \\
 & Senti4SD & 0.46 & 74\% & 1\% & 25\%  \\
 & SentiCR & 0.34 & 73\% & 1\% & 26\%  \\
 & SentiStrengthSE & 0.67 & 83\% & 0\% & 17\%  \\
    \bottomrule
    \end{tabular}%
  \label{tab:agreement_with_manual_labelling}%
\end{table}%

\subsubsection{Results}\label{sec:rq-cross-platform-performance-results} 
In \tbl\ref{tab:cross_platform_performance}, we show the performance of the two shallow learning SD tools (Senti4SD, SentiCR) and the rule-based tool SentiStrengthSE in the 
last three columns. In \tbl\ref{tab:cross_platform_performance_drop}, the last two columns show the percentage drop in performance of the shallow ML-based tools 
in cross-platform settings compared to their within platform settings.  As SentiStrengthSE does not need to be retrained, it does not have any performance drop in cross-platform settings and thus skipped in \tbl\ref{tab:cross_platform_performance_drop}.

\bf{Across all the SD tools for SE in our study, SentiStrengthSE is the best performer in five out of the six cross-platform settings, while BERT4SentiSE is the best performer in one cross-platform setting.} While BERT4SentiSE \it{considerably} outperforms SentiStrengthSE within platform settings, 
SentiStrengthSE \it{marginally} outperforms BERT4SentiSE in five out of the six cross-platform settings. SentiStrengthSE has a Macro F1-score of 0.78 
compared to 0.77 of BERT4SentiSE, when BERT4SentiSE is trained in SO and then tested in GitHub (F1-score difference = 0.01). Overall, the difference in Macro F1 scores 
between  SentiStrengthSE and BERT4SentiSE across the five cross-platform settings where SentiStrengthSE outperforms BERT4SentiSE is between 0.01 - 0.02 except when BERT4SentiSE 
is trained in Jira and tested on GitHub (Macro F1 difference is 0.13). 
The one cross-platform setting where BERT4SentiSE outperforms 
SentiStrengthSE is when it is trained in GitHub and tested in SO, where the difference in Macro F1-score is 0.06. Therefore, the difference in performance 
between SentiStrengthSE and BERT4SentiSE in cross-platform settings is at most 0.06 except when BERT4SentiSE is trained in a dataset other than Jira. Therefore, the performance of 
a supervised tool like BERT4SentiSE and whether it can outperform 
a rule-based tool like SentiStrengthSE in cross-platform settings can depend on the type of training dataset.

\bf{Between the shallow learning and deep learning SD tools for SE, BERT4SentiSE is the best performer in five out of six cross-platform settings, while Senti4SD 
is the best performer in one cross-platform setting.} The difference in Macro F1-score between BERT4SentiSE and Senti4SD 
is between 0.04 - 0.16 across the five cross-platform settings where BERT4SentiSE outperforms Senti4SD. Senti4SD outperforms BERT4SentiSE in one cross-platform setting (Macro F1 score of 0.67 vs. 0.65, i.e., difference = 0.02) 
when the tools were trained in Jira and tested in GitHub. BERT4SentiSE also outperforms Senti4SD in all three within platform settings. BERT4SentiSE 
also outperforms other tools across the three within platform settings. The findings confirm that similar to other domains for text classification, in SE also transformer pre-trained tools like BERT4SentiSE are superior to other supervised SD tools. BERT4SentiSE is also slightly inferior to rule-based tool SentiStrengthSE and thus could be a sensible choice both for within and cross-platform settings.    

\bf{RNN4SentiSE exhibits most severe and mild errors among the supervised SD tools for within and cross-platform settings.} In \tbl\ref{tab:agreement_with_manual_labelling}, 
we show how each of the five supervised SD tools in our study agrees with the manual labels during within and cross-platform settings. We also report the agreement of rule-based SentiStrengthSE and manual labels, but as it does not require re-training, its cross-platform agreement does not change. We compute the agreement using two metrics: Cohen $\kappa$~\cite{cohen1968weighted} and percent agreement. We compute the disagreement between a tool's prediction and manual label as severe (when a sentence with a manual label as positive is predicted as negative and vice-versa) and mild (when a sentence with a neutral label is predicted as non-neutral and vice-versa). Among the SD tools, the two-deep learning SD 
tools (BERT4SentiSE and SentiMoji) show the least amount of severe and mild disagreements with the manual labels. 
In \tbl\ref{tab:agreement_with_manual_labelling}, we show the agreements and disagreements for within and cross-platform settings. Overall, 
for any given test dataset, the amount of severe and mild disagreements increase for a tool from within to cross-platform settings. However, 
similar to RQ1, where we noted that BERT4SentiSE and SentiMoji suffer mainly from mild errors in cross-platform settings, all the supervised SD tools (except RNN4SentiSE) 
show similar pattern with manual label in \tbl\ref{tab:agreement_with_manual_labelling}. Thus the supervised SD tools can benefit 
more from fixing the mild errors during classification in cross-platform settings.    

\textbf{The two best performing deep learning SD tools (BERT4SentiSE and SentiMoji) have the lowest severe disagreement with manual
annotation among the supervised SD tools.} In Table \ref{tab:agreement_with_manual_labelling}, we report the
paired comparisons with the tools' predictions with manual labeling. In
cross-platform settings, when it is trained on Stack Overflow dataset and tested
on GitHub dataset, both BERT4SentiSE and SentiMoji has the highest severe
disagreement (3\% and 2\% respectively) as well as the highest mild disagreement (30\% and 33\%
respectively) with manual labeling. 
We find that most of
the severe disagreements (61\%) are because of subjectivity in annotation  or
examples lacking proper context. For example, a sample in the GitHub dataset
(e.g., ``It's quite a bad practice to do so :)'')  both Bert4SentiSE and
SentiMoji was able to understand the underlying context of this sentence correctly, i.e., the `:)' emoji did not confuse the DL tools and both of them
predicted negative sentiment for this sample. However, for the particular
example, the oracle is `positive', i.e., the human annotator probably had some more
context which prompted the annotator to annotate this, or this is because of
subjectivity among annotators (subjectivity in annotation is discussed further under RQ$_4$).

\begin{tcolorbox}[flushleft upper,boxrule=1pt,arc=0pt,left=0pt,right=0pt,top=0pt,bottom=0pt,colback=white,after=\ignorespacesafterend\par\noindent]
\bf{RQ$_2$ Do the deep learning SD tools perform better than non-deep learning SD tools in cross-platform settings?}
Across all the SD tools for SE in our study, SentiStrengthSE is the best performer with narrow margin with BERT4SentiSE in five out of the six cross-platform settings and BERT4SentiSE is the 
best performer in the other cross-platform setting in our three studied datasets. Between the shallow learning and deep learning SD tools for SE, 
BERT4SentiSE is the best performer in five out of six cross-platform settings, while Senti4SD 
is the best performer in one cross-platform setting.
\end{tcolorbox}

\subsection{RQ$_3$ What are the error categories of the tools in the cross-platform settings?}\label{subsec:error_analysis}\label{sec:rq-cross-platform-error}

\subsubsection{Motivation} The findings from RQ$_1$ and RQ$_2$ show that all of the supervised SD tools for SE suffer from a drop in performance in cross-platform settings. From RQ$_3$, We find that the more similar two datasets are, the less performance drop we can expect to have from them in cross-platform 
settings. However, it does not entirely explain the performance drop and why an SD predicts incorrectly for a given sentence while another SD tool predicts correctly.
Understanding error types for such sentences can offer further insights into the future improvements of the tools in cross-platform settings.

\subsubsection{Approach} We manually analyze error categories for the misclassified sentences in cross-platform settings. 
We perform the analysis in three steps. \ul{First}, we randomly select 400 sentences from
the 4,188 misclassified sentences where both
BERT4SentiSE and SentiMoji were wrong. This sample size is statistically significant with a 95\%
confidence level and five intervals. We then manually label each sentence to an
error category from Novielli et al.~\cite{Novielli-SEToolCrossPlatform-MSR2020},
who produced the categories for their study on the effectiveness of shallow
SE-specific SD tools in cross-platform settings. In this study, we used the same
seven error categories used by Novielli et al., but we changed one category name `Polar
facts' to `Domain specific' because what the authors considered as polar facts
are actually software domain-specific jargons that have some inherent desirable
or undesirable polarity (e.g., ``bugs'', ``issues'', ``resolved''). \rev{The last two authors 
participated in the error analysis process. They discussed the guideline of
categorizing the errors into different categories. The authors conducted joint
sessions in multiple iterations to determine the error category for each of the
400 sentences in our sample. During the joint sessions, the two authors labeled each of the 400 
sentences together, instead of labeling the sentences separately. This setup ensured that the two authors 
discussed together for each sentence and were able to enforce/clarify the understanding of the error codes during the labeling process. 
In addition, the two authors also consulted the first author during the labeling process, when the clarifications warranted the consultation of all the 
authors. Due to the nature of the joint sessions, the authors did not need to record the level of agreement/disagreements between themselves.}  
\ul{Second}, we analyze our error categories with those from Novielli et al.~\cite{Novielli-SEToolCrossPlatform-MSR2020}, who 
analyzed the error categories for the two shallow learning SD tools (Senti4SD and SentiCR) in the three datasets for cross-platform 
settings, where both tools were wrong. \ul{Third}, we analyze a random sample of 70 sentences where the rule-based 
tool SentistrengthSE was wrong in the three datasets. The purpose of the analysis in steps 2 and 3 was to determine whether the error categories might differ depending on the types of the tools (i.e., deep learning vs. shallow learning vs. rule-based).

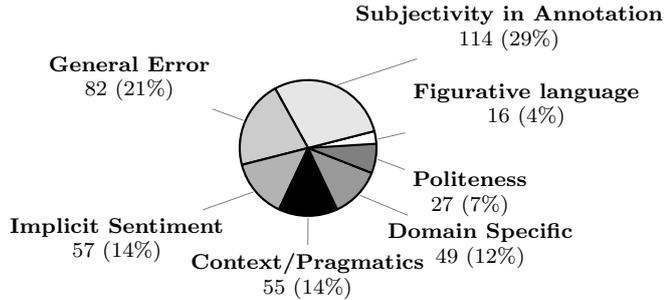
\begin{figure}[h]
	\centering\begin{tikzpicture}[scale=0.3]-
    \pie[
        /tikz/every pin/.style={align=center},
        text=pin, number in legend,
        explode=0.0,
        color={black!0, black!10, black!20, black!30,, black!40, black!50, black!60, black!70},
        ]
        {
            4/\bf{Figurative language}\\ 16 (4\%),
            29/\bf{Subjectivity in Annotation}\\ 114 (29\%),
            21/\bf{General Error}\\  82 (21\%),
            14/\bf{Implicit Sentiment}\\ 57 (14\%),
            14/\bf{Context/Pragmatics}\\ 55 (14\%),
            12/\bf{Domain Specific} \\ 49 (12\%),
            7/\bf{Politeness} \\ 27 (7\%)
        }
    \end{tikzpicture}
	\caption{Distribution of error categories by two deep learning tools (BERT4SentiSE and SentiMoji) in our 400 samples}
	\vspace{-2mm}
	\label{fig:misclassification_chart_dl}
\end{figure}


\subsubsection{Results} \bf{Most (29\%) of the errors in our sample where both BERT4SentiSE and SentiMoji are wrong are due to `Subjectivity in Annotation', i.e., 
inconsistency in the manual labeling.} In \fig\ref{fig:misclassification_chart_dl}, we show a pie chart to show the percentage distribution of seven error categories that we observed in the 400 samples of sentences where both BERT4SentiSE and SentiMoji were wrong across the six cross platform settings.  Most of the misclassifications (29\%) were observed due to
`Subjectivity in Annotation'. The second most observed category is `General
Error' (21\%). This distribution is slightly different from the distribution
reported by Novielli et al.~\cite{Novielli-SEToolCrossPlatform-MSR2020} for the
shallow ML-based SD tools, where `General Error' was observed the most (68\%).
As we discuss below, a general error occurs due to the inability of the tools to
process sentences (e.g., emoticons). One of the deep learning tools in our
study, SentiMoji, is specifically designed to handle emojis better. This is why
we find much lower coverage of misclassifications related to `General Errors' in
our sample. We now discuss the error categories with examples below.

{\ul{Subjectivity in Annotation (29\%).}} The biggest cause of misclassification is subjectivity in annotation. Human labeling is always subjective to human annotators, even if there is a standard guideline on labeling a dataset~\cite{Novielli-SEToolCrossPlatform-MSR2020}. For example, \emt{+1 looks good.} is labeled as neutral and \emt{+1 great job pat.} is labeled as positive in the same Jira dataset. This is because sometimes a conservative rater might label sentences with mild emotions as neutral~\cite{Novielli-SEToolCrossPlatform-MSR2020}. In particular, subjectivity is quite prevalent for sentences with politeness. For example, this sentence is manually labeled as neutral, but both tools (BERT4SentiSE and SentiMoji) labeled it as positive, \emt{Thanks you for quick response}.

{\ul{General errors (21\%).}} The inability of the tools to identify the linguistic cues can result in their misclassification of the polarity labeling. As we noted above, this error category is much less observed in our sample than Novielli et al.~\cite{Novielli-SEToolCrossPlatform-MSR2020}. Two deep learning tools perform relatively better in this category because both are pre-trained, which helps generalize the context of a sentence even when a sentence is not processed properly. In addition, SEngMoji is based on DeepMoji~\cite{deepMoji_2017}, which is good at detecting emoticons like  ``:)'' and ``:smiley:''. However, we still find misclassified cases in the datasets. For example, we find this error in GitHub dataset:\emt{Ok, this can be removed :-)}. The error happened when both tools were trained on Jira and tested on GitHub. Both tools provide correct polarity labels when they are trained using the SO dataset. We find that in the Jira dataset, there are sentences like this \emt{Vadim you're keeping us busy :).} or this \emt{Oh my stupid:)} or this \emt{Stupid user error :)}. The three sentences are manually labeled as non-positive, but the two tools label those as positive. Therefore, while we put such misclassifications under the general error category, another category would have been inconsistency in labeling datasets.

{\ul{Implicit Sentiment (14\%)}}. The presence of implicit sentiment can confuse any SD tools, especially when the unit of analysis is a sentence with very few words (i.e., without much context). The two deep learning SD tools most misclassify such cases as neutral because the sentences do not contain any lexical cues. For example, the following two sentences are manually labeled as negative, while both SD tools label those as neutral: \begin{inparaenum}
\item \emt{This is not what I build a framework for.}
\item \emt{25 hours gone.}
\end{inparaenum}

{\ul{Context/Pragmatics (14\%)}.} The lack of proper context or pragmatics in sentences, especially in short sentences, can make it challenging for the tools to determine the correct polarity labels. This can also be problematic when a manual labeling process also took cues from surrounding sentences of a given sentence. As Uddin et al.~\cite{Uddin-OpinionValue-TSE2019} show, lack of context can be problematic for datasets based on SO, if manual annotation of a sentence is done by looking at the entire post of the sentence in SO. For example, the following sentences are labeled as neutral by the tools but are labeled as negative in the benchmarks: \begin{inparaenum}
\item \emt{huh ... i thought i did resolve this.} (in Jira)
\item \emt{How about fixing hashcode too?} (in GitHub).
\end{inparaenum}

{\ul{Domain Specificity (12\%)}} errors occur due to the presence of SE domain-specific jargons in a sentence, which an SD tool has not observed during training. Some SE-specific jargons have inherently desirable or undesirable emotions such as ``bug'' as negative and ``patch accepted'' as positive polarity. This sentence is manually labeled as negative, but the tools predicted it as neutral: \emt{My patch wouldn't compile}. Another example is \emt{`killed that process and restarted, and it worked!}. Here, the polarity in manual labeling is positive because, in the SE context, the issue was fixed after restarting, which is a desirable outcome.
To make matters worse, these SE-specific jargons are inconsistently labeled across datasets, such as bug-related sentences in the Jira dataset are always labeled with positive or negative polarity~\cite{Novielli-SEToolCrossPlatform-MSR2020, Novielli-BenchmarkStudySentiSE-MSR2018, Novielli-SESentimentCP-EMSE2021}. However, in the other two datasets, bug-related discussions are labeled as neutral unless there is a clear expression of emotion. The following is manually labeled as negative but is predicted as neutral by both tools: \emt{The visibility is missing, and the default value seems weird}.

{\ul{Politeness (7\%).}} type errors occur due to two reasons \begin{inparaenum}[(1)]
\item sentences conveying politeness are labeled inconstantly across datasets. For example,  politeness is considered neutral in SO and GitHub datasets (e.g., \emt{Thank you})~\cite{Novielli-SEToolCrossPlatform-MSR2020}. On the other hand, in the Jira dataset, thanking expression is labeled as positive polarity (e.g., \emt{Thanks Ashish For committing the patch.}),
\item sentences with politeness are most vulnerable to subjective annotation, especially in cross-platform datasets (as we noted above under subjective annotation category).
\end{inparaenum}

{\ul{Figurative language (4\%)}.}  errors occur due to the presence of sarcasm or expletives in sentences. For example, the following sentence is labeled as positive in the GitHub dataset, but both tools predicted it as neutral in cross-platform setting: \emt{There is a backdoor in this commit, i'll give a cookie to the one who finds it first:P}. The following sentences are labeled as neutral in the benchmarks, but the tools predicted those as neutral: \begin{inparaenum}
\item \emt{because hack shit methods must die.} (GitHub),
\item \emt{I used to be the latter, but then I went to law school and got a degree in this shit.} (Jira). 
\end{inparaenum}

\begin{figure}[h]
	\centering\begin{tikzpicture}[scale=0.3]-
    \pie[
        /tikz/every pin/.style={align=center},
        text=pin, number in legend,
        explode=0.0,
        color={black!0, black!10, black!20, black!30,, black!40, black!50, black!60, black!70},
        ]
        {
            68/\bf{General Error}  214 (68\%),
            11/\bf{Subjectivity in Annotation} 35 (11\%),
            8/\bf{Domain Specific (Polar facts)}  25 (8\%),
            6/\bf{Politeness}   19 (6\%),
            5/\bf{Implicit Sentiment} 16 (5\%),   
            2/\bf{Figurative language} 6 (2\%),
            2/\bf{Context/Pragmatics} 6 (2\%)
        }
    \end{tikzpicture}
	\caption{Distribution of error categories by Shallow SD tools (SentiCR and Senti4SD) as reported by Novielli et al.~\cite{Novielli-SEToolCrossPlatform-MSR2020}}
	\vspace{-2mm}
	\label{fig:misclassification_sentisdsenticr}
\end{figure}
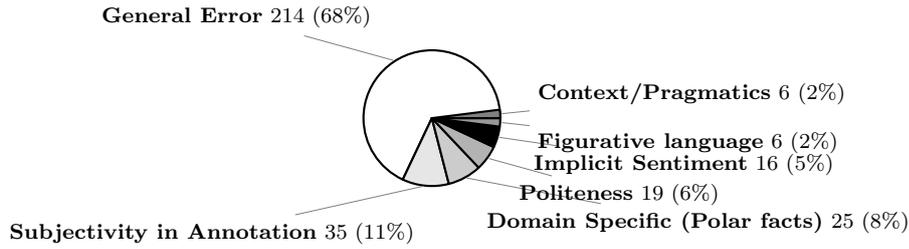

\bf{The distribution of error categories in the two deep learning tools differs \it{slightly} from those observed for shallow SD tools in cross-platform settings.} 
\fig\ref{fig:misclassification_sentisdsenticr} shows the error categories in 320 sentences where both of the two shallow learning-based SD tools (Senti4SD and SentiCR) 
are wrong. We report the results from Novielli et al.~\cite{Novielli-SEToolCrossPlatform-MSR2020}, i.e., we did not analyze the 320 sentences. 
The most frequent category for the two shallow learning SD tools is `General Error', which was the second most frequent category in \fig\ref{fig:misclassification_chart_dl}, i.e., 
error categories observed for the two deep learning SD tools. The most observed error category for the two deep learning SD tools in \fig\ref{fig:misclassification_chart_dl} 
is `Subjectivity in Annotation', which is the second most observed category for the two shallow learning SD tools in \fig\ref{fig:misclassification_sentisdsenticr}. 
The difference in the distribution of error categories between the shallow learning and deep learning SD tools is due to two reasons: 
\begin{inparaenum}[(1)]
\item General errors mainly occur due to the inability of a tool to analyze non-textual contents that are useful to detect polarity (e.g., emoji). 
The two deep learning SD tools, especially SentiMoji are pre-trained on a large corpus of emojis found in Twitter. Therefore, the two deep learning SD tools 
are more resilient towards such general errors during polarity labeling in cross-platform settings.
\item The two deep learning SD tools, especially BERT4SentiSE is based on BERT. BERT is better designed to infer contexts from a sentence due to the novel use 
of transformer architecture and attention layer. Thus, BERT4SentiSE can recognize the underlying context from training data, which becomes problematic 
for sentences that are similar between the training and testing datasets but that are labeled differently between the two datasets due to the subjectivity 
in manual labeling. Thus, the deep learning SD tools suffer more from such inconsistency.
\end{inparaenum}     

\begin{figure}[h]
	\centering\begin{tikzpicture}[scale=0.3]-
    \pie[
        /tikz/every pin/.style={align=center},
        text=pin, number in legend,
        explode=0.0,
        color={black!0, black!10, black!20, black!30,, black!40, black!50, black!60, black!70},
        ]
        {
            2/\bf{Figurative language} 2 (2\%),
            39/\bf{Domain Specific}  36 (39\%),
            19/\bf{Politeness}   18 (19\%),
            16/\bf{Subjectivity in Annotation} 15 (16\%),
            14/\bf{General Error}  13 (14\%),
            5/\bf{Context/Pragmatics} 5 (5\%),
            4/\bf{Implicit Sentiment} 4 (4\%)
        }
    \end{tikzpicture}
	\caption{Distribution of error categories by the rule-based tool (SentiStrengthSE) in our 93 samples}
	\vspace{-2mm}
	\label{fig:misclassification_sentistrengthse}
\end{figure}
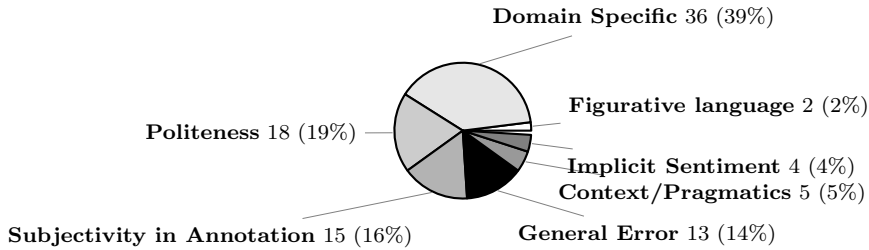

\textbf{Lack of understanding of context and underlying domain specificity contributes to the most error in rule-based SD tool SentistrengthSE.} 
In our three datasets, SentiStrengthSE misclassifies 3,470 sentences, out of which we randomly sample 
93 sentences for manual analysis of error categories. This sample is statistically significant with a 95\% confidence level with 10 confidence interval. 
Figure \ref{fig:misclassification_sentistrengthse} shows the highlight
of the error misclassification for SentiStrengthSE. 
From our
error analysis, we find that the main cause for misclassification for the
lexicon-based tool SentiStrengthSE is failure to understand domain-specific
context (39\%). PTM DL-based SD tools perform quite well in this category (only
12\% errors).  Some examples of errors by SentistrengthSE related to `domain specificity' are:  ``Confirm, this breaks compile. :( (GitHub)'', ``ROT13. It's
super efficient too! (Stack Overflow)''). Inconsistent labeling guideline is
the second biggest cause of errors (19\%): politeness such as ``Thanks'',
``Thank you, some\_name'' is always predicted as positive polarity by
SentiStrengthSE but in Jira dataset, politeness is not annotated as positive
unless there is a clear expression of emotions. The third categories of errors
belong to subjective annotation (16\%). The fourth error category is General categories (14\%), i.e., inability to understand linguistic
cues (e.g., ``+1  Patch looks nice.'' ``:+1: for speed!",  ``ya +1 for clearer
code comment :D") this type of language cues are not detected by the
SentiStrengthSE. 

\textbf{When both the two best performing deep learning SD tools (BERT4SentiSE and SentiMoji) have severe disagreement with manual labeling, 
most of the time, it is due to subjectivity in manual labeling.}
From Table \ref{tab:agreement_with_manual_labelling} we see in cross-platform
settings when trained on Stack Overflow and test on GitHub dataset both the
pretrained DL-based SD tools SentiMoji and BERT4SentiSE have the highest severe
disagreement with manual labeling (2\% and 3\% respectively). In order to
better understand the issues, we investigate a little further. We find 74 samples
in GitHub dataset where but BERT4SentiSE and SentiMoji have severe disagreement
with the manual labeling, i.e., if the manual labeling for a particular
sentence is positive, then both the SD tools have predicted negative and vice
versa. After manually categorizing these 74 samples into the seven error
categories, we find that 45 samples (i.e., 61\%) actually have subjective
annotation. That means we do agree with the manual labeling assigned to
those samples, and rather we think the tools' predictions are correct. For
example, ``God I am awful at this :)'' and ``It's quite a bad practice to do so
:)'' are labeled as positive, but both the tools predicted negative for this and
we agree with this. These errors can be fixed by following standard annotation
guidelines.
We also find that in 16 (i.e., 22\%) cases, both tools have made failed to understand
implicit sentiments in sentences. For example,``Thanks for this, but things are
about to explode for me. :( (GitHub, negative)'' and ``yea,right..silly me for
not checking another commits..sorry : (GitHub, positive)''. We find 10 (i.e., 14\%) cases the tools make mistakes because of not properly
understanding software domain concepts. For example, ``Great, again
client freezes ....''. Both the tools have predicted positive for the sample
where the correct prediction should be negative. Our analysis shows the tools
failed to understand SE-domain specific terms such as ``fix'', ``hang'',
``difficult to use a library''. 

\begin{figure}[h]
\centering
\includegraphics[scale=0.46]{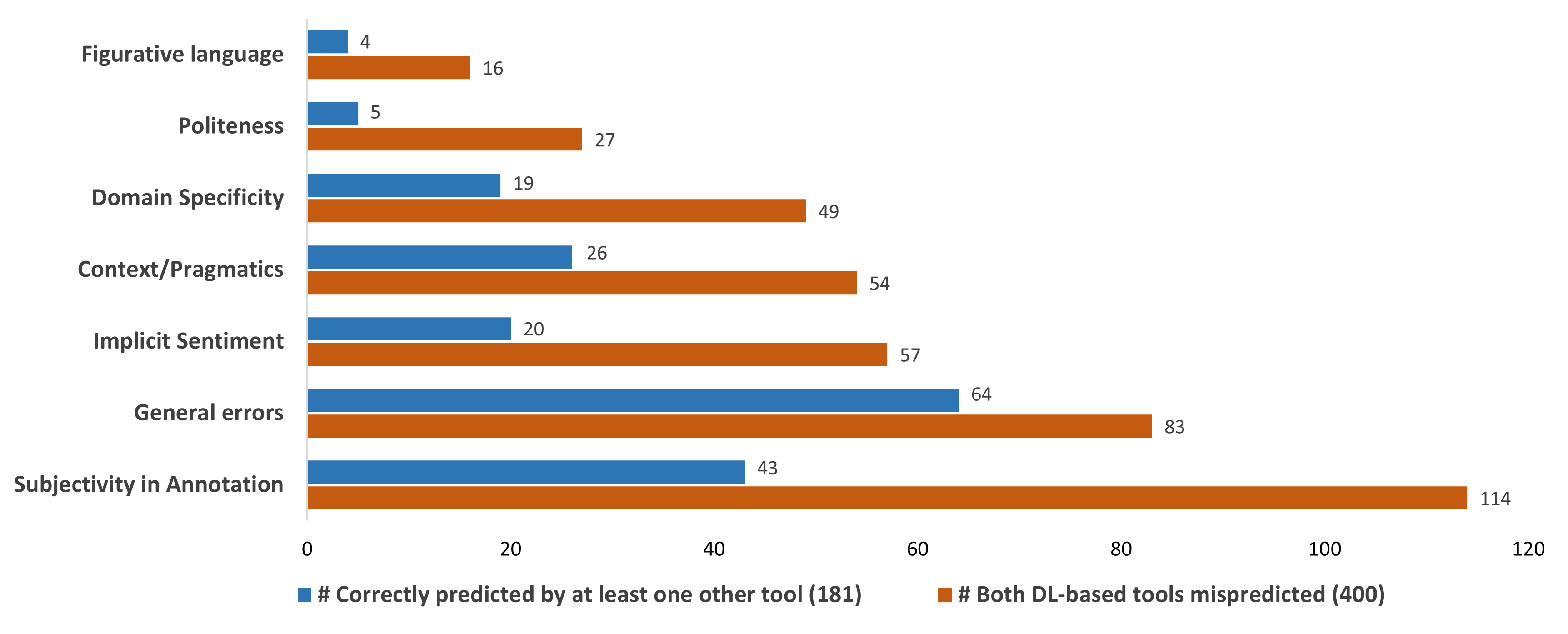}
\caption{Misclassifications by both BERT4SentiSE and SentiMoji in \fig\ref{fig:misclassification_chart_dl} that can be corrected by at least one non-deep learning SD tool}
\label{fig:distribution_error_complement}
\end{figure}
\bf{Around 45\% of the misclassified sentences where both deep learning SD tools (BERT4SentiSE, SentiMoji) are wrong, can be corrected by the 
non-deep learning SD tools.}  \fig\ref{fig:distribution_error_complement} shows whether and how each of the 400 sentences in our sample from \fig\ref{fig:misclassification_chart_dl} 
where both of the deep learning SD tools (BERT4SentiSE, SentiMoji), can be corrected by the other three non-deep learning SD tools in our study. 
Most of the general errors by the two deep learning SD tools can be corrected by the non-deep learning SD tools. We found that this is because of the data pre-processing error, e.g., the in-built Tokenizer module in the deep learning SD tools was wrong in some cases to handle emoticons. The shallow learning SD tool SentiCR uses a pre-defined dictionary of emoticons that compensate for many such errors. Overall, the 
deep learning SD tools could benefit from the correct classifications from the non-deep learning SD tools across all seven error categories.


\begin{tcolorbox}[flushleft upper,boxrule=1pt,arc=0pt,left=0pt,right=0pt,top=0pt,bottom=0pt,colback=white,after=\ignorespacesafterend\par\noindent]
\bf{RQ$_3$ What are the error categories of the tools in the cross-platform settings?}
Subjectivity in annotation is responsible for the most amount of errors in our
sample of 400 cases where both BERT4SentiSE and SentiMoji were wrong in
cross-platform settings. We also observed inconsistent cases between datasets
(i.e., similar sentences labeled differently between datasets). General errors are the most prevalent for sentences where 
both shallow learning SD tools (Senti4SD, SentiCR) are wrong. For the rule-based tool SentistrengthSE, lack of understanding of underlying context/domain 
contributed to the most amount of errors.   
\end{tcolorbox}

\subsection{RQ$_4$ To what extent the deep learning and non-deep learning sentiment detection tools can complement each other in cross-platform settings?}\label{sec:cross-platform-agreement}
\subsubsection{Motivation}
Our findings from RQ1 and RQ2 denote that the supervised SD tools suffer from a drop in performance in cross-platform settings. Our manual analysis of misclassifications by the tools in RQ4 shows that 
the errors can be seven types with general errors, subjectivity in annotation, and domain specificity 
being the most prevalent error categories across the tools.  
The SD tools are designed using different techniques and based on different platforms, e.g., SentistrengthSE was designed by primarily analyzing code reviews in Jira while 
SentiMoji uses emojis collected from forums like Twitter. Intuitively, given the diversity of how the tools are designed, the tools can complement each other. Indeed, \fig\ref{fig:distribution_error_complement} in RQ4 shows that 
the non-deep learning SD tools can complement the deep learning SD tools by correcting many of the misclassified cases 
where both of the top two deep learning SD tools (BERT4SentiSE and SentiMoji) are wrong. 
Therefore, it is important to know whether we could reduce the drop in performance in the SD tools 
in cross-platform settings by allowing the tools to complement each other, e.g., by taking 
the final polarity decision of a given sentence based on the polarity labels from multiple SD tools. 


\begin{sidewaystable}
\vspace{15cm}
\parbox{.32\linewidth}{
\renewcommand{\thetable}{\arabic{table}a}
    \tiny
  \centering
  \caption{Agreement between tools. Test in GitHub (G). Mji = SentiMoji Brt = BERT4SentiSE K = Cohen $\kappa$}
    \begin{tabular}{clrr|rr}
    \toprule
    \multirow{2}{*}{\bf{Data}} & \multirow{2}{*}{\bf{T1 $\leftrightarrow$ T2}} &  \multicolumn{2}{c}{\textbf{Agreement}} & \multicolumn{2}{c}{\bf{Disagree}}\\
    \cmidrule{3-6}
    & & {\bf{$\kappa$}} & {\textbf{P}} & \textbf{S} & \textbf{M} \\
    \midrule
\multirow{15}{*}{G (W)} 
 & Mji vs Brt & 0.86 & 91\% & 1\% & 8\%  \\
 & Mji vs Rnn & 0.52 & 68\% & 7\% & 25\%  \\
 & Mji vs S4D & 0.83 & 89\% & 1\% & 10\%  \\
 & Mji vs SCr & 0.76 & 85\% & 2\% & 13\%  \\
 & Mji vs SSt & 0.68 & 79\% & 2\% & 18\%  \\
 & Brt vs Rnn & 0.53 & 69\% & 7\% & 24\%  \\
 & Brt vs S4D & 0.85 & 90\% & 2\% & 8\%  \\
 & Brt vs SCr & 0.74 & 83\% & 2\% & 14\%  \\
 & Brt vs SSt & 0.68 & 80\% & 2\% & 18\%  \\
 & Rnn vs S4D & 0.52 & 69\% & 7\% & 25\%  \\
 & Rnn vs SCr & 0.43 & 63\% & 7\% & 31\%  \\
 & Rnn vs SSt & 0.44 & 63\% & 6\% & 31\%  \\
 & S4D vs SCr & 0.74 & 83\% & 2\% & 14\%  \\
 & S4D vs SSt & 0.67 & 79\% & 3\% & 18\%  \\
 & SCr vs SSt & 0.63 & 77\% & 3\% & 21\%  \\
\midrule

\multirow{15}{*}{S(C)} 
 & Mji vs Brt & 0.72 & 83\% & 1\% & 15\%  \\
 & Mji vs Rnn & 0.58 & 76\% & 2\% & 22\%  \\
 & Mji vs S4D & 0.51 & 70\% & 3\% & 27\%  \\
 & Mji vs SCr & 0.40 & 68\% & 2\% & 30\%  \\
 & Mji vs SSt & 0.57 & 74\% & 2\% & 25\%  \\
 & Brt vs Rnn & 0.59 & 76\% & 2\% & 22\%  \\
 & Brt vs S4D & 0.52 & 70\% & 4\% & 26\%  \\
 & Brt vs SCr & 0.39 & 66\% & 3\% & 31\%  \\
 & Brt vs SSt & 0.61 & 76\% & 2\% & 22\%  \\
 & Rnn vs S4D & 0.43 & 65\% & 4\% & 31\%  \\
 & Rnn vs SCr & 0.30 & 62\% & 3\% & 35\%  \\
 & Rnn vs SSt & 0.48 & 68\% & 3\% & 29\%  \\
 & S4D vs SCr & 0.45 & 68\% & 2\% & 30\%  \\
 & S4D vs SSt & 0.55 & 72\% & 3\% & 25\%  \\
 & SCr vs SSt & 0.45 & 68\% & 2\% & 31\%  \\
\midrule

\multirow{15}{*}{J(C)} 
 & Mji vs Brt & 0.78 & 90\% & 0\% & 10\%  \\
 & Mji vs Rnn & 0.54 & 81\% & 1\% & 19\%  \\
 & Mji vs S4D & 0.60 & 82\% & 1\% & 18\%  \\
 & Mji vs SCr & 0.65 & 87\% & 0\% & 13\%  \\
 & Mji vs SSt & 0.46 & 69\% & 0\% & 30\%  \\
 & Brt vs Rnn & 0.56 & 81\% & 1\% & 18\%  \\
 & Brt vs S4D & 0.60 & 81\% & 1\% & 18\%  \\
 & Brt vs SCr & 0.67 & 87\% & 0\% & 13\%  \\
 & Brt vs SSt & 0.51 & 71\% & 1\% & 28\%  \\
 & Rnn vs S4D & 0.44 & 74\% & 2\% & 24\%  \\
 & Rnn vs SCr & 0.45 & 79\% & 1\% & 20\%  \\
 & Rnn vs SSt & 0.36 & 63\% & 2\% & 34\%  \\
 & S4D vs SCr & 0.52 & 79\% & 1\% & 20\%  \\
 & S4D vs SSt & 0.48 & 70\% & 2\% & 29\%  \\
 & SCr vs SSt & 0.39 & 66\% & 0\% & 34\%  \\

    \bottomrule
    \end{tabular}%
  \label{tab:agreement_test_github}%
}
\hfill
\parbox{.32\linewidth}{
\addtocounter{table}{-1}
\renewcommand{\thetable}{\arabic{table}b}
   \tiny
  \centering
  \caption{Agreement between tools. Test in SO (S). S4D = Senti4SD, Scr = SentiCR, P = Percent Agreement}
    \begin{tabular}{clrr|rr}
    \toprule
    \multirow{2}{*}{\bf{Data}} & \multirow{2}{*}{\bf{T1 $\leftrightarrow$ T2}} &  \multicolumn{2}{c}{\textbf{Agreement}} & \multicolumn{2}{c}{\bf{Disagree}}\\
    \cmidrule{3-6}
    & & {\bf{$\kappa$}} & {\textbf{P}} & \textbf{S} & \textbf{M} \\
    \midrule

\multirow{15}{*}{ S (W)} 
 & Mji vs Brt & 0.86 & 91\% & 1\% & 8\%  \\
 & Mji vs Rnn & 0.70 & 80\% & 3\% & 17\%  \\
 & Mji vs S4D & 0.81 & 88\% & 1\% & 12\%  \\
 & Mji vs SCr & 0.78 & 85\% & 2\% & 12\%  \\
 & Mji vs SSt & 0.72 & 81\% & 1\% & 17\%  \\
 & Brt vs Rnn & 0.71 & 81\% & 2\% & 17\%  \\
 & Brt vs S4D & 0.83 & 88\% & 1\% & 11\%  \\
 & Brt vs SCr & 0.78 & 85\% & 3\% & 12\%  \\
 & Brt vs SSt & 0.72 & 81\% & 1\% & 18\%  \\
 & Rnn vs S4D & 0.70 & 80\% & 3\% & 17\%  \\
 & Rnn vs SCr & 0.65 & 77\% & 4\% & 19\%  \\
 & Rnn vs SSt & 0.63 & 75\% & 3\% & 21\%  \\
 & S4D vs SCr & 0.78 & 85\% & 3\% & 12\%  \\
 & S4D vs SSt & 0.73 & 82\% & 1\% & 17\%  \\
 & SCr vs SSt & 0.70 & 81\% & 2\% & 17\%  \\

\midrule

\multirow{15}{*}{ G (C)} 
 & Mji vs Brt & 0.79 & 86\% & 1\% & 13\%  \\
 & Mji vs Rnn & 0.53 & 70\% & 4\% & 27\%  \\
 & Mji vs S4D & 0.70 & 80\% & 3\% & 17\%  \\
 & Mji vs SCr & 0.62 & 75\% & 5\% & 20\%  \\
 & Mji vs SSt & 0.67 & 79\% & 1\% & 20\%  \\
 & Brt vs Rnn & 0.53 & 69\% & 5\% & 26\%  \\
 & Brt vs S4D & 0.73 & 82\% & 2\% & 16\%  \\
 & Brt vs SCr & 0.60 & 74\% & 5\% & 21\%  \\
 & Brt vs SSt & 0.69 & 79\% & 1\% & 20\%  \\
 & Rnn vs S4D & 0.54 & 69\% & 4\% & 26\%  \\
 & Rnn vs SCr & 0.40 & 61\% & 6\% & 33\%  \\
 & Rnn vs SSt & 0.48 & 66\% & 4\% & 30\%  \\
 & S4D vs SCr & 0.58 & 73\% & 4\% & 23\%  \\
 & S4D vs SSt & 0.65 & 77\% & 1\% & 22\%  \\
 & SCr vs SSt & 0.56 & 71\% & 4\% & 25\%  \\
\midrule

\multirow{15}{*}{ J (C)} 
 & Mji vs Brt & 0.66 & 82\% & 0\% & 18\%  \\
 & Mji vs Rnn & 0.47 & 75\% & 1\% & 23\%  \\
 & Mji vs S4D & 0.60 & 81\% & 1\% & 19\%  \\
 & Mji vs SCr & 0.56 & 80\% & 1\% & 19\%  \\
 & Mji vs SSt & 0.49 & 68\% & 0\% & 31\%  \\
 & Brt vs Rnn & 0.43 & 70\% & 2\% & 28\%  \\
 & Brt vs S4D & 0.57 & 77\% & 1\% & 22\%  \\
 & Brt vs SCr & 0.52 & 75\% & 1\% & 25\%  \\
 & Brt vs SSt & 0.60 & 75\% & 0\% & 25\%  \\
 & Rnn vs S4D & 0.42 & 72\% & 2\% & 27\%  \\
 & Rnn vs SCr & 0.33 & 70\% & 1\% & 29\%  \\
 & Rnn vs SSt & 0.36 & 60\% & 3\% & 36\%  \\
 & S4D vs SCr & 0.51 & 77\% & 1\% & 22\%  \\
 & S4D vs SSt & 0.46 & 66\% & 1\% & 33\%  \\
 & SCr vs SSt & 0.40 & 63\% & 1\% & 36\%  \\


    \bottomrule
    \end{tabular}%
  \label{tab:agreement_test_so}%
}
\hfill
\parbox{.32\linewidth}{
  \addtocounter{table}{-1}
  \renewcommand{\thetable}{\arabic{table}c}
  \tiny
  \centering
  \caption{Agreement between tools. Test in Jira (J). SSt = SentistrengthSE, C, W = Cross, Within-Platform}
    \begin{tabular}{clrr|rr}
    \toprule
    \multirow{2}{*}{\bf{Data}} & \multirow{2}{*}{\bf{T1 $\leftrightarrow$ T2}} &  \multicolumn{2}{c}{\textbf{Agreement}} & \multicolumn{2}{c}{\bf{Disagree}}\\
    \cmidrule{3-6}
    & & {\bf{$\kappa$}} & {\textbf{P}} & \textbf{S} & \textbf{M} \\
    \midrule

\multirow{15}{*}{ J (W)} 
 & Mji vs Brt & 0.85 & 93\% & 0\% & 7\%  \\
 & Mji vs Rnn & 0.68 & 86\% & 0\% & 14\%  \\
 & Mji vs S4D & 0.72 & 87\% & 0\% & 13\%  \\
 & Mji vs SCr & 0.81 & 91\% & 0\% & 9\%  \\
 & Mji vs SSt & 0.75 & 87\% & 0\% & 13\%  \\
 & Brt vs Rnn & 0.68 & 85\% & 1\% & 14\%  \\
 & Brt vs S4D & 0.72 & 87\% & 0\% & 13\%  \\
 & Brt vs SCr & 0.79 & 90\% & 0\% & 10\%  \\
 & Brt vs SSt & 0.75 & 87\% & 0\% & 13\%  \\
 & Rnn vs S4D & 0.61 & 83\% & 0\% & 17\%  \\
 & Rnn vs SCr & 0.70 & 86\% & 1\% & 14\%  \\
 & Rnn vs SSt & 0.63 & 81\% & 1\% & 18\%  \\
 & S4D vs SCr & 0.70 & 86\% & 0\% & 13\%  \\
 & S4D vs SSt & 0.67 & 83\% & 1\% & 16\%  \\
 & SCr vs SSt & 0.77 & 88\% & 1\% & 11\%  \\
\midrule

\multirow{15}{*}{ G (C)} 
 & Mji vs Brt & 0.81 & 89\% & 1\% & 10\%  \\
 & Mji vs Rnn & 0.56 & 72\% & 3\% & 25\%  \\
 & Mji vs S4D & 0.72 & 84\% & 1\% & 14\%  \\
 & Mji vs SCr & 0.75 & 86\% & 1\% & 13\%  \\
 & Mji vs SSt & 0.75 & 86\% & 1\% & 13\%  \\
 & Brt vs Rnn & 0.57 & 73\% & 3\% & 24\%  \\
 & Brt vs S4D & 0.73 & 85\% & 1\% & 14\%  \\
 & Brt vs SCr & 0.75 & 87\% & 1\% & 13\%  \\
 & Brt vs SSt & 0.75 & 86\% & 1\% & 13\%  \\
 & Rnn vs S4D & 0.52 & 70\% & 3\% & 27\%  \\
 & Rnn vs SCr & 0.50 & 69\% & 2\% & 28\%  \\
 & Rnn vs SSt & 0.54 & 71\% & 2\% & 27\%  \\
 & S4D vs SCr & 0.68 & 83\% & 1\% & 16\%  \\
 & S4D vs SSt & 0.70 & 83\% & 1\% & 16\%  \\
 & SCr vs SSt & 0.76 & 87\% & 1\% & 12\%  \\
\midrule

\multirow{15}{*}{ S (C)} 
 & Mji vs Brt & 0.80 & 89\% & 1\% & 10\%  \\
 & Mji vs Rnn & 0.70 & 84\% & 1\% & 15\%  \\
 & Mji vs S4D & 0.61 & 79\% & 1\% & 19\%  \\
 & Mji vs SCr & 0.35 & 70\% & 1\% & 29\%  \\
 & Mji vs SSt & 0.77 & 87\% & 1\% & 12\%  \\
 & Brt vs Rnn & 0.72 & 84\% & 1\% & 14\%  \\
 & Brt vs S4D & 0.60 & 79\% & 2\% & 19\%  \\
 & Brt vs SCr & 0.36 & 71\% & 1\% & 28\%  \\
 & Brt vs SSt & 0.77 & 87\% & 1\% & 12\%  \\
 & Rnn vs S4D & 0.56 & 77\% & 2\% & 22\%  \\
 & Rnn vs SCr & 0.30 & 67\% & 1\% & 32\%  \\
 & Rnn vs SSt & 0.70 & 83\% & 1\% & 16\%  \\
 & S4D vs SCr & 0.41 & 76\% & 1\% & 23\%  \\
 & S4D vs SSt & 0.59 & 78\% & 2\% & 20\%  \\
 & SCr vs SSt & 0.34 & 70\% & 1\% & 29\%  \\
    \bottomrule
    \end{tabular}%
  \label{tab:agreement_test_jira}%
 }
\end{sidewaystable}

\begin{figure*}[t]
\centering
\captionsetup[subfigure]{labelformat=empty}
\pgfplotstableread{
1	.78 .78 .80 .79 .75 .82 .79 .80      .80 .78 .80 .81 .76 .82 .78 .80    .85 .75 .81 .85 .77 .84 .80 .82     .85 .78 .82 .85 .77 .84 .81 .83    .79 .61 .70 .75 .73 .74 .73 .72    .76 .61 .68 .72 .71 .70 .70 .69
2	.78 .80 .81 .68 .69 .81 .80 .81      .80 .80 .81 .80 .79 .82 .81 .81    .79 .81 .83 .78 .68 .82 .80 .83     .80 .80 .82 .84 .75 .82 .80 .82    .79 .69 .72 .76 .61 .75 .73 .74    .76 .67 .69 .74 .62 .72 .70 .71
}\datatable
\pgfplotscreateplotcyclelist{xcolor}{%
red!60, orange, yellow, blue!50, brown, violet, purple, black!60
}

\edef\mylst{"a", "b", "c", "d", "e", "f", "g", "h"}

      \subfloat[Test on GitHub (F1 score)]
      {
      \resizebox{2.2in}{!}{%
      \begin{tikzpicture}
      
        	\begin{axis}[
        	xtick=data,
        	xticklabels={Trained on SO\\(Baseline: SentiStrengthSE), Trained on Jira\\(Baseline: SentiStrengthSE)},
        	xticklabel style={align=center},
        	enlarge y limits=false,
        	enlarge x limits=0.5,
            axis y line=none,
            axis x line*=bottom,
            axis y line*=left,
        	ymin=0,ymax=1,
        	ybar,
        	bar width=0.6cm,
        	width=5.0in,
        	height = 2.3in,
        	ymajorgrids=false,
         	yticklabel style={font=\LARGE},
        	xticklabel style={font=\LARGE, /pgf/number format/fixed},	
        	major x tick style = {opacity=0},
        	minor x tick num = 1,    
        	minor tick length=1ex,
        	legend columns=2, 
            legend style={
                /tikz/column 2/.style={
                    column sep=5pt,
                },
                },
            legend style={draw=none},
        	legend style={at={(0.5,1.40)},
        	legend style={font=\LARGE},
        	legend image post style={scale=2.0},
            anchor=north,legend columns=-1
            },
            nodes near coords style={font=\LARGE, text=black},
            nodes near coords style={rotate=90,  anchor=west}, %
        	nodes near coords =\pgfmathprintnumber{\pgfplotspointmeta},
        	cycle list name = xcolor,
            every axis plot/.append style={fill,draw=black!80}%
        	]
        	\addplot table[x index=0,y index=1] \datatable;
        	\addplot table[x index=0,y index=2] \datatable;
        	\addplot table[x index=0,y index=3] \datatable;
        	\addplot table[x index=0,y index=4] \datatable;
            \addplot table[x index=0,y index=5] \datatable;
        	\addplot table[x index=0,y index=6] \datatable;
        	\addplot table[x index=0,y index=7] \datatable;
        	\addplot table[x index=0,y index=8] \datatable;


            \legend{Baseline,
            		 Majority-All, , ,
                     Majority-Shallow ML,
                     Majority-PMT DL + Rule, , ,
            		 }
        	\end{axis}
    	\end{tikzpicture}
     	}
    	\label{presentation-inconsistency}
    } 
    \subfloat[Test on GitHub (Precision)]
      {
       \resizebox{2.2in}{!}{%
      \begin{tikzpicture}
        	\begin{axis}[
        	xtick=data,
        	xticklabels={Trained on SO\\(Baseline: SentiStrengthSE), Trained on Jira\\(Baseline: SentiStrengthSE)},
        	xticklabel style={align=center},
        	enlarge y limits=false,
        	enlarge x limits=0.50,
            axis y line=none,
            axis x line*=bottom,
            axis y line*=left,
        	ymin=0,ymax=1,
        	ybar,
        	bar width=0.6cm,
        	width=5.0in,
        	height = 2.3in,
        	ymajorgrids=false,
         	yticklabel style={font=\LARGE},
        	xticklabel style={font=\LARGE, /pgf/number format/fixed},	
        	major x tick style = {opacity=0},
        	minor x tick num = 1,    
        	minor tick length=1ex,
        	legend columns=2, 
            legend style={
                /tikz/column 2/.style={
                    column sep=5pt,
                },
                },
            legend style={draw=none},
        	legend style={at={(0.5,1.40)},
        	legend style={font=\LARGE},
        	legend image post style={scale=2.0},
            anchor=north,legend columns=-1
            },
            nodes near coords style={font=\LARGE, text=black},
            nodes near coords style={rotate=90,  anchor=west}, %
        	nodes near coords =\pgfmathprintnumber{\pgfplotspointmeta},
        	cycle list name = xcolor,
            every axis plot/.append style={fill,draw=black!80}%
        	]
        	\addplot table[x index=0,y index=9] \datatable;
        	\addplot table[x index=0,y index=10] \datatable;
        	\addplot table[x index=0,y index=11] \datatable;
        	\addplot table[x index=0,y index=12] \datatable;
            \addplot table[x index=0,y index=13] \datatable;
        	\addplot table[x index=0,y index=14] \datatable;
        	\addplot table[x index=0,y index=15] \datatable;
        	\addplot table[x index=0,y index=16] \datatable;

            \legend{, , Majority-All (Except RNN4SentiSE),
            		 Majority-PMT DL, , ,
                     Majority-Shallow ML + Rule,
                     Majority-BTPC
            		}
        	\end{axis}
    	\end{tikzpicture}
     	}
    	\label{presentation-inconsistency}
    } \\
    
    \subfloat[Test on Stack Overflow (SO) (F1 score)]
      {
       \resizebox{2.2in}{!}{%
      \begin{tikzpicture}
        	\begin{axis}[
        	xtick=data,
        	xticklabels={Train on GitHub\\(Baseline: BERT4SentiSE), Train on Jira\\(Baseline: BERT4SentiSE)},
        	xticklabel style={align=center},
        	enlarge y limits=false,
        	enlarge x limits=0.5,
            axis y line=none,
            axis x line*=bottom,
            axis y line*=left,
        	ymin=0,ymax=1,
        	ybar,
        	bar width=0.6cm,
        	width=5.0in,
        	height = 2.3in,
        	ymajorgrids=false,
         	yticklabel style={font=\LARGE},
        	xticklabel style={font=\LARGE, /pgf/number format/fixed},	
        	major x tick style = {opacity=0},
        	minor x tick num = 1,    
        	minor tick length=1ex,
        	legend columns=2, 
            legend style={
                /tikz/column 2/.style={
                    column sep=5pt,
                },
                },
            legend style={draw=none},
        	legend style={at={(0.5,1.40)},
        	legend style={font=\LARGE},
            anchor=north,legend columns=-1
            },
            nodes near coords style={font=\LARGE, text=black},
            nodes near coords style={rotate=90,  anchor=west}, %
        	nodes near coords =\pgfmathprintnumber{\pgfplotspointmeta},
        	cycle list name = xcolor,
            every axis plot/.append style={fill,draw=black!80}%
        	]
        	\addplot table[x index=0,y index=17] \datatable;
        	\addplot table[x index=0,y index=18] \datatable;
        	\addplot table[x index=0,y index=19] \datatable;
        	\addplot table[x index=0,y index=20] \datatable;
            \addplot table[x index=0,y index=21] \datatable;
        	\addplot table[x index=0,y index=22] \datatable;
        	\addplot table[x index=0,y index=23] \datatable;
        	\addplot table[x index=0,y index=24] \datatable;

        	\end{axis}
    	\end{tikzpicture}
     	}
    	\label{presentation-inconsistency}
    } 
    \subfloat[Test on Stack Overflow (SO) (Precision)]
      {
       \resizebox{2.2in}{!}{%
      \begin{tikzpicture}
        	\begin{axis}[
        	xtick=data,
        	xticklabels={Train on GitHub\\(Baseline: BERT4SentiSE), Train on Jira\\(Baseline: BERT4SentiSE)},
        	xticklabel style={align=center},
        	enlarge y limits=false,
        	enlarge x limits=0.50,
            axis y line=none,
            axis x line*=bottom,
            axis y line*=left,
        	ymin=0,ymax=1,
        	ybar,
        	bar width=0.6cm,
        	width=5.0in,
        	height = 2.3in,
        	ymajorgrids=false,
         	yticklabel style={font=\LARGE},
        	xticklabel style={font=\LARGE, /pgf/number format/fixed},	
        	major x tick style = {opacity=0},
        	minor x tick num = 1,    
        	minor tick length=1ex,
        	legend columns=2, 
            legend style={
                /tikz/column 2/.style={
                    column sep=5pt,
                },
                },
            legend style={draw=none},
        	legend style={at={(0.5,1.40)},
        	legend style={font=\LARGE},
            anchor=north,legend columns=-1
            },
            nodes near coords style={font=\LARGE, text=black},
            nodes near coords style={rotate=90,  anchor=west}, %
        	nodes near coords =\pgfmathprintnumber{\pgfplotspointmeta},
        	cycle list name = xcolor,
            every axis plot/.append style={fill,draw=black!80}%
        	]
        	\addplot table[x index=0,y index=25] \datatable;
        	\addplot table[x index=0,y index=26] \datatable;
        	\addplot table[x index=0,y index=27] \datatable;
        	\addplot table[x index=0,y index=28] \datatable;
            \addplot table[x index=0,y index=29] \datatable;
        	\addplot table[x index=0,y index=30] \datatable;
        	\addplot table[x index=0,y index=31] \datatable;
        	\addplot table[x index=0,y index=32] \datatable;

        	\end{axis}
    	\end{tikzpicture}
     	}
    	\label{presentation-inconsistency}
    } \\
    
    \subfloat[Test on Jira (F1 score)]
      {
       \resizebox{2.2in}{!}{%
      \begin{tikzpicture}
        	\begin{axis}[
        	xtick=data,
        	xticklabels={Train on GitHub\\(Baseline: SentiStrengthSE), Train on SO\\(Baseline: SentiStrengthSE)},
        	xticklabel style={align=center},
        	enlarge y limits=false,
        	enlarge x limits=0.5,
            axis y line=none,
            axis x line*=bottom,
            axis y line*=left,
        	ymin=0,ymax=1,
        	ybar,
        	bar width=0.6cm,
        	width=5.0in,
        	height = 2.3in,
        	ymajorgrids=false,
         	yticklabel style={font=\LARGE},
        	xticklabel style={font=\LARGE, /pgf/number format/fixed},	
        	major x tick style = {opacity=0},
        	minor x tick num = 1,    
        	minor tick length=1ex,
        	legend columns=2, 
            legend style={
                /tikz/column 2/.style={
                    column sep=5pt,
                },
                },
            legend style={draw=none},
        	legend style={at={(0.5,1.40)},
        	legend style={font=\LARGE},
            anchor=north,legend columns=-1
            },
            nodes near coords style={font=\LARGE, text=black},
            nodes near coords style={rotate=90,  anchor=west}, %
        	nodes near coords =\pgfmathprintnumber{\pgfplotspointmeta},
        	cycle list name = xcolor,
            every axis plot/.append style={fill,draw=black!80}%
        	]
        	\addplot table[x index=0,y index=33] \datatable;
        	\addplot table[x index=0,y index=34] \datatable;
        	\addplot table[x index=0,y index=35] \datatable;
        	\addplot table[x index=0,y index=36] \datatable;
            \addplot table[x index=0,y index=37] \datatable;
        	\addplot table[x index=0,y index=38] \datatable;
        	\addplot table[x index=0,y index=39] \datatable;
        	\addplot table[x index=0,y index=40] \datatable;

        	\end{axis}
    	\end{tikzpicture}
     	}
    	\label{presentation-inconsistency}
    } 
    \subfloat[Test on Jira (Precision)]
      {
       \resizebox{2.2in}{!}{%
      \begin{tikzpicture}
        	\begin{axis}[
        	xtick=data,
        	xticklabels={Train on GitHub\\(Baseline: SentiStrengthSE), Train on SO\\(Baseline: SentiStrengthSE)},
        	xticklabel style={align=center},
        	enlarge y limits=false,
        	enlarge x limits=0.50,
            axis y line=none,
            axis x line*=bottom,
            axis y line*=left,
        	ymin=0,ymax=1,
        	ybar,
        	bar width=0.6cm,
        	width=5.0in,
        	height = 2.3in,
        	ymajorgrids=false,
         	yticklabel style={font=\LARGE},
        	xticklabel style={font=\LARGE, /pgf/number format/fixed},	
        	major x tick style = {opacity=0},
        	minor x tick num = 1,    
        	minor tick length=1ex,
        	legend columns=2, 
            legend style={
                /tikz/column 2/.style={
                    column sep=5pt,
                },
                },
            legend style={draw=none},
        	legend style={at={(0.5,1.40)},
        	legend style={font=\LARGE},
            anchor=north,legend columns=-1
            },
            nodes near coords style={font=\LARGE, text=black},
            nodes near coords style={rotate=90,  anchor=west}, %
        	nodes near coords =\pgfmathprintnumber{\pgfplotspointmeta},
        	cycle list name = xcolor,
            every axis plot/.append style={fill,draw=black!80}%
        	]
        	\addplot table[x index=0,y index=41] \datatable;
        	\addplot table[x index=0,y index=42] \datatable;
        	\addplot table[x index=0,y index=43] \datatable;
        	\addplot table[x index=0,y index=44] \datatable;
            \addplot table[x index=0,y index=45] \datatable;
        	\addplot table[x index=0,y index=46] \datatable;
        	\addplot table[x index=0,y index=47] \datatable;
        	\addplot table[x index=0,y index=48] \datatable;

        	\end{axis}
    	\end{tikzpicture}
     	}
    	\label{presentation-inconsistency}
    } \\
    
\caption{Performance comparison of majority voting based ensemble tools with the SD tools (All- All of the SD tools except baseline DL-RNN4SEntiSE, BTPC-Best SD tool from PMT DL-based (BERT4SentiSE), Shallow ML-based (Senti4SD), Rule-based (SentiStrengthSE)}

\label{fig:ensemble_result}
\end{figure*}
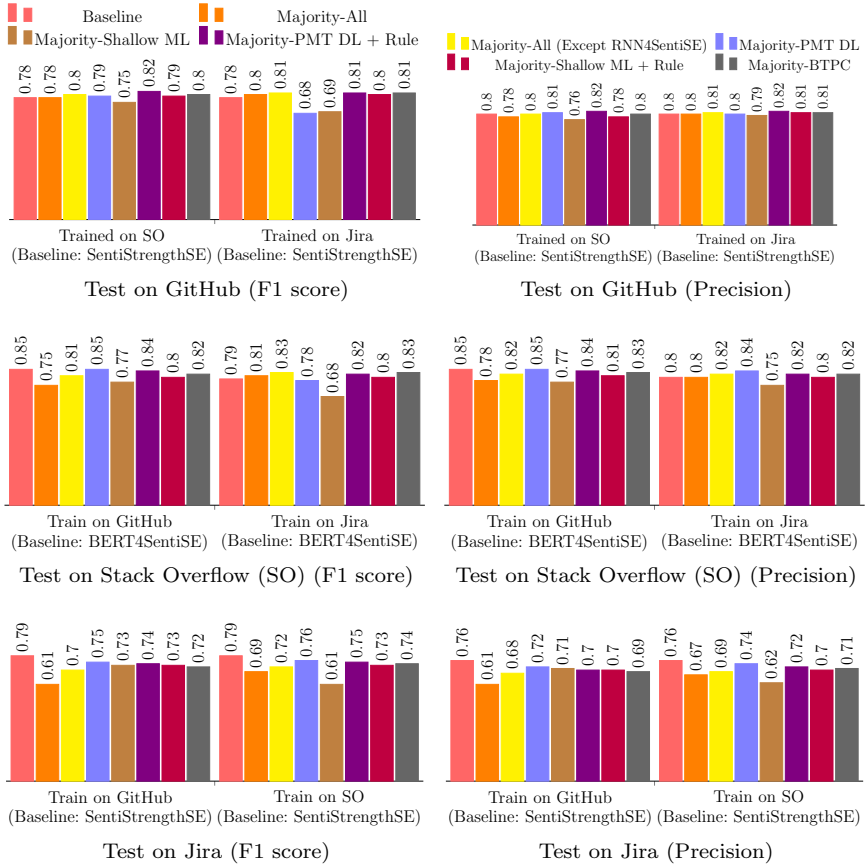

\subsubsection{Approach} We investigate different combinations of the tools within a majority-based 
voting technique to understand whether and how the tools can complement each other. The purpose is to learn whether such combinations can reduce the drop in performance of the tools in cross-platform settings. 
This means that if we have three tools (e.g., all three deep learning SD tools) as a 
combination, we first take the polarity labels of the three tools for a given sentence and then assign the sentence the 
polarity label that is found in the most number of the tools (i.e., at least 2 tools). 
To understand the different combinations of the tools, we compute the agreement and disagreement 
between the tools in our study for both within and cross-platform settings across the three studied datasets. 
We compute the agreements between two tools for a given setting and a dataset using Cohen $\kappa$ and Percent Agreement. We compute the disagreements using two metrics: Severe and Mild. The metrics are explained in RQ2 (\tbl\ref{tab:agreement_with_manual_labelling}).
Based on the analysis, we investigate seven different combinations of the tools in our majority voting technique:
\begin{itemize}
  \item Majority-All: Majority voting based on SentiMoji, BERT4SentiSE, RNN4SentiSE, Senti4SD, SentiCR, and SentiStrengthSE.
  \item Majority-All except RNN4SentiSE.
  \item Majority-PMT DL: The two best performing deep learning SD tools ( SentiMoji and BERT4SentiSE). 
  \item Majority-Shallow ML: The two best performing shallow learning SD tools (Senti4SD and SentiCR).
  \item Majority-PMT DL + Rule: The two best performing deep learning SD tools plus the rule-based SD tool: (SentiMoji, BERT4SentiSE, and SentiStrengthSE).
  \item Majority-Shallow ML + Rule: The two shallow learning SD tools plus the rule-based SD tool (Senti4SD, SentiCR, and SentiStrengthSE).
  \item Majority-BTPC: The best performing deep learning, shallow learning, and rule-based SD tools: (BERT4SentiSE, Senti4SD, and SentiStrengthSE).
\end{itemize}   
We compute and report the precision, recall and F1-score of each combination across the six cross-platform settings. 

\subsubsection{Results}
\textbf{The two best performing deep learning SD tools (BERT4SentiSE, SentiMoji) have the highest level of agreement and lowest level disagreement between them among all the six SE-specific SD tools in our study.} In Table~\ref{tab:agreement_test_github},~\ref{tab:agreement_test_so}, and~\ref{tab:agreement_test_jira}, we report the paired comparisons of agreement of
each pair of classifiers. It shows that the SEntiMoji and BERT4SentiSE have the
highest agreement between themselves. They also have the highest percentage of
perfect agreement and the lowest percentage of severe disagreement. Both
SentiMoji and BERT4SentiSE are pre-trained models. They already have an
understanding of languages. These tools are retrained for specific tasks. Both
of them have high performance in terms of correctly predicting polarity.
Conversely, RNN4SentiSE and SentiCR have the lowest agreement between them.

\textbf{A majority voting based ensemble based on the two best performing deep learning SD tools (BERT4SentiSE + SentiMoji) or the two best performing deep learning SD tools (BERT4SentiSE + SentiMoji) 
+ Rule-based tool (SentistrengthSE) provides slightly better performance than the stand-alone SD tools in cross-platform settings.}
\fig\ref{fig:ensemble_result} shows the performance of the seven majority voting-based
combinations in our six cross-platform settings. For each setting, we report the performance of each of the seven combinations (hereafter called as ensemblers) and 
the best performing stand-along SD (denoted as the baseline in \fig\ref{fig:ensemble_result}). For each setting, we report two performance metrics: Precision and Macro F1-score. For example, the topmost row in \fig\ref{fig:ensemble_result} 
shows the Macro F1 score and precision for the baselines and the seven ensemblers when they are tested on GitHub after they are trained on either SO or Jira. The leftmost eight bars show the Macro F1 scores of baseline + seven ensemblers when the tools 
are trained on SO and tested on GitHub. We also note the name of baseline under the eight bars, which in this case is SentistrengthSE. 
In this case, we see that an ensemble of the two best performing deep learning SD tools (BERT4SentiSE, SentiMoji) and SentistrengthSE 
offers the best Macro F1-score of 0.82 out of the baseline and the ensemblers. Overall, in terms of Macro F1-scores, 
an ensemble-based tool is the best performer in three cross-platform settings, which a stand-alone SD tool (i.e., a baseline) 
is the best performer in the other three cross-platform settings. When a baseline 
is the best performer, its SentistrengthSE in two settings (tested on Jira after trained on SO or GitHub) and 
BERT4SentiSE in one setting (trained on GitHub, tested on SO). One of the reasons for the superiority of SentistrengthSE for the 
Jira dataset is that the rules for SentiStrengthSE were developed by studying Jira
platform \cite{Ortu-EmotionalSideJira-MSR2016}. Therefore, the tool might have some advantage
over other tools it is tested on in our Jira dataset. Among the three settings where the majority voting 
based ensemblers offered the best F1-scores, it is always the combination of the two best performing deep learning 
SD tools (BERT4SentiSE, SentiMoji) and SentistrengthSE. 

\bf{The difference in performance between the best performing ensembler and the baseline (i.e., best performing stand-alone tool) in each cross-platform setting 
is almost negligible.} According to \fig\ref{fig:ensemble_result}, the majority based ensemblers are 
the best performers in three cross-platform settings: when tested on GitHub after it is trained on SO (Macro F1-score of the best ensembler 0.82) or 
Jira (F1 = 0.81) and when trained on Jira and then tested on SO (F1 = 0.82). The best performing baselines in those three settings are SentistrengthSE for the first two cases (F1 = 0.78) and BERT4SentiSE in the last setting (F1 = 0.79). Therefore, the maximum gain in performance we can get by using an ensembler over a baseline in those three settings is 0.03. Therefore, majority-based voting of the six tools in cross-platform settings is not 
practical, when we get only a gain of 0.03 in Macro F1-score.

\begin{tcolorbox}[flushleft upper,boxrule=1pt,arc=0pt,left=0pt,right=0pt,top=0pt,bottom=0pt,colback=white,after=\ignorespacesafterend\par\noindent]
 \bf{RQ$_4$ To what extent the deep learning and non-deep learning sentiment detection tools can complement each other in cross-platform settings?}
The two best performing deep learning SD tools (BERT4SentiSE, SentiMoji) have
the highest level of agreement and lowest level disagreement between them among
all the six SE-specific SD tools in our study. A majority voting based ensemble
based on the two best performing deep learning SD tools (BERT4SentiSE +
SentiMoji) or the two best performing deep learning SD tools (BERT4SentiSE +
SentiMoji) + Rule-based tool (SentistrengthSE) provides slightly better
performance than the stand-alone SD tools in cross-platform settings. The
difference in performance between the best performing ensembler and the baseline
(i.e., best performing stand-alone tool) in each cross-platform setting is
almost negligible.
\end{tcolorbox}

\section{Discussions}\label{sec:discussion}
In this section, we discuss major themes that emerged during our study.

\subsection{Can similarity analysis between the studied datasets explain the drop in performance 
of the deep learning SD tools in the datasets in cross-platform settings?}\label{sec:rq-cross-platform-similarity}
Across the supervised SD tools for SE in our three studied datasets, we observe 
a drop in performance in cross-platform settings. While the two deep learning SD tools (BERT4SentiSE and SentiMoji) offer better performance than the 
other supervised SD tools in the majority of cross-platform settings, their performance still varies across the datasets. 
For example, as we noted in RQ$_1$, the tools suffered the most performance drop when the tools were trained in Jira and then tested in the other two datasets. Intuitively, the more similar the textual contents are between the pair of datasets, the less drop we can expect in performance for a tool in those two datasets in cross-platform settings. Therefore, it 
is important to know whether there is a correlation between content similarity of the studied datasets and the performance 
of the studied tools in cross-platform settings.

\subsubsection{Approach} For this analysis, we focus on the two best performing deep learning SD tools for SE from RQ$_1$, i.e., BERT4SentiSE and SentiMoji. 
For each tool and for each cross-platform setting involving two datasets (i.e., one dataset for training and another for testing). we analyze whether there is any correlation between the performance 
of the tool in cross-platform settings and the textual similarity between the two datasets. 
We compute the similarity
between two datasets using two standard metrics, Cosine similarity, and Jensen-Shannon Divergence. We discuss below how we use the two metrics to compute the similarity between two given datasets.

\begin{inparaenum}[(1)]
\item\bf{Cosine similarity~\cite{Manning-IRIntroBook-Cambridge2009}} helps to
find text similarity between two documents irrespective of the size of the
documents.
For each dataset, we take the top 10,000 unigrams after removing stopwords (we
used stopwords from the Python `nltk' library). We then combine those to create a
unified vocabulary list ($V$) for all the datasets. We use $V$ to create a
vectorized representation of each dataset based on the frequency of the words of
each sentence in the dataset. For any two datasets (e.g., A, B), we fed their
resulting vectors into \eq\ref{eq:consine} to compute the similarity between the two
datasets. We use cosine\_similarity method from Sklearn library\footnote{\url{https://scikit-learn.org/stable/modules/generated/sklearn.metrics.pairwise.cosine_similarity.html}}.

\begin{equation}\label{eq:consine}
Cosine_{Similarity}(A,B) = \frac{A.B}{|A||B|}   
\end{equation}

\item \bf{Jensen-Shannon Divergence~\cite{fuglede2004jensen} ($D_{JS}$)} is a
symmetric and smoothed version of Kullback-Leibler Divergence $D_{KL}$,
$D_{KL}(A||B) = \sum_{i}^{N} a_i log \frac{a_i}{b_i}$. $D_{KL}$ is a probability
distribution of A and B. While $D_{KL}$ is undefined when any element $a_i \in
A$, has zero probability. Whereas, $D_{JS}$ is a probability distribution of
every word in vocabulary $V$. The metric is originally proposed to calculate the
similarity between two datasets A and B as follows:
\begin{equation}\label{eq:jsd}
D_{JS}(A,B) = \frac{1}{2} [D_{KL}(A||B)+D_{KL}(B||A)]    
\end{equation} With \eq\ref{eq:jsd}, a value of $D_{JS}$ = 0, it signifies the
datasets are most similar a value of 1.0 denotes they are completely different.
Given that Cosine similarity adopts an opposite notion for similarity (i.e., 1
as completely similar), we change \eq\ref{eq:jsd} as follows:
\begin{equation}\label{eq:jsd}
D_{JS}(A,B) = 1- \frac{1}{2} [D_{KL}(A||B)+D_{KL}(B||A)]    
\end{equation} Thus in our case, when $D_{JS}$ = 1.0, it signifies the datasets
are most similar and when $D_{JS}$ = 0.0 it signifies that the datasets are
completely diverse from each other. We compute the metric for each pair of
datasets (A, B) as follows. Similar to cosine similarity approach we create a
vectorized representation of each dataset. For any two datasets (e.g., A, B), we
fed their resulting vectors into \eq\ref{eq:jsd} to compute $D_{JS}$ similarity
between the two datasets.
\end{inparaenum} 


We compute the correlation between the similarity in the two datasets and the performance of a tool in cross-platform settings using 
Mann-Whitney (MW) U Test~\cite{mcknight2010mann}. MW test is non-parametric, which is suitable for our data where data is not normally distributed. We report the significance of correlation using p-value and the effect size using Cliff's $\delta$ value. 
We interpret the $\delta$ following Romano et al.~\cite{Romano-EffectSize-2006}:
\begin{equation}
  \tr{Effect Size} =\left\{
  \begin{array}{@{}ll@{}}
    {negligible~(N)}, & \tr{if}\ |\delta| \leq 0.147 \\
    {small~(S)}, & \tr{if}\ 0.147 < |\delta| \leq 0.33 \\
    {medium~(M)}, & \tr{if}\ 0.33 < |\delta| \leq 0.474 \\
    {large~(L)}, & \tr{if}\ 0.474 < |\delta| \leq 1 
  \end{array}\right.
\end{equation} The effect size categories (negligible, small, medium, 
or large) quantify the differences between the distributions.

\subsubsection{Results}  
\begin{figure*}[h]
\centering
\captionsetup[subfigure]{labelformat=empty}
\pgfplotstableread{

1	.88  .97 .75 .77     .88  .97  .82 .85      .87  .96  .77 .77
2	.87  .96 .62 .65     .85  .93  .64 .76      .85  .93  .77 .78
}\datatable
      \subfloat[Test on GitHub]
      {
      \resizebox{1.5in}{!}{%
      \begin{tikzpicture}
        	\begin{axis}[
        	xtick=data,
        	xticklabels={Trained on SO, Trained on Jira},
        	enlarge y limits=false,
        	enlarge x limits=0.55,
            axis y line=none,
            axis x line*=bottom,
            axis y line*=left,
        	ymin=0,ymax=1,
        	ybar,
        	bar width=0.7cm,
        	width=3.6in,
        	height = 2.3in,
        	ymajorgrids=false,
         	yticklabel style={font=\LARGE},
        	xticklabel style={font=\LARGE, /pgf/number format/fixed},
        	major x tick style = {opacity=0},
        	minor x tick num = 1,
        	minor tick length=1ex,
        	legend columns=2,
            legend style={
                /tikz/column 2/.style={
                    column sep=5pt,
                },
                },
        	legend style={at={(0.5,1.40)},
        	legend style={font=\LARGE},
        	legend style={draw=none},
        	legend image post style={scale=2.0},
            anchor=north,legend columns=-1
            },
            nodes near coords style={font=\huge},
            nodes near coords style={rotate=90,  anchor=west}, %
        	nodes near coords =\pgfmathprintnumber{\pgfplotspointmeta}
        	]
        	\addplot[draw=black!80, fill=black!5] table[x index=0,y index=1] \datatable;
        	\addplot[draw=black!80, fill=black!30] table[x index=0,y index=2] \datatable;
        	\addplot[draw=black!80, fill=black!70] table[x index=0,y index=3] \datatable;
        	\addplot[draw=black!80, fill=black!100] table[x index=0,y index=4] \datatable;

            \legend{JS Divergence,
            		 Cosine Similarity,
            		 }
        	\end{axis}
    	\end{tikzpicture}
     	}
    	\label{presentation-inconsistency}
    }
    \subfloat[Test on Stack Overflow (SO)]
     {
      \resizebox{1.5in}{!}{%
      \begin{tikzpicture}
        	\begin{axis}[
        	xtick=data,
        	xticklabels={Train on GitHub, Train on Jira},
        	enlarge y limits=false,
        	enlarge x limits=0.55,
            axis y line=none,
            axis x line*=bottom,
            axis y line*=left,
        	ymin=0,ymax=1,
        	ybar,
        	bar width=0.7cm,
        	width=3.6in,
        	height = 2.3in,
        	ymajorgrids=false,
         	yticklabel style={font=\LARGE},
         	xticklabel style={font=\LARGE, /pgf/number format/fixed},
        	major x tick style = {opacity=0},
        	minor x tick num = 1,
        	minor tick length=1ex,
        	legend columns=2,
            legend style={
                /tikz/column 2/.style={
                    column sep=5pt,
                },
                },
        	legend style={at={(0.5,1.40)},
        	legend style={font=\LARGE},
        	legend style={draw=none},
        	legend image post style={scale=2.0},
            anchor=north,legend columns=-1
            },
            nodes near coords style={font=\huge},
            nodes near coords style={rotate=90,  anchor=west}, %
        	nodes near coords =\pgfmathprintnumber{\pgfplotspointmeta}
        	]
        	\addplot[draw=black!80, fill=black!5] table[x index=0,y index=5] \datatable;
        	\addplot[draw=black!80, fill=black!30] table[x index=0,y index=6] \datatable; 
        	\addplot[draw=black!80, fill=black!60] table[x index=0,y index=7] \datatable;   
            \addplot[draw=black!80, fill=black!100] table[x index=0,y index=8] \datatable;  
            \legend{,,
            		 SEntiMoji F1,
            		 BERT4SentiSE F1
            		 }
        	\end{axis}
    	\end{tikzpicture}
    	}
    	\label{signature-inconsistency}
    }
    \subfloat[Test on Jira]
     {
      \resizebox{1.5in}{!}{%
      \begin{tikzpicture}
        	\begin{axis}[
        	xtick=data,
        	xticklabels={Train on GitHub, Train on SO
        	},
        	enlarge y limits=false,
        	enlarge x limits=0.55,
            axis y line=none,
            axis x line*=bottom,
            axis y line*=left,,
        	ymin=0,ymax=1.08,
        	ybar,
        	bar width=0.7cm,
        	width=3.4in,
        	height = 2.3in,
        	ymajorgrids=false,
         	yticklabel style={font=\LARGE},
         	xticklabel style={font=\LARGE, /pgf/number format/fixed},
        	major x tick style = {opacity=0},
        	minor x tick num = 1,
        	minor tick length=1ex,
        	legend style={
         	at={(0.6,0.98)},
            anchor=north,legend columns=-1
            },
            nodes near coords style={font=\huge},
            nodes near coords style={rotate=90,  anchor=west}, %
        	nodes near coords =\pgfmathprintnumber{\pgfplotspointmeta}
        	]
        	\addplot[draw=black!80, fill=black!5] table[x index=0,y index=9] \datatable;
        	\addplot[draw=black!80, fill=black!30] table[x index=0,y index=10] \datatable; 
        	\addplot[draw=black!80, fill=black!60] table[x index=0,y index=11] \datatable; 
            \addplot[draw=black!80, fill=black!100] table[x index=0,y index=12] \datatable; 

        	\end{axis}
    	\end{tikzpicture}
    	}
    	\label{fig:status-inconsistency}
    }

\caption{Similarity scores between datasets vs Macro F1-scores per tool (BERT4SentiSE \& SentiMoji) in cross platform settings. 
There are three 
sets of bar charts, one set each for each of the three datasets. For the first set, `Test on GitHub' denotes that the the tool BERT4SentiSE is trained on the 
other SE-specific two datasets and also the non-SE dataset (i.e., Twitter) and then tested on GitHub. The legends for `Test on GitHub' are: Train on SO (i.e., Stack Overflow), 
Train on Jira, and Train on Twitter (for brevity, `train on' phrase is taken out from the last two). For each `Train on <dataset>', there are three bars: the first bar for `Train on SO' 
(i.e., the white bar) denotes the JS divergence score between the train and test datasets (in this case between SO and GitHub), the second bar 
denotes (i.e., the gray bar) denotes the cosine similarity score between the train and test datasets (i.e., between SO and GitHub), and the 
third bar (i.e., black bar) denotes F1-score of BERT4SentiSE in such cross-platform settings (i.e., train on SO and then test on GitHub).
}
\vspace{-3mm}
\label{fig:diversity_performance}
\end{figure*}
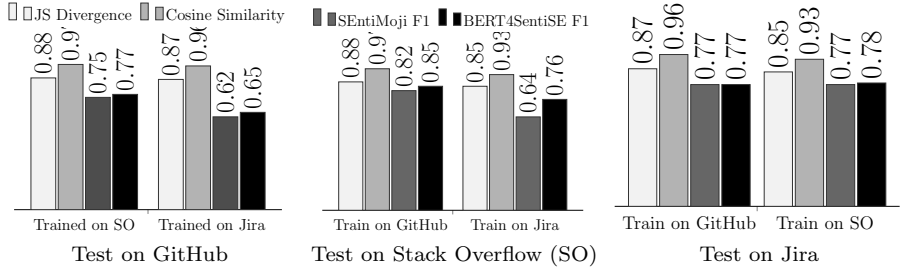

\bf{There is a statistically significant correlation
between content similarity of the datasets and their Macro F1 score in
cross-platform settings.} In
\tbl\ref{tab:corrSimAcc}, we report the p-value of the correlation analysis between the dataset similarity and tool performance values using MW U test. We also show the effect size.
\begin{table}[t]
  \centering
  \caption{Correlation between dataset similarity values and Macro-F1 scores in cross-platform settings}
    \begin{tabular}{lrr|rr}\toprule
          & \multicolumn{2}{c}{\bf{BERT4SentiSE}} & \multicolumn{2}{c}{\bf{SentiMoji}} \\
          \cmidrule{2-5}
          & {p-value} & Effect Size     & {p-value} & Effect Size \\
          \midrule
    Jansen Shannon    & 0.04*  & Large (0.6) & 0.003** & Large (1.0) \\
    Cosine & 0.002** & Large (1.0) & 0.003** & Large (1.0) \\
    \bottomrule
    \multicolumn{5}{l}{Signif. codes:  ***$p < 0.001$, **$p < 0.01$, *$p < 0.05$}
    \end{tabular}%
  \label{tab:corrSimAcc}%
\end{table}%
In \fig\ref{fig:diversity_performance}, we show the
similarity scores between the datasets and compare the Macro F1-scores between the datasets for the two best performing
deep learning SD tools from RQ$_1$ (i.e., BERT4SentiSE and SentiMoji). For
example, in the first block (i.e., Test on GitHub), we have two sets of bars:
similarity scores between GitHub and SO datasets and Macro F1-scores of
BERT4SentiSE and SentiMoji for the cross-platform setting (i.e., trained on SO
and tested on GitHub). The Jansen similarity scores between the datasets are at least
0.87 and at least 0.93 for cosine similarity. Therefore, while datasets are produced from different platforms, they
have a considerable number of similar words. Thus the more similar two datasets are, the less performance drop 
we can expect in cross-platform settings. 

\begin{table}[h]
\newcolumntype{C}{>{\centering\arraybackslash}p{9em}}
  \centering
  
  \caption{Distribution of dataset content similarity vs drop in performance (Macro F1-score) in Cross-Platform Settings compared to Within Platform Settings for BERT4SentiSE and SentiMoji}
    \tiny
    \begin{tabular}{lccc|CC}\toprule
    {\textbf{Test}} & {\textbf{Train}} & \multicolumn{2}{c}{\textbf{Similarity Score}} & \multicolumn{2}{c}{\textbf{\% Performance drop}} \\
    \cmidrule{3-6}
    {\textbf{Dataset}} & \textbf{Dataset} & {\textbf{JS divergence}} & {\textbf{Cosine distance}} & \textbf{SentiMoji} & \textbf{Bert4SentiSE} \\
    \midrule
    \multicolumn{1}{l}{\multirow{2}[0]{*}{GitHub}} & SO    & 0.88  & 0.97  & 17    & 16 \\
          & Jira  & 0.87  & 0.96  & 31    & 29 \\
          \midrule
    \multicolumn{1}{l}{\multirow{2}[0]{*}{SO}} & GitHub & 0.88  & 0.97  & 5     & 3 \\
          & Jira  & 0.85  & 0.93  & 26    & 14 \\
          \midrule
    \multicolumn{1}{l}{\multirow{2}[0]{*}{Jira}} & GitHub & 0.87  & 0.96  & 5     & 7 \\
          & SO    & 0.85  & 0.93  & 5     & 6 \\
          \bottomrule
    \end{tabular}%
  \label{tab:sim_vs_drop}%
\end{table}%

\bf{The significance analysis cannot explain the varying degrees of performance between Jira and non-Jira datasets in cross-platform settings.} Our findings in RQ$_1$ 
show the largest performance drop in the two deep learning SD tools when trained in the Jira dataset and tested in the other two datasets. 
The significance analysis in \tbl\ref{tab:corrSimAcc} suggests that this performance drop could be attributed to the similarity (or lack of thereof) between Jira and the other two datasets. However, \fig\ref{fig:diversity_performance} shows that the Jira dataset has almost similar content similarity scores with the SO and GitHub 
as do SO and GitHub with each other. Moreover, the tools suffer a much bigger drop in performance only when they 
are trained in Jira. For example, BERT4SentiSE shows a Macro F1-score of 0.65 when trained in Jira and tested in GitHub, but it shows 
an F1-score of 0.77 when trained in GitHub and then tested in Jira. In \tbl\ref{tab:sim_vs_drop}, we 
show the percentage drop in Macro F1-score compared to within platform setting for each cross-platform settings. For example, the first two rows in 
\tbl\ref{tab:sim_vs_drop} show that when trained in the SO dataset and then tested in the GitHub dataset, the drop in performance in BERT4SentiSE is 16\% compared to the performance 
of BERT4SentiSE when trained and tested in GitHub. However, when BERT4SentiSE is trained in the Jira dataset and then tested in the GitHub dataset, the drop in performance 
is 29\%. The cosine similarity between GitHub and SO is 0.97 and between GitHub and Jira is 0.96. 
When the tool was trained in GitHub and tested in SO, the drop was 3\%, but the drop was 14\% when the tool was trained in Jira and tested in SO. 
The cosine similarity between SO and GitHub is 0.97, and between SO and Jira is 0.93. Therefore, the drop in performance while using the Jira 
dataset could be due to reasons other than content similarity. 

Instead, the the distribution of the polarity classes in the Jira dataset could offer an alternative explanation (see 
\tbl\ref{tab:benchmark_dataset}). While both GitHub and SO are more balanced towards the three polarity classes (around 27-35\% of sentences per polarity class), the Jira dataset is imbalanced with only 19\% positive and 14\% negative classes. {This means that a tool trained using the Jira dataset 
does not have enough information of the two polarity classes compared to when it is trained 
in the other two datasets.} The imbalance in the Jira dataset may also have
introduced bias in the polarity labeling (e.g., by showing more preference
towards neutral classes), contributing to the high percentage of mild disagreements in 
\tbl\ref{tab:agreement_with_manual_labelling}. This happened despite each tool is designed to 
handle the class imbalance problem in model training.


\begin{table}[t]
  \centering
  \normalsize
  \caption{Distribution of False Neutral and False Non-neutral prediction in cross-platform settings. (False Neutral = Prediction is neutral for non-neutral, False Non-neutral = Prediction is non-neutral for neutral sentences)}
  \resizebox{\columnwidth}{!}{%
    \begin{tabular}{ll|cc|cc|cc}\toprule
    {\textbf{SD}} & \textbf{\% Correct} & \multicolumn{2}{c}{\textbf{Train in GitHub}} & \multicolumn{2}{c}{\textbf{Train in SO}} & \multicolumn{2}{c}{\textbf{Train in Jira}} \\
    \cmidrule{3-8}
    {\textbf{Tool}} & \textbf{Misclassified} & {\textbf{Test SO}} & {\textbf{Test Jira}} & {\textbf{Test GitHub}} & {\textbf{Test Jira}} & {\textbf{Test GitHub}} & {\textbf{Test SO}} \\
    \midrule
\multirow{2}{*}{\textbf{SentiMoji}} &False Neutral &61 &22 &75 &28 &\textbf{94} &\textbf{97} \\
&False Non-neutral &34 &74 &15 &69 &\textbf{2} &\textbf{3} \\
\midrule
\multirow{2}{*}{\textbf{BERT4SentiSE}} &False Neutral &41 &23 &69 &26 &\textbf{91} &\textbf{88} \\
&False Non-neutral &52 &72 &17 &69 &\textbf{3} &\textbf{10} \\

          \bottomrule
    \end{tabular}%
    }
  \label{tab:rq-mild-error-analysis}%
\end{table}%

\textbf{When trained on the imbalanced Jira dataset (67\% neutral), the most dominant error case (around 90\%) is that all the supervised tools predict neutral polarity for non-neutral sentences.}
From Table~\ref{tab:cross_platform_performance}, we can see that when trained in Jira dataset and tested on non-Jira datasets, both SentiMoji and BERT4SentiSE show high precision and low recall for both the non-neutral classes. At the same time, these tools exhibit high recall and low precision for the neutral class. For example, when trained on the Jira dataset and tested on GitHub dataset, BERT4SentiSE shows a precision of 0.90 and recall of 0.43 for the negative class. The tool shows a precision of 0.91 and a recall of 0.51 for the positive class. At the same time, they report a high recall of .97 and low precision of 0.59 for the neutral class. 
This observation is consistent with SentiMoji in cross-platform settings.
Table~\ref{tab:rq-mild-error-analysis} presents details about the drop in performance when the supervised SD tools are trained on cross-platform settings. 
Around 95\% to 99\% misclassifications fall under mild errors (i.e., the sum of `False Neutral' and `False Non-neutral'). Around 88\% to 97\% of the errors are due to `False Neutral' prediction, i.e., the manual label is non-neutral, but the tool predicts as neutral. For example, when BERT4SentiSE is trained on the Jira dataset and tested on GitHub, it has around 29\% performance drop in terms of macro F1-score and 94\% of its total errors are mild error, and 6\% are severe errors. Our analysis shows that in that settings, 91\% of the errors fall under the `False Neutral' category, and only 3\% of errors fall under the `False Non-neutral' category. Novelli et al.~\cite{Novielli-SEToolCrossPlatform-MSR2020} also reported a similar performance drop in cross-platform settings for the supervised tools, and the deep-learning-based SD tools outperformed all non-DL based SD tools in within-platform settings even for the imbalanced Jira dataset. As mentioned in \sec\ref{sec:studied-tools}, for this study, we used the best model configuration, hyperparameters, and settings for the studied supervised tools reported by the original authors in respective studies.

\subsection{Cross-Platform Similarity vs Accuracy}\label{sec:sim-vs-acc-twitter}
\begin{figure*}[h]
\centering
\captionsetup[subfigure]{labelformat=empty}
\pgfplotstableread{

1	.88  .97  .77     .88  .97   .85      .87  .96   .77
2	.87  .96  .65     .85  .93   .76      .85  .93   .78
3	.74  .74 .51      .74  .76   .65      .72  .72   .57
}\datatable
      \subfloat[Test on GitHub]
      {
       \resizebox{1.6 in}{!}{%
      \begin{tikzpicture}
        	\begin{axis}[
        	xtick=data,
        	xticklabels={Train on SO,  Jira,  Twitter},
        	enlarge y limits=false,
        	enlarge x limits=0.23,
            axis y line=none,
            axis x line*=bottom,
            axis y line*=left,
        	ymin=0,ymax=1,
        	ybar,
        	bar width=0.7cm,
        	width=3.6in,
        	height = 2.3in,
        	ymajorgrids=false,
         	yticklabel style={font=\LARGE},
        	xticklabel style={font=\LARGE, /pgf/number format/fixed},	
        	major x tick style = {opacity=0},
        	minor x tick num = 1,    
        	minor tick length=1ex,
        	legend columns=2, 
            legend style={
                /tikz/column 2/.style={
                    column sep=5pt,
                },
                },
        	legend style={at={(0.5,1.40)},
        	legend style={font=\LARGE},
        	legend style={draw=none},
        	legend image post style={scale=2.0},
            anchor=north,legend columns=-1
            },
            nodes near coords style={font=\huge},
            nodes near coords style={rotate=90,  anchor=west}, %
        	nodes near coords =\pgfmathprintnumber{\pgfplotspointmeta}
        	]
        	\addplot[draw=black!80, fill=black!5] table[x index=0,y index=1] \datatable;
        	\addplot[draw=black!80, fill=black!30] table[x index=0,y index=2] \datatable;
        	\addplot[draw=black!80, fill=black!70] table[x index=0,y index=3] \datatable;
        
            \legend{JS Divergence
            		 }
        	\end{axis}
    	\end{tikzpicture}
     	}
    	\label{presentation-inconsistency}
    }
    \subfloat[Test on Stack Overflow (SO)]
     {
      \resizebox{1.6 in}{!}{%
      \begin{tikzpicture}
        	\begin{axis}[
        	xtick=data,
        	xticklabels={Train on GitHub,  Jira,  Twitter},
        	enlarge y limits=false,
        	enlarge x limits=0.23,
            axis y line=none,
            axis x line*=bottom,
            axis y line*=left,
        	ymin=0,ymax=1,
        	ybar,
        	bar width=0.7cm,
        	width=3.6in,
        	height = 2.3in,
        	ymajorgrids=false,
         	yticklabel style={font=\LARGE},
         	xticklabel style={font=\LARGE, /pgf/number format/fixed},	
        	major x tick style = {opacity=0},
        	minor x tick num = 1,    
        	minor tick length=1ex,
        	legend columns=2, 
            legend style={
                /tikz/column 2/.style={
                    column sep=5pt,
                },
                },
        	legend style={at={(0.5,1.40)},
        	legend style={font=\LARGE},
        	legend style={draw=none},
        	legend image post style={scale=2.0},
            anchor=north,legend columns=-1
            },
            nodes near coords style={font=\huge},
            nodes near coords style={rotate=90,  anchor=west}, %
        	nodes near coords =\pgfmathprintnumber{\pgfplotspointmeta}
        	]
        	\addplot[draw=black!80, fill=black!5] table[x index=0,y index=4] \datatable; 
        	\addplot[draw=black!80, fill=black!30] table[x index=0,y index=5] \datatable; 
        	\addplot[draw=black!80, fill=black!70] table[x index=0,y index=6] \datatable;   
            \legend{,
                     Cosine Similarity
            		 }
        	\end{axis}
    	\end{tikzpicture}
    	}
    	\hspace{-1mm}
    	\label{signature-inconsistency}
    }
    \subfloat[Test on Jira]
     {
      \resizebox{1.6 in}{!}{%
      \begin{tikzpicture}
        	\begin{axis}[
        	xtick=data,
        	xticklabels={Train on SO,  Jira,  Twitter},
        	enlarge y limits=false,
        	enlarge x limits=0.23,
            axis y line=none,
            axis x line*=bottom,
            axis y line*=left,,
        	ymin=0,ymax=1.08,
        	ybar,
        	bar width=0.7cm,
        	width=3.6in,
        	height = 2.3in,
        	ymajorgrids=false,
         	yticklabel style={font=\LARGE},
         	xticklabel style={font=\LARGE, /pgf/number format/fixed},	
        	major x tick style = {opacity=0},
        	minor x tick num = 1,    
        	minor tick length=1ex,
        	legend columns=2, 
            legend style={
                /tikz/column 2/.style={
                    column sep=5pt,
                },
                },
        	legend style={at={(0.5,1.40)},
        	legend style={font=\LARGE},
        	legend style={draw=none},
        	legend image post style={scale=2.0},
            anchor=north,legend columns=-1
            },
            nodes near coords style={font=\huge},
            nodes near coords style={rotate=90,  anchor=west}, %
        	nodes near coords =\pgfmathprintnumber{\pgfplotspointmeta}
        	]
        	\addplot[draw=black!80, fill=black!5] table[x index=0,y index=7] \datatable;
        	\addplot[draw=black!80, fill=black!30] table[x index=0,y index=8] \datatable; 
        	\addplot[draw=black!80, fill=black!70] table[x index=0,y index=9] \datatable; 
            \legend{, ,
            		 BERT4SentiSE F1
            		 }
            
        	\end{axis}
    	\end{tikzpicture}
    	}
    	\label{fig:status-inconsistency}
    }

\caption{Similarity scores between Twitter and other datasets vs Macro F1-scores of BERT4SentiSE in cross platform settings. There are three 
sets of bar charts, one set each for each of the three datasets. For the first set, `Test on GitHub' denotes that the the tool BERT4SentiSE is trained on the 
other SE-specific two datasets and also the non-SE dataset (i.e., Twitter) and then tested on GitHub. The legends for `Test on GitHub' are: Train on SO (i.e., Stack Overflow), 
Train on Jira, and Train on Twitter (for brevity, `train on' phrase is taken out from the last two). For each `Train on <dataset>', there are three bars: the first bar for `Train on SO' 
(i.e., the white bar) denotes the JS divergence score between the train and test datasets (in this case between SO and GitHub), the second bar 
denotes (i.e., the gray bar) denotes the cosine similarity score between the train and test datasets (i.e., between SO and GitHub), and the 
third bar (i.e., black bar) denotes F1-score of BERT4SentiSE in such cross-platform settings (i.e., train on SO and then test on GitHub).}

\vspace{-3mm}
\label{fig:diversity_performance_twitter}
\end{figure*}
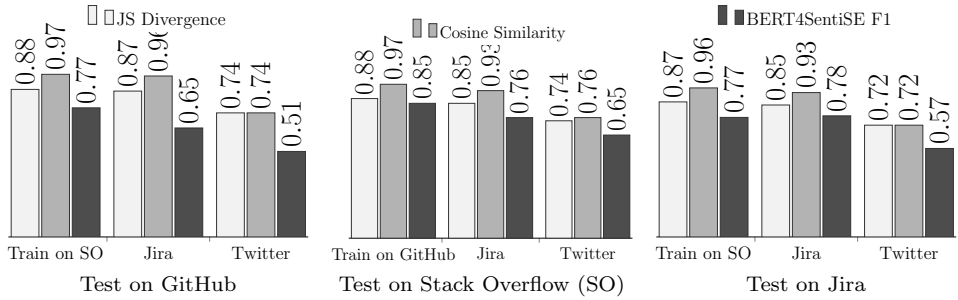

In \sec\ref{sec:rq-cross-platform-similarity}, we observed that the three SE datasets we used in this study have
considerable similarity in terms of textual contents (at least 85\%) and that
there is a statistically significant positive correlation between the similarity
between two datasets and the cross-platform performance of the best performing
deep learning SD tool BERT4SentiSE for the two datasets. To better understand
whether this observation is justified, we also checked the similarity of the
three SE tools against a non-SE dataset. The dataset is called
`tweeter-airline-sentiment'. It is available in
Kaggle\footnote{https://www.kaggle.com/crowdflower/twitter-airline-sentiment}.
This dataset has 14,640 tweets from consumers (and their sentiment polarity
labels) about six major airlines from the USA. We trained BER4SentiSE using this
Twitter dataset and then tested the model on our three SE datasets. We also
computed the similarity between the Twitter dataset and our three SE datasets
using the two metrics we introduced in RQ$_2$, i.e., Jensen Shannon Divergence
and Cosine Similarity. In \fig\ref{fig:diversity_performance_twitter}, we show
three blocks of bars. The first block shows the similarity of the GitHub dataset
with the other three datasets (Twitter, GitHub, Jira) using six bars (two each
for the two similarity metrics). The first block also shows the Macro F1-score
of BERT4SentiSE while the tool is tested on the GitHub dataset after training on
the other three datasets. The second block shows similar data for the `Test on SO
dataset', and the third block shows data for the `Test on Jira' dataset. Across all
three test datasets, the Twitter dataset shows the lowest similarity scores
among all datasets (at most 0.74). The cross-platform F1 scores in the three SE
datasets while trained on Twitter are also significantly low (0.51 - 0.65 while
training on Twitter compared to at least 0.77 while training on an SE dataset).
These results confirm that the SE datasets are indeed more similar to each other
compared to their similarity with a non-SE dataset like Twitter. The increased
similarity does indeed impact the performance of the tool BERT4SentiSE in
cross-platform settings for SE datasets.

\subsection{Learning Curves of the Tools}\label{subsec:learning_curve}
\begin{figure}
\subfloat[Github]{\includegraphics[scale=0.5]{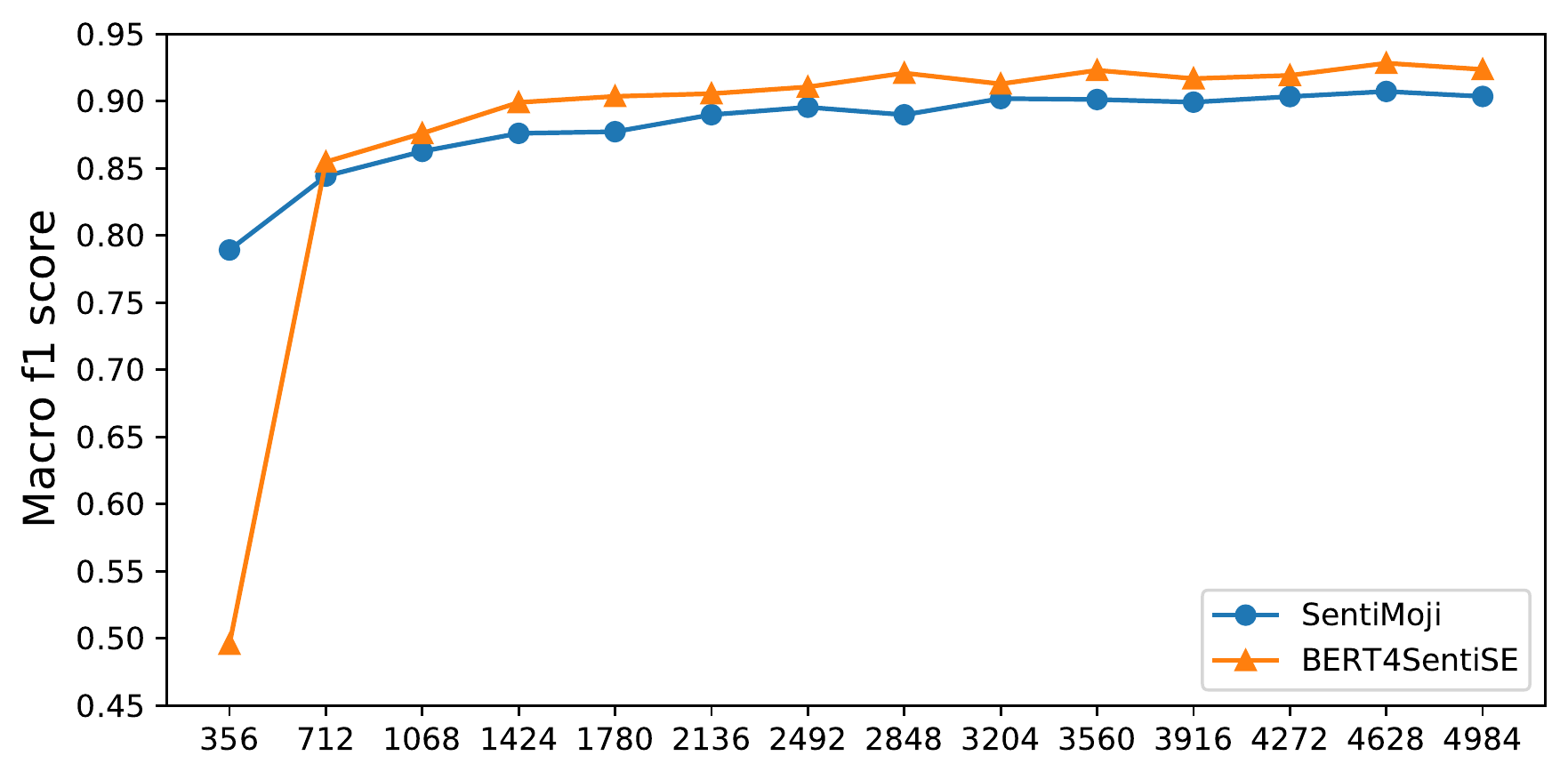}} \\
\subfloat[Stack Overflow]{\includegraphics[scale=0.5]{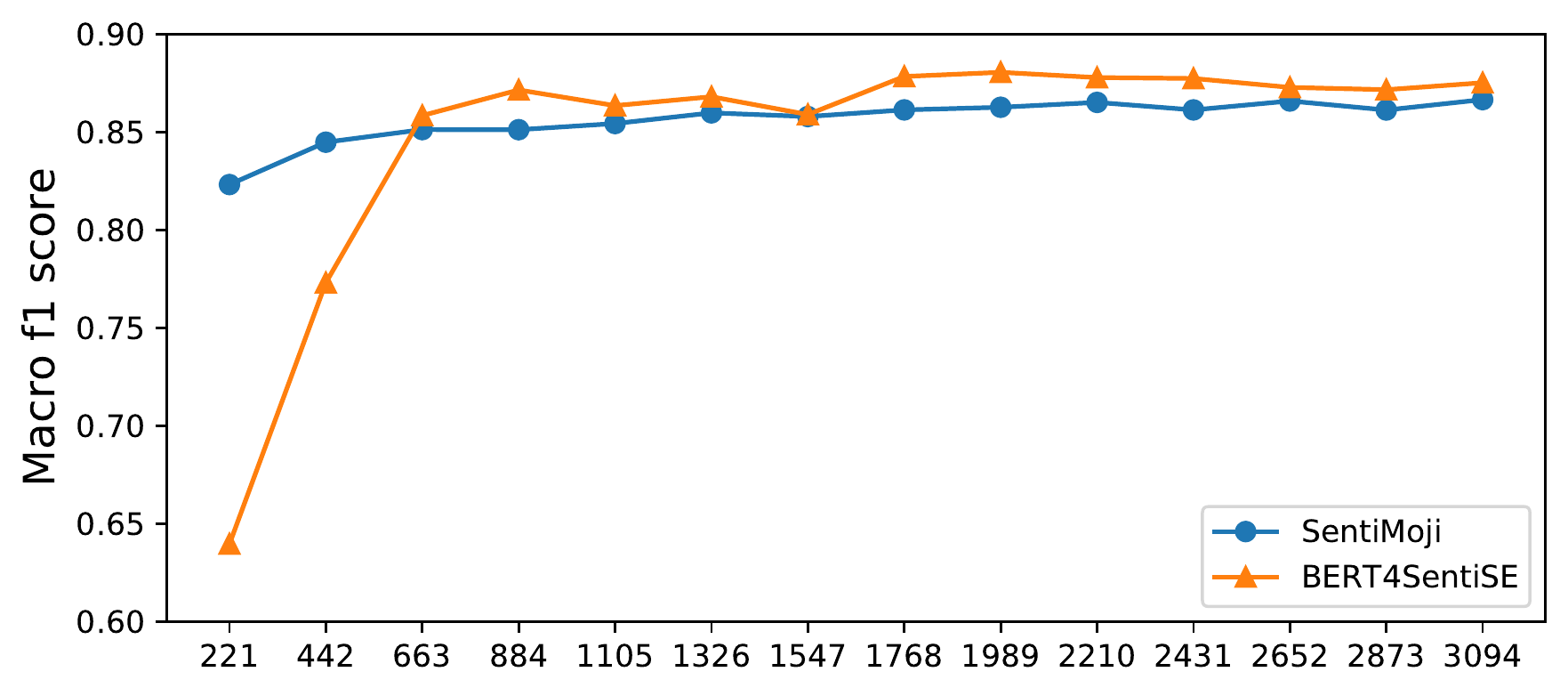}} \\
\subfloat[Jira]{\includegraphics[scale=0.5]{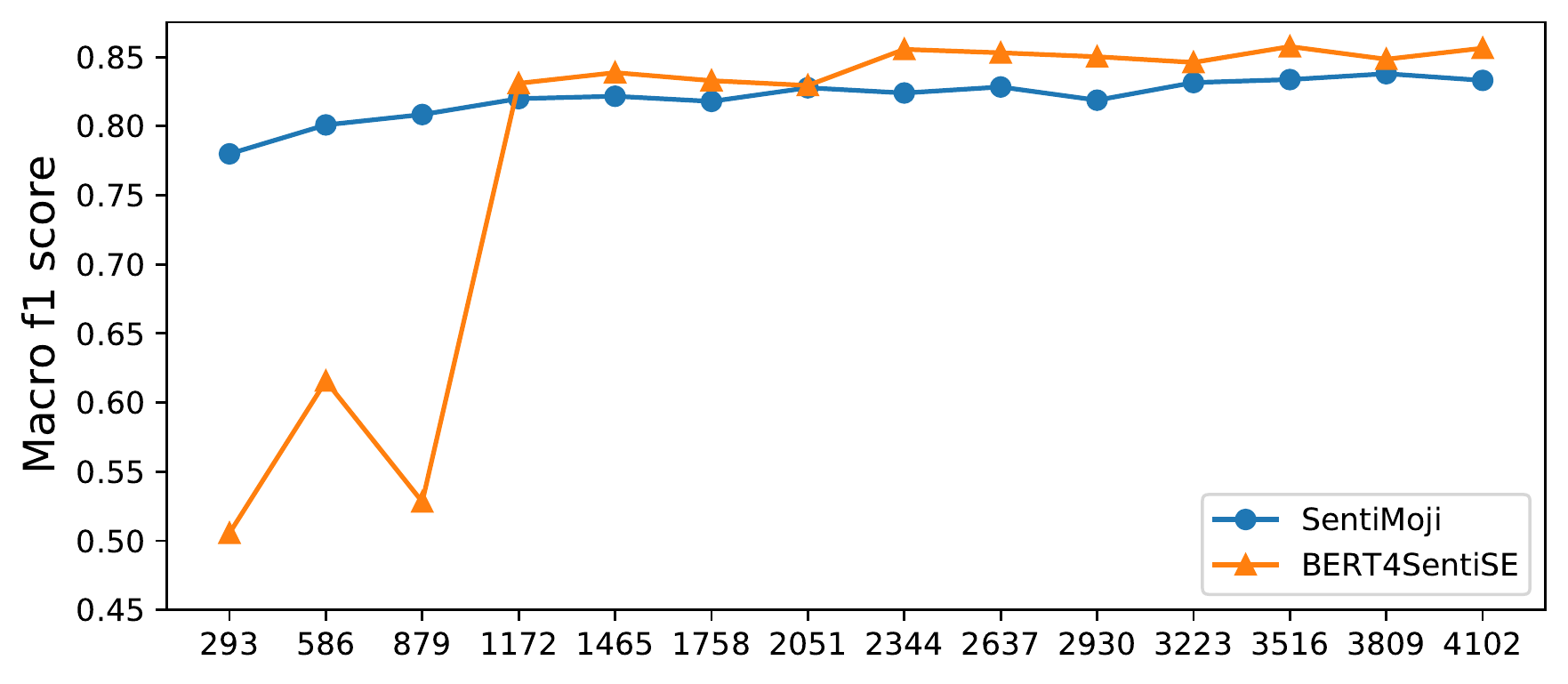}} \\
\caption{Learning Curves of SentiMoji \& BERT4SentiSE for within platform settings on (a) Github, (b) Stack Overflow, and (C) Jira}
\label{fig:learningCurve}
\end{figure} Our performance analysis of the tools finds that the deep learning SD tool BERT4SentiSE is the best performer across all three datasets in within platform settings. It is also the best among the supervised SD tools for cross-platform settings. Another deep learning SD tool, SentiMoji, shows the second-best performance in within platform settings across the datasets. SentiMoji also outperforms the shallow learning SD tools in the majority of cross-platform settings. These two deep learning SD tools are suitable for deployment in wide-range datasets, especially when the tools can be retrained in each dataset. It is thus important to know how much training data they need to provide reasonably well performance. In general, deep learning tools need more data than shallow learning tools for better performance. However, given that both tools use pre-trained embedding, the tools may not need a large amount of training data. In \fig\ref{fig:learningCurve}, we show three charts, one each for the three datasets. For each chart, we show how the tools perform as we increase the size of our training data. In order to generate the learning curves, we test each tool with the same held-out test set of 30\% samples after training it by incrementally using more training samples. At first, we sample 5\% of the training set using stratified sampling, which preserves the class distributions, and then we increment the training set by 5\% and test the models using the already held out test set. A similar method was used by Novielli et al.~\cite{Novielli-SEToolCrossPlatform-MSR2020} to plot the learning curves of Senti4SD and SentiCR in the same three datasets.

Across the three datasets, SentiMoji shows better performance (Macro F1-score) while using 5-10\% of the training data. However, after that, BERT4SentiSE starts to outperform SentiMoji in two datasets (GitHub and SO). For the other dataset (Jira), BERT4SentiSE starts to outperform after using around 20-25\% of the training data. Both tools show more than 0.80-0.85 Macro F1-score by using 20-25\% of the training data. The increase in the performance in the tools is less rapid after that. The tools show the best F1-scores of above 0.9 for GitHub and 0.85-0.88 for SO and Jira after they are trained on 30-40\% of the data, after which the increase in performance is negligible. This means that the two tools are sensible options in real-world scenarios where creating a large-scale benchmark data may not be possible, but a smaller size (e.g., 500 - 1500 labeled sentences) might not be challenging.




\subsection{Distribution of Error Categories}\label{sec:error-cat-dist}
In \tbl\ref{tab:misclassification}, we show the distribution of the six error categories that we reported in RQ$_3$ across the three datasets (GitHub, SO, and Jira). Recall that the sample size was 400 sentences for which both of the two best performing deep learning SD tools (BERT4SentiSE and SentiMoji) were wrong. This is a random sample. We see almost a similar proportion of the misclassified cases across the datasets based on the actual representation in the datasets. Overall, most misclassified cases are found in the Jira dataset, where the two tools performed the worst. 
\begin{table}[t]
  \centering
  \caption{Distribution of error categories in the 400 manually analyzed sample across the three datasets for tools BERT4SentiSE and SentiMoji in cross-platform settings}
    \begin{tabular}{lccc}
    \toprule{}
    \textbf{Category} & \textbf{GitHub} & \textbf{SO} & \textbf{Jira}\\
    \midrule

    Subjectivity in Annotation & 59 & 76 & 44  \\
    General Error & 11 & 34 & 67 \\
    Implicit Sentiment & 4 & 13 & 49 \\
    Context/Pragmatics & 23 & 16 & 29 \\
    Domain Specific & 14 & 16 & 38 \\
    Politeness & 18 & 23 & 8 \\
    Figurative language &  9 & 6 & 11 \\
    \midrule
    Sum &  138 & 184 & 246 \\

    \bottomrule
    \end{tabular}%
  \label{tab:misclassification}%
\end{table}%
Among the seven error categories, one category was observed the most when the tools were trained on the GitHub dataset: Subjectivity in Annotation. This error was found the most in the Jira dataset in cross-platform settings, i.e., the Jira dataset has inconsistencies in manual labeling (RQ$_3$). One category was observed the most when the tools were trained on the SO dataset: Politeness. This is because we observe fewer sentences relate to politeness in the SO dataset. As such, the tools fail to learn much about the politeness-related polarities, although both GitHub and Jira contain much more non-neutral sentences with politeness.

\begin{figure}[h]
\centering
\includegraphics[scale=0.6]{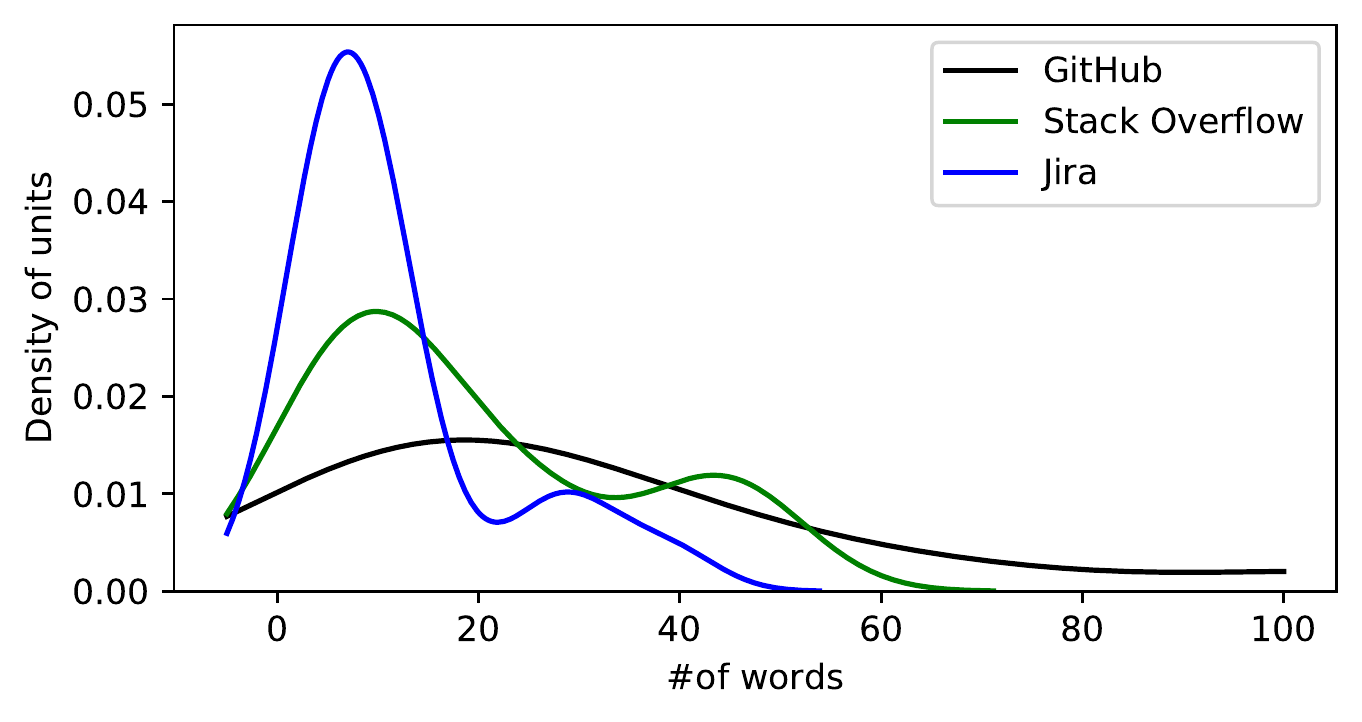}
\caption{Distribution of word counts per sentence}
\label{fig:word_counts}
\end{figure}

The other seven categories are observed the most when the tools were trained on the Jira dataset: \begin{inparaenum}
\item General Error, 
\item Implicit Sentiment, 
\item Context/Pragmatics, 
\item Domain Specificity, and 
\item Figurative Language.
\end{inparaenum} There are several reasons for the higher concentration of error categories. First, the Jira dataset does not have many sentences with emojis compared to the other two datasets, which means the tools could not learn notations in words related to emojis properly. Second, the Jira dataset also has inconsistencies in labeling emojis, as we noted in RQ$_3$, which could confuse the tools during learning and then during their prediction in cross-platform settings. Third, most of the sentences in Jira are shorter than the sentences in the other two datasets, thereby offering fewer contexts during the learning of the tools. In \fig\ref{fig:word_counts}, we show the percentage of sentences in each dataset with a number of words. For example, we have around 6\% sentences with less than five words in the Jira dataset, while in the SO dataset, we have around 2\% sentences with less than five words. This is why we see more errors in the tools related to context/pragmatics while those were trained using the Jira dataset. Therefore, the Jira dataset could benefit from more consistent labeling and richer data points (e.g., longer sentences).


\subsection{The \it{Issues} with Polarity for Politeness}\label{sec:issue-with-polarity}
As we noted in RQ$_3$, the sentences containing politeness are not consistently labeled in the datasets. For example, in the Jira dataset, we showed that similar sentences are labeled as neutral or positive in different places. The presence of politeness seems to have confused the human coders of the three datasets, making the supervised tools error-prone while predicting the labels of the sentences with politeness. To investigate the extent of this problem, we conducted a simple experiment by first picking a sample of sentences with politeness and determining how many of those got misclassified across the datasets in cross-platform settings. First, we find a sample of sentences with politeness with lexical cues like searching for three keywords in the sentences: ``thanks'', ``thank you'', and ``sorry''. In \tbl\ref{tab:politeness_misclassification}, we show the distribution of such sentences in each dataset. Across the three datasets, there are around 2,700 sentences containing politeness (i.e., 15.5\% of all sentences in the three datasets). Out of the three datasets, Jira has the highest number of politeness, around 1,386 (24\%), followed by GitHub with 928 sentences (13\%) and SO with 430 sentences (10\%). 
The last three columns in \tbl\ref{tab:politeness_misclassification} show the percentage of sentences with politeness in each dataset that got misclassified in cross-platform settings while we used the best performing deep learning SD tool in our study, i.e., BERT4SentiSE. For example, when the tool was trained and tested on Jira (i.e., within platform setting), 37.5\% of the sentences with politeness were misclassified. However, while the tool was trained on Jira but tested on GitHub, the only 4.1\% of sentences with politeness in Github were misclassified and only 0.8\% were misclassified for SO dataset. In fact, the difference between the ratios of misclassified cases in the Jira dataset from the other datasets is quite large. When BERT4SentiSE is trained on GitHub and tested on Jira, 36.2\% of the sentences with politeness in Jira remain misclassified. When the tool is trained in SO and tested on Jira, 36.3\% of the sentences remain misclassified. The analysis results confirm our previous observation that to improve the cross-platform performance of the tools in our three studied datasets; we could benefit from reducing inconsistencies in the labeling of politeness in the Jira dataset.


\begin{table}[t]
\newcolumntype{C}{>{\centering\arraybackslash}p{4em}}
  \centering
   \caption{Distribution of sentences with politeness and their misclassifications by BERT4SEntiSE and SEntiMoji in cross-platform settings across the datasets}
    \begin{tabular}{lcc|CCC}
    \toprule
    \multirow{2}{1.2cm}{\bf{Trained on}} & \multicolumn{2}{c}{\textbf{Politeness}} & \multicolumn{3}{c}{\bf{\% of Misclassified while tested on}}  \\ \cmidrule{2-6}
     & \# & \% & Jira & GitHub & SO \\ 
    \midrule
    Jira       & 1,386 & 24\% & 37.5\% & 4.1\% &  0.8\% \\
    GitHub       & 928 & 13\% & 36.2\% &  9.3\% &  1.5\%\\
    SO  &         430 & 10\% &  36.3\% &  6.0\% &  0.8\% \\

    
    \bottomrule
    \end{tabular}%
  \label{tab:politeness_misclassification}%
\end{table}%






\subsection{The Superiority of SentistrengthSE in Cross-Platform Settings}\label{sec:superiority-sentistrengthse}
\rev{In RQ2 (see \sec\ref{sec:rq-cross-platform-performance}), we find that the rule-based SD tool for SE SentistrengthSE outperforms all the other 
supervised shallow and DL tools for SE in five out of the six cross-platform settings, narrowly beating the deep learning SD tool Bert4SentiSE in four out of the five 
settings. The finding motivated us to do the misclassification analysis of the tools in RQ3 (see \sec\ref{sec:rq-cross-platform-error}). In \fig\ref{fig:misclassification_chart_dl}, 
we show the distribution of 400 sentences where both BERT4SentiSE and SentiMoji were wrong. In \fig\ref{fig:misclassification_sentistrengthse}, we show the distribution 
of 93 sentences where SentistrengthSE was wrong. For the deep learning SD tools, the most frequent error category is `Subjectivity in Annotation' (29\% of cases). For SentistrengthSE, 
the most frequent error category is `Domain Specificity' (39\% of cases). Intuitively, this finding makes sense in that it is not possible for a rule-based tool like SentistrengthSE to 
learn about particular domain specific insight from a dataset, given it cannot be retrained on the dataset. A supervised SD tool like BERT4SentiSE can learn about 
domain specificity, when it is trained on a dataset. Given that our cross-platform datasets are considerably similar (see \sec\ref{sec:rq-cross-platform-similarity}), the 
supervised SD tools are better positioned than the rule-based SD tool to handle such domain specificity both in cross- and within-platform settings. 
For example, consider this sentence in the Jira dataset \emt{Bug Flavio more? Seriously}. 
The manual label is `negative', which is correct given that the underlying context is about 
the presence of `more bug' in the system. Supervised tools like BERT4SentiSE were correct to label the sentence as 
`negative'. However, SentistrengthSE labels it as neutral because 
it is unable to underlying context. However, while such training can help the supervised SD tools to better handle domain specificity, it can also become a problem for such supervised tools if the datasets 
have problems with the manual labeling that could lead to the error `Subjectivity in Annotation'. Inconsistency in manual labeling in the dataset makes a supervised 
SD tool confused, because what it learns during training on a given sentence (e.g., `thank you') may not be applicable in the test dataset, if the same/similar 
sentences are labeled as differently between the train and test datasets. For example, in \sec\ref{sec:issue-with-polarity}, we provided examples of sentences 
containing politeness and showed evidence of how they are inconsistently labeled across the datasets. For a rule-based SD tool like SentistrengthSE, such 
inconsistencies can also result in some misclassification like when a manual label deviates from the rules. One scenario would be when 
the tool always considers `thank you' should be labeled as positive, while the human coders sometimes labeled it positive and some other times neutral. However, 
subjectivity in annotation can extend beyond the cases of politeness and thus can be more detrimental to the supervised SD tools, given the tools are trained and tested 
on the datasets and thus are more exposed to the underlying nuances.}

\section{Implications of Findings}\label{sec:implications}
The findings from this study can guide the following stakeholders in software engineering (SE):
\begin{inparaenum}[(1)]
\item \bf{SE Data Scientists} in the software engineering team to determine the feasibility of an SD tool for SE artifacts across diverse SE data repositories,
\item \bf{SE Managers} to determine the best possible SD tool for SE artifacts to learn about organizational sentiments from SE data repositories,
\item \bf{SE Architects} to design an SE pipeline that can include SE-specific SD tools to analyze SE data repositories,
\item \bf{SE Researchers} to improve the performance of the SE-specific SD tools from diverse SE data sources (i.e., cross-platform settings). 
\end{inparaenum} We discuss the implications below by referring to our results from \sec\ref{sec:result} and \sec\ref{sec:discussion}.

\nd\bf{\ul{$\bullet$ SE Data Scientists.}} Technology companies like Microsoft are now including data scientists in almost each SE team~\cite{amershi_se_case_study_2019}. 
The data scientists can work on developing SD tools to analyze the diverse SE data repositories in an organization. Indeed, software analytics using 
sentiment detection has become an important field to support diverse development tasks like software library selection, bug fixing, and so on~\cite{Uddin-OpinionValue-TSE2019,Uddin-OpinerEval-ASE2017}. 
The data scientists in an SE can use the findings from our study to decide whether and how deep learning SD tools can be used in cross-platform settings. In particular, the data scientists can benefit from the following insights from our study:

\begin{inparaenum}[(1)]
\item \ib{Dataset similarity vs. class imbalance problem.} Our analysis in RQ3 finds that there is a statistically significant positive correlation between the similarity of any two datasets in our study and the performance of the two best performing deep learning SD tools (BERT4SentiSE and SentiMoji) in the two datasets in cross-platform settings. We further validate this assumption in \sec\ref{sec:sim-vs-acc-twitter}, where 
we show that for a non-SE dataset like Twitter, the similarity value as well performance of the tools drop 
when we use the dataset to train a tool and then test it on our SE datasets. However, as we noted in RQ$_3$, this observation fails to explain the drop in performance in the tools when they are trained on the Jira 
dataset and then tested on the other two datasets. As we noted in RQ3, the Jira dataset is considerably imbalanced for the two 
polarity classes (positive, negative) compared to the other two datasets. Therefore, data scientists in an SE team can 
use the above two insights as follows. \begin{inparaenum}[(a)]
\item Make the SD datasets more balanced across the three polarity classes, and 
\item The more similar two SE datasets are, the less drop in performance we can expect in the two tools (BERT4SentiSE, SentiMoji) in cross-platform settings.
\end{inparaenum}

\item \ib{Domain-specificity vs cross-platform performance drop.} The rule-based tool SentistrengthSE outperforms 
the best performing supervised SD tool BERT4SentiSE in five out of the six cross-platform settings. Four of the settings involve 
the Jira dataset, e.g., using Jira to train a tool and then test it on other two datasets or use other two datasets to train a tool and then 
test it on the Jira dataset. As we noted above, the Jira dataset has a class imbalance problem, which means that the supervised tool 
may not get enough information of the imbalanced classes when it is trained on the Jira dataset. Our analysis of the dataset shows that there in the Jira dataset, there is
around less than 5\% examples with less than five words. These short sentences
sometimes do not contain enough contextual information. These two shortcomings (i.e., class imbalance and short sentences) in the Jira dataset 
contributed to the inferior performance of the two supervised SD tools when trained on Jira and tested on other datasets. Therefore as discussed in Section (\ref{sec:rq-cross-platform-similarity}) despite configuring the SD tools with the best performing configuration (model settings, hyperparameters, pre-processing guideline) as mentioned by the authors for within-platform settings, the tools made around 94\% to 99\% of its errors by predicting neutral for non-neutral sentences when they were trained on Jira and tested on non-Jira datasets. Moreover, the Jira dataset has the also the most number of errors related to `General Error' among all the datasets (see \sec\ref{sec:error-cat-dist}) for the two deep learning tools BERT4SentiSE and 
SentiMoji. This means that the two tools failed to properly process the Jira dataset during the testing of the dataset. The rule-based 
tool SentistrengthSE is developed by creating rules based on the Jira dataset created by Ortu et
    al.~\cite{Ortu-EmotionalSideJira-MSR2016}. Therefore, SentistrengthSE can also suffer from a lack of understanding of underlying contexts in cross-platform settings. 
    Indeed, 
    our error analysis in RQ4 finds that the most number of errors in SentistrengthSE in the three datasets is due to the inability of the tool 
    to determine the underlying contexts in the data (i.e., domain specificity). Therefore, data scientists in SE need to investigate further 
    SE-specific SD tools 
    before we claim the superiority of 
    a rule-based tool like SentistrengthSE in cross-platform settings.

    \item \ib{Pre-trained transformer vs. Recurrent neural network architecture in deep learning SD tools.} 
    \rev{In our study, we investigated three deep learning SD tools for SE: BERT4SentiSE~\cite{Biswas-ReliableSentiSEBERT-ICSME2020}, SentiMoji~\cite{Chen-SentiEmoji-FSE2019}, and RNN4SentiSE~\cite{Biswas-SentiSEWordEmbedding-MSR2019}. 
    These three tools are selected, because they were found to be the best performing deep learning SD tools for SE during the time of our analysis. 
    While BERT4SentiSE is the best performer among the supervised SD tools in our three datasets, 
    both within and cross-platform settings, another deep learning SD tool, RNN4SentiSE, is the worst performer among all tools. Our 
    analysis of the data and the RNN4SentiSE show that this is not due to any specific obvious shortcomings in the tool itself. 
    Instead, this is due to how the tool is designed and used. For example, BERT4SentiSE uses the pre-trained transformer architecture of BERT. 
    SentiMoji uses the Bi-LSTM architecture and is also pre-trained on a large corpus of Twitter and other emoji data. RNN4SentiSE 
    uses simple RNN architecture, and it is not pre-trained either. Indeed, in previous studies of within-platform settings, BERT4SentiSE was found to outperform 
    RNN4SentiSE~\cite{Biswas-ReliableSentiSEBERT-ICSME2020}. As such, it is intuitive that BERT4SentiSE outperformed RNN4SentiSE in cross-platform settings as well - 
    having observed otherwise would have raised questions about underlying model, data, and even the study methodology. 
    Indeed, indeed success of pre-trained models over other domains is reported 
    across multiple textual classification tasks. As we noted in RQ2, the best performing supervised SD tool in our study, BERT4SentiSE, also 
    shows an almost negligible difference in performance (0.01 - 0.03) from the best performing rule-based tool SentistrengthSE, while BERT4SentiSE 
    outperforming SentistrengthSE considerably for within platform settings. Therefore, SE data scientists can 
    focus on improving SD tools like BERT4SentiSE, like increasing the number of hidden layers, assessing different loss/optimization parameters, etc.}   
    
%

\end{inparaenum}

\nd\bf{\ul{$\bullet$ SE Architects.}} Given the architects in an SE are responsible for designing a software system, they 
can use the findings from our study to decide on the appropriate tool or combination of tools into a Machine Learning (ML) software pipeline 
that supports the analysis of sentiment polarity in diverse SE datasets. In particular, the following findings can assist them in the decision-making process.  

\begin{inparaenum}[(1)]
\item \ib{The trade-off between the best and a good enough SE-specific SD tool across within and cross-platform settings.} The embedding of an ML pipeline 
into a software development life-cycle can be challenging and complicated due to the constant need of 
updating the model and the data used to train and test the model~\cite{Alamin-MLSASDLCQA-TCAS2021}. 
For an SE architect tasked with embedding a SE-specific SD tool into a software system can be 
challenging given the diversity of data in software repositories and the drop in performance 
of the SE-specific SD tools in cross-platform settings as observed by our study (RQ1-RQ2). 
This means that an SE architect will have to consider different combinations of the tools depending on the 
type of data. Such a setup can become more difficult when each tool needs to be updated with new data to ensure 
the tool does not drift and decay over time. In such cases, a good enough tool across all platforms might be desirable, especially 
if the tool does not show a drastic drop in performance in cross-platform settings. The best performing deep learning 
SD tool BERT4SentiSE outperforms other tools across the 
within platform settings. The tool also is the second-best performer by a narrow margin, behind only the rule-based tool SentistrengthSE, 
across the cross-platform settings. Therefore, BERT4SentiSE could be the best choice for an SE architect 
to use it in a software development life-cycle where the sentiment polarity of diverse SE datasets needs to be analyzed in a production 
environment.    

\item \ib{Learning curve vs. data necessity.} The success of supervised SD tools depends largely on the 
volume and quality of data. Intuitively, the more data and representative data we have, the better we can 
expect the supervised SD tool can be to detect sentiment polarity labels. The use of transfer learning principles into 
the pre-trained deep learning SD tools can reduce the reliance on large volume of data, as long as the diverse 
platforms also have vocabularies/patterns learned from domain-neutral data. Indeed, as we find in \sec\ref{subsec:learning_curve}, 
the two best performing pre-trained deep learning SD tools in our study (BERT4SentiSE, SentiMoji) can achieve the best performance for within platform 
settings just by training on 30\% of all training data. Given data collection and maintenance can complicate a software development 
life-cycle greatly, SE architects can benefit from using tools like BERT4SentiSE and SentiMoji in SE-specific software products.
\end{inparaenum}

\nd\bf{\ul{$\bullet$ SE Researchers.}} We have come a long way towards having no SE-specific SD tools in 2016 to having several high-performing 
SE-specific SD tools that can outperform cross-domain SD tools in diverse SE datasets. However, the drop in performance of the SE-specific SD tools 
in cross-platform settings warrants the need for further research in this area to improve the SE-specific SD tools in cross-platform settings. 
In particular, our study findings can assist SE researchers working on this area as follows.   

\begin{inparaenum}[(1)]
    \item \ib{Needs for standard guidelines on the creation of sentiment
    polarity benchmarks.} \rev{In this study, we used the same three polarity benchmarks (i.e., datasets) that 
    were used by Novielli et al.~\cite{Novielli-SEToolCrossPlatform-MSR2020,Novielli-SESentimentCP-EMSE2021} in their investigation 
    of shallow learning SD tools in cross-platform settings. We decided to use the same three benchmarks, so that we could compare our study findings with 
    those from Novielli et al.~\cite{Novielli-SEToolCrossPlatform-MSR2020,Novielli-SESentimentCP-EMSE2021}. Besides the study of Novielli et al.~\cite{Novielli-SEToolCrossPlatform-MSR2020,Novielli-SESentimentCP-EMSE2021} and 
    our study, the three datasets were used in multiple studies and published papers so far in SE~\cite{Biswas-ReliableSentiSEBERT-ICSME2020,Biswas-SentiSEWordEmbedding-MSR2019,Novielli-BenchmarkStudySentiSE-MSR2018}.
    Our manual analysis and comparison of error categories in the misclassified cases 
    by the two best performing deep learning SD tools in cross-platform setting shows that  
    subjective annotation in manual labeling was the major reason for misclassification by the two tools. This 
    happened because similar sentences were labeled differently by human coders in a given dataset. This means that 
    the human coders were not instructed on specific annotation guidelines. Subjectivity in annotation also happened 
    when similar sentences between two datasets are labeled differently by human coders in the two datasets. This means 
    that two datasets were created by following two different sentiment polarity annotation guidelines. Therefore, a tool trained in 
    one dataset can be wrong in another dataset when it tried to label a similar sentence based on the training instances. 
    Therefore, future SD research in SE should focus on the standardization of the guideline for annotation. Based on our findings, 
    we also call for fixing current datasets so that they all can follow similar annotation guidelines.}

\begin{figure}[h]
\centering
\includegraphics[scale=0.55]{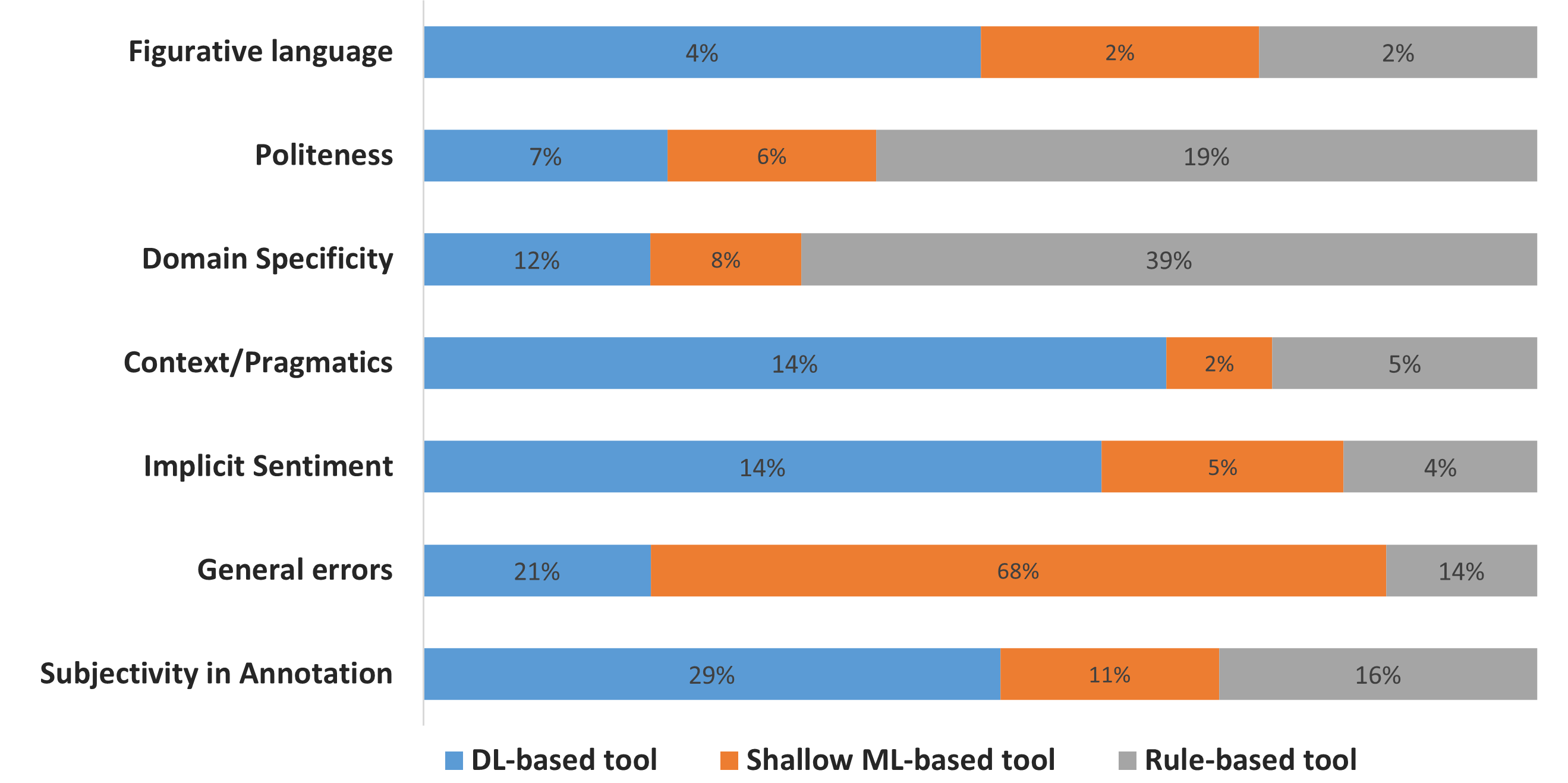}
\caption{Comparison of misprediction categories of DL-based vs shallow ML-based vs Rule-based SD tools.}
\label{fig:distribution_error_complement}
\vspace{-5mm}
\end{figure}
\item \ib{Finding proper ensemble tool to utilize the strengths of the different SD tools.} Our error analysis in RQ4 shows that 
the distribution of error categories differs depending on the type of the SD tools (i.e., deep learning vs shallow learning vs rule-based). 
In \fig\ref{fig:distribution_error_complement}, we summarize the distribution of error categories (in percentages) from RQ4 across the three types of 
tools we SD tools we studied in our six cross-platform settings. The summary is created by taking the data from the three pie charts in 
\figs\ref{fig:misclassification_chart_dl}, \ref{fig:misclassification_sentisdsenticr}, and \ref{fig:misclassification_sentistrengthse}. The deep learning SD 
tools (BERT4SentiSE, SentiMoji) show most errors due to subjectivity in manual annotation, the shallow learning SD tools (Senti4SD, SentiCR) show most errors due to general errors, while 
the rule-based SD tool (SentistrengthSE) shows the most amount error due to lack of understanding of domain-specific contexts. Intuitively, the three types of 
tools thus can complement each other for a given dataset. This was the motivation behind our RQ5, where we tried seven different ensembles of the tools 
based on the principles of majority voting. However, the majority-based ensemble tool outperformed the best performing stand-alone SD tools in only three out of 
six cross-platform settings, and that too in a narrow margin. Therefore, future research in this area can investigate ensemble techniques better than 
simply taking the majority of the voting.
    
    \item \ib{The need for `more' transfer learning.} From our experiments, we see
    the pre-trained deep learning SD tool BERT4SentiSE performs better than shallow
    ML-based tools SentiCR and Senti4SD. Rule-based tool SentiStrengthSE
    outperforms it by a little margin. BERT4SentiSE uses the BERT architecture, which is based on 
    the principles of transfer learning. This means the BERT4SentiSE benefits from the learning of domain-neutral 
    word embedding in BERT from large corpus data (e.g., Wikipedia, English novels, etc.). Indeed, transfer learning 
    has shown promise to generalize supervised classification models across multiple domains and to make those models 
    more efficient for a given domain with a little amount of training data from the given domain. Indeed, in \sec\ref{subsec:learning_curve}, 
    we find that BERT4SentiSE can reach the optimal performance level for a given dataset by simply using around 30\% of all training data from that 
    dataset. Transfer learning is a constantly evolving paradigm in natural language processing with new techniques being developed 
    rapidly (e.g., RoBERTa\cite{liu2019roberta}, XLNet\cite{Yang-Xlnet-Arxiv2020}, Multi-lingual LASER models, etc.). Therefore, SE research in SD tools can benefit from creating more 
    SD tools for SE based on transfer learning techniques. 
    
\end{inparaenum}

\nd\bf{\ul{$\bullet$ SE Managers.}} Managers in SE teams and organizations need to assess the productivity and mental health of the fellow employees. 
Automated analysis of sentiments expressed by the employees in the SE repositories can be helpful to the manager to get a real-time and comprehensive as well as organization-wide understanding of the employees. Our findings from this study can guide the SE managers while using the SE-specific SD tools for such analysis as follows.  

\begin{inparaenum}[(1)]
    \item \ib{Assessment of developer productivity using sentiment analysis in cross-platform settings.} 
The productivity in a team may depend on developers' sentiments in/of their
diverse development
activities~\cite{Ortu-AreBulliesMoreProductive-MSR205,Guzman-EmotionalAwareness-FSE2013,Pletea-SecurityEmotionSE-MSR2014,Mika-MiningValenceBurnout-MSR2016}. 
Therefore, SE managers in an organization can utilize the sentiment polarity analysis in the diverse SE repositories of the organization to 
learn about the mental health, productivity, and team cohesion in the organization. However, as we have seen in this study, SE datasets can 
be diverse, and an SD tool trained in one dataset suffers a drop in performance when it is tested on another dataset. 
Therefore, SE managers should train and test their SD tool for SE in a given platform instead of applying the SD tool in cross-platform settings. 
Otherwise, the findings obtained may not give an accurate picture of the sentiments across the organization.

    \item \ib{The trade-off between majority voting vs. stand-along tool.} Our error analysis in RQ4 and in \fig\ref{fig:distribution_error_complement} show that 
    the SD tools show different errors in cross-platform settings depending on their types (e.g., deep learning vs shallow learning vs rule-based). However, 
    a majority-based ensemble could not outperform the stand-alone best performing SD tools in 50\% of the cross-platform settings. Therefore, SE managers may refrain from using a majority-based ensemble tool to mine sentiment polarity from SE repositories. 
\end{inparaenum}
\section{Threats to Validity}\label{sec:threats_to_validity}
 In this Section we discuss the threats to validity of our study following the guideline for empirical studies \cite{wohlin2012experimentation}.

\nd\bf{Construct Validity} threats relate to potential errors in the research methodology. In this study, we used the same three benchmark datasets used by Novielli et al. ~\cite{Novielli-SEToolCrossPlatform-MSR2020} and by others for similar studies\cite{Chen-SentiEmoji-FSE2019, Lin-PatternBasedOpinionMining-ICSE2019}. During our error analysis of the deep learning SD tools, we followed the similar guideline of related studies~\cite{Novielli-SEToolCrossPlatform-MSR2020, opiner_2019}. We also observed the same error categories that were previously reported by Novielli et al.~\cite{Novielli-SEToolCrossPlatform-MSR2020}. Another issue can be reporting the performance that may not be standard. In order to mitigate this threat, following standard reporting practices in this domain, we report both Macro and micro scores and Cohen Kappa agreement values.

\nd\bf{Internal Validity} threats concern about the authors' bias while conducting the study. It refers to errors that can occur due to not using optimal configuration parameters for sentiment detection tools. For each tool in our study, we used the code provided by the authors to run and fine-tune their performance (e.g., hyper-parameter tuning). In within platform settings, we adopted 10-fold cross-validation settings and trained and tested each tool ten times to mitigate variance of the model's prediction. In order to mitigate the coding errors, we used widely used open-source libraries (e.g., SciKit-learn ~\cite{website:python-scikit-learn}) to compute the metrics presented in this study. Another area of bias can be the human bias while categorizing the misclassification of the SD tools. Two authors discussed and participated in the error categorization process to reduce human biases. We also share our manual categorization with our replication package.

\nd\bf{External Validity} threats refer to the generalizability of the findings. For this study, we used three datasets that represent all the available datasets for sentiment detection for the software engineering domain and used in similar studies.  We used datasets of around 17K units from three popular and widely used collaboration platforms: question-answering (Stack Overflow), collaborative development platform (GitHub), issue tracking platform (Jira). Transposing the results to another new SE platform may introduce new insights regarding the cross-platform adoption of the tools. In summary, our study shows promising results for PTM DL-based tools for sentiment detection tasks because compared to other tools, they show promising results to understand and generalize SE-domain context. 



\section{Related Work}\label{sec:related_work}

\rev{Our study in this paper takes motivation from the recent studies by
Novielli et al.~\cite{Novielli-SEToolCrossPlatform-MSR2020,Novielli-SESentimentCP-EMSE2021}.  Novielli et al.~\cite{Novielli-SEToolCrossPlatform-MSR2020,Novielli-SESentimentCP-EMSE2021} studied 
the effectiveness of shallow learning
SD tools in cross-platform settings. We extend this body of SE knowledge by
analyzing the effectiveness of the three most recent deep learning SD tools for
SE in cross-platform settings. Unlike Novielli et al.~\cite{Novielli-SEToolCrossPlatform-MSR2020,Novielli-SESentimentCP-EMSE2021}, we also examine the 
benefits of creating ensembles of the different SD tools for SE in cross-platform settings. 
In \sec\ref{sec:introduction}, we summarized that an in-depth and systematic study of the effectiveness of the deep learning SD tools for SE in cross-platform 
  settings is warranted due to the following reasons: \begin{inparaenum}
  \item Consistency in superiority of performance. We wanted to know whether deep learning SD tools outperform shallow learning SD tools in cross-platform settings as they do for within platform settings.
  \item Pre-training setup of the models. We assumed that the pre-training setup in deep learning models like BERT4SentiSE could show that the model could outperform all other models in both cross-platform settings as it does for within-platform settings.
  \end{inparaenum} 
  Our overarching goal was to find a model that could offer reliable performance both in cross and within-platform settings, so that we do not have to train the model for each 
  platform to get reliable sentiment labels. This goal is further motivated due to the findings from Novielli et al.~\cite{Novielli-SEToolCrossPlatform-MSR2020} that shallow learning 
  SD tools show a considerable drop in performance in cross-platform settings, and as such are not as reliable for cross-platform settings as they are for within-platform settings.
  Therefore, similar to Novielli et al.~\cite{Novielli-SEToolCrossPlatform-MSR2020}, we use the same three benchmark datasets. In addition, we pick the 
  best performing shallow learning SD tools and unsupervised SD tools from Novielli et al.~\cite{Novielli-SEToolCrossPlatform-MSR2020} for comparison with our three deep learning SD 
  tools for SE. Our findings through our four research questions offer new insights over the findings of Novielli et al.~\cite{Novielli-SEToolCrossPlatform-MSR2020} as follows.
\begin{itemize}[leftmargin=10pt]
  \item The deep learning SD tools like BERT4SentiSE show a slightly lower drop in performance in cross-platform settings than the shallow learning SD tools. Therefore, 
  the deep learning SD tools can be more reliable than the shallow learning SD tools in cross-platform settings, when it is not possible to train the tools in a new platform.
  \item Overall, the deep learning SD tool BERT4SentiSE is the best performer in five out of the six cross-platform settings, when compared against shallow learning SD tools. However, 
  the tool is the second best performer to the unsupervised SD tool SentistrengthSE in only two out of the six cross-platform settings. This finding is consistent with Novielli et al.~\cite{Novielli-SEToolCrossPlatform-MSR2020}, 
  who also reported the superiority of SentistrengthSE over the shallow learning SD tools. However, in their case SentistrengthSE outperformed the shallow learning SD tools in all of the six cross-platform settings.
  \item Our labeling of the misclassification cases by the tools finds that the deep learning tools mostly suffer from the subjectivity in annotation in the datasets for their 
  drop in performance, i.e., it is the problem with how inconsistently the sentiment polarities are labeled during the creation of the datasets. 
  This finding differs from the misclassification analysis of shallow learning SD tools by Novielli et al.~\cite{Novielli-SEToolCrossPlatform-MSR2020}, who reported the majority of 
  misclassified cases for the shallow learning SD tools are related to the presence of general errors like the inability of the tool to understand particular syntax (e.g., emoji) in the datasets. 
  This finding motivated us to see if the deep learning SD tools can complement the shallow learning SD tools in an ensemble setting, e.g., whether a majority voting of sentiment polarity 
  of a given sentence out of all the available tools may offer better performance than the individual tools.
  \item In RQ4, we report that a simple majority-based ensemble of the tools does not offer better performance than the individual tools in cross-platform setting. Novielli et al.~\cite{Novielli-SEToolCrossPlatform-MSR2020}
  did not study the effectiveness of an ensemble of the individual SD tools in cross-platform settings. 
  \item Unlike Novielli et al.~\cite{Novielli-SEToolCrossPlatform-MSR2020}, we also check whether the similarity between two datasets may be used to understand when/how an SD tool for SE could be more/less useful in cross-platform settings. We analyzed this systematically and with detailed results in  
  \sec\ref{sec:rq-cross-platform-similarity}. 
  We showed that the cross-platform datasets for SE do indeed have high degree of similarity between themselves, yet the SD tools suffer from performance drop in cross-platform settings. To further understand whether the drop in performance any correlation with the similarity between datasets could have, in  \sec\ref{sec:sim-vs-acc-twitter}, 
  we further checked the similarity of the three SE tools against a non-SE dataset and found that the SE datasets are indeed more similar to each other than to the non-SE datasets. The findings led us to conclude that the SE datasets have inconsistency among themselves in terms of how sentiment polarities are labeled, which then confuses the SD tools in cross-platform settings (because the tools cannot reliably apply knowledge gained from one dataset to another dataset).       
\item In \sec\ref{sec:error-cat-dist}, we show how the SD tools could have more issues while trained in one dataset (e.g., Jira) than others during cross-platform setup. 
In \sec\ref{sec:issue-with-polarity}, we show concrete evidence of how inconsistencies in SD labels could affect due to the presence of politeness in the texts (e.g., `thank you' was labeled as neutral in some and positive on others, etc.). These findings 
could be further investigated in future studies to determine how such SE datasets could be fixed/improved, so that SD tools could be better applied in cross-platform settings.  
\end{itemize}}

Other related works for sentiment analysis in SE research can broadly be divided into two categories: 
\bf{Studies} of sentiments in diverse SE repositories and by developing SE-specific SD \bf{Tools} for different SE datasets.

\bf{\ul{Studies.}} A significant number of studies are conducted to analyze the prevalence and impact of sentiments in SE repositories and development scenarios, such as IT tickets~\cite{Castaldi-SentimentAnalysisITSupport-MSR2016}, collaborative distributed team communications~\cite{Guzman-SentimentAnalysisGithub-MSR2014}, Twitter feeds~\cite{Guzman-NeedleInHaystackTwitterSoftware-RE2016},  commit logs~\cite{Sinha-DeveloperSentimentCommitLog-MSR2016}, security-related discussions in GitHub~\cite{Pletea-SecurityEmotionSE-MSR2014}, API reviews~\cite{Uddin-OpinionValue-TSE2019,Uddin-OpinerEval-ASE2017,Uddin-OpinerReviewToolDemo-ASE2017}, and code reviews~\cite{Ikram-SentimentCodeReview-IST2019}. Studies also combined sentiment analysis with other techniques to improve domain-specific classification, such as to classify app reviews~\cite{Maalej-AutomaticClassificationAppReviews-RE2016,Panichella-ClassifyAppUserReview-ICSME2015}. While sentiment analysis focuses on polarity, emotion analysis focuses on finer-grained expressions such as anger, love, etc. Emotion analysis is used to determine team cohesion, such as the impact of bullies in an SE team~\cite{Ortu-AreBulliesMoreProductive-MSR205}, the relation of VAD (Valence, Arousal, Dominance) scores~\cite{Warriner-VadLexicons-BRM2013} with productivity and burn-out in SE teams~\cite{Mika-MiningValenceBurnout-MSR2016}, and the usage of expressed emotions to prioritize improvements or raise team awareness~\cite{Guzman-EmotionalAwareness-FSE2013,Gachechiladze-AngerDirectionCollaborativeSE-ICSENIER2017}. Overall, analysis of sentiments and emotions is an increasingly popular field in SE~\cite{Novielli-IntroToSpIssueAffectSE-JSS2019,Novielli-SentimentEmotionSE-IEEESoftware2019}

\bf{\ul{Tools.}} SE-specific sentiment tools are needed, because off-the-shelf sentiment detection tools from other domains are not very accurate~\cite{Jongeling-SentimentAnalysisToolsSe-ICSME2015,Jongeling-SentimentNegative-EMSE2017}. Subsequently, a number of sentiment and emotion detectors and benchmarks were developed for SE~\cite{Islam-Deva-ACMSAC2018,Islam-SentistrengthSE-MSR2017, Lin-PatternBasedOpinionMining-ICSE2019,Uddin-OpinionValue-TSE2019,Calefato-Senti4SD-EMSE2017,Ahmed-SentiCRNIER-ASE2017,Murgia-DoDevelopersFeelEmotion-MSR2014,Ortu-EmotionalSideJira-MSR2016,Chen-SentiEmoji-FSE2019,Islam-Deva-ACMSAC2018,Calefato-EmoTxt-ACII2017,Maipradit-SentimentClassificationNGram-IEEESW2019}. Largely, sentiment detection efforts for SE can be divided into three categories: rule-based, shallow learning-based, and deep learning-based. In this paper, we studied the most recently introduced three deep learning SD tools in cross-platform settings and compared their performance against the best performing shallow learning and rule-based SD tools for SE.

In our recently published paper~\cite{Uddin2022SentimentSEHybrid}, we reported an empirical study of the ensemble of sentiment detection tools for software engineering. 
We studied both shallow learning and deep learning SD tools. We checked for a given dataset, where the ensemble of the tools 
can offer better performance. In this paper, as part of RQ4 (see \sec\ref{sec:cross-platform-agreement}), we also checked whether the 
ensemble of the SD tools for SE can offer better performance. However, in this paper we checked the performance of the tools in cross-platform 
settings, whereas in our published paper we checke the performance of the tools for within platform settings. In addition, 
the three deep learning SD tools studied in this paper (BERT4SentiSE, RNN4SentiSE, and SentiMoji) were not investigated in our 
previously published paper~\cite{Uddin2022SentimentSEHybrid}.  

\section{Conclusions}\label{sec:conclusion}
The diversity in SE platforms can make it challenging to deploy a supervised
SE-specific SD tool across all the platforms without training it in each
platform. Given that it is expensive and sometimes not possible to train an SD tool in each SE platform, it
can be useful to find an SD tool that can perform reasonably well in both within
and cross-platform settings. We studied three recently introduced deep learning
SD tools for SE in three SE benchmark datasets from three SE platforms. We find that BERT4SentiSE is the best
performer in within platform settings, while it is also the best among the
supervised SD tools for SE in cross-platform settings. BERT4SentiSE is slightly
outperformed by lexicon-based tool SentistrengthSE in cross-platform settings,
but BERT4SentiSE outperforms SentistrengthSE for within platform settings by a
large margin. The pre-trained deep learning-based tools show promise in understanding the software engineering context 
from a small training dataset. However, inconsistent annotation and lack of standard annotation guidelines for the 
benchmark dataset are hindering future improvement in this domain. Our future research will focus
on improving the SE datasets for SD (e.g., to fix the inconsistencies in polarity labeling).

\begin{small}
\bibliographystyle{abbrv}
\bibliography{consolidated}
\end{small}

\end{document}